\documentclass[a4paper,12pt]{article}
\usepackage{jheppub_modified_green}

\usepackage{epsfig}
\usepackage{amsfonts}
\usepackage{amssymb,esint}
\usepackage{enumitem}
\usepackage{amsthm}
\usepackage{subfig}
\usepackage{bm}
\usepackage{hyperref}
\usepackage{mathrsfs}
\usepackage{bbm}
\usepackage{cancel}
\usepackage{xcolor}
\usepackage{relsize}
\usepackage{float, slashed, graphicx, amssymb, amsmath}
\usepackage{shuffle}
\usepackage{pgfplots}
\usepackage{tikz-feynman}
\tikzfeynmanset{compat=1.1.0}
\tikzfeynmanset{graviton/.style={circle, draw=green!60, fill=green!5, very thick, minimum size=7mm}}
\tikzfeynmanset{codot/.style={/tikz/shape=circle,/tikz/fill=white,/tikz/minimum size=0.1cm,/tikz/inner sep=1.8pt}}
\tikzfeynmanset{myblob/.style={/tikz/shape=rectangle,/tikz/fill=red,/tikz/minimum size=0.2cm,/tikz/inner sep=1.8pt} }
\tikzfeynmanset{ghc/.style={/tikz/shape=circle,/tikz/fill=white,/tikz/minimum size=0.05cm,} }
\tikzfeynmanset{HV/.style={/tikz/shape=circle,/tikz/fill={rgb:black,1;white,2},/tikz/minimum size=0.1cm,} }
\tikzfeynmanset{GR/.style={/tikz/shape=ellipse,/tikz/fill={rgb:black,1;white,2},/tikz/minimum size=0.3cm,} }
\tikzfeynmanset{GR2/.style={/tikz/shape=ellipse,/tikz/fill={rgb:black,1;white,2},/tikz/minimum width = 1.2cm, 
    /tikz/minimum height = 3.4cm} }
\tikzset{box/.pic={\filldraw[fill=black]  (0,0) circle (2.5pt); \filldraw [fill=black] (0.5,0) circle (2.5pt); \draw [line width=5pt] (0,0) -- (0.5,0);}}
\usetikzlibrary{arrows.meta} 
\usetikzlibrary{calc}
\usetikzlibrary{decorations.pathmorphing}
\usetikzlibrary{decorations.pathreplacing}
\usetikzlibrary{decorations.markings}
\tikzset{
   vector2/.style={decorate, decoration={snake, amplitude=1pt, segment length=6pt}, draw,double},
   vector/.style={decorate, decoration={snake, amplitude=1pt, segment length=6pt}, draw},
	provector/.style={decorate, decoration={snake,amplitude=2.5pt}, draw},
	antivector/.style={decorate, decoration={snake,amplitude=-2.5pt}, draw},
    fermion/.style={draw=black, postaction={decorate},
        decoration={markings,mark=at position .55 with {\arrow[draw=black]{>}}}},
    fermionbar/.style={draw=black, postaction={decorate},
        decoration={markings,mark=at position .55 with {\arrow[draw=black]{<}}}},
    fermionnoarrow/.style={draw=black},
    gluon/.style={decorate, draw=black,
        decoration={coil,amplitude=4pt, segment length=5pt}},
    scalar/.style={dashed,draw=black, postaction={decorate},
        decoration={markings,mark=at position .55 with {\arrow[draw=black]{>}}}},
    scalarbar/.style={dashed,draw=black, postaction={decorate},
        decoration={markings,mark=at position .55 with {\arrow[draw=black]{<}}}},
    scalarnoarrow/.style={dashed,draw=black},
    electron/.style={draw=black, postaction={decorate},
        decoration={markings,mark=at position .55 with {\arrow[draw=black]{>}}}},
	bigvector/.style={decorate, decoration={snake,amplitude=4pt}, draw},
}
\tikzset{cross/.style={cross out, draw, 
         minimum size=2*(#1-\pgflinewidth), 
         inner sep=0pt, outer sep=0pt}}

\tikzstyle{block} = [draw, rectangle, 
    minimum height=3em, minimum width=6em]

\usepackage[T1]{fontenc} 

\newcommand{\tr}{\text{tr}}

\def\sc#1{\overline{#1}}
\makeatletter
\newcommand \UPlus {\mathop {\operator@font \uplus }\limits }
\makeatother
\makeatletter
\newcommand \Bigcup {\mathop {\operator@font \bigcup }\limits }
\makeatother
  \def\LabelNote#1{}
 \def\Label#1{\label{#1}%
  \smash{\hbox to\phipt{\raise1ex\hbox{\tiny[#1]}\hss}}}
  
  \def\mdot{{\cdot}}

\newcommand{\cM}{\mathcal{M}}

\newcommand{\cO}{\mathcal{O}}

\newcommand{\eps}{\epsilon}
\newcommand{\vareps}{\varepsilon}
\newcommand{\mb}{\bar m}
\newcommand{\vb}{\bar v}
\newcommand{\pb}{\bar p}

\newcommand{\ta}{\tilde{a}}
\newcommand{\cpm}{\varrho}

\def\nn{\nonumber}

\newcommand{\black}{\color{black}}

\def\spa#1.#2{\left\langle#1\,#2\right\rangle}
\def\spb#1.#2{\left[#1\,#2\right]}

\def\be{\begin{equation}}
\def\ee{\end{equation}}
\def\bea{\begin{eqnarray}}
\def\eea{\end{eqnarray}}  
\allowdisplaybreaks

\newcommand{\npre}{\mathcal{N}}

\newcommand{\tF}{\widetilde F}

\usepackage{amssymb,amsmath}
\usepackage{mathtools} 
\usepackage{cancel} 
\usepackage{graphicx} 

\usepackage{hyperref}
\hypersetup{
	colorlinks=true,
	linktoc=page,
	citecolor=americanrose,
	linkcolor=cadmiumgreen,
	urlcolor=blue} 
\definecolor{americanrose}{rgb}{1.0, 0.01, 0.24}
\definecolor{cadmiumgreen}{rgb}{0.0, 0.42, 0.24}

\usepackage[colorinlistoftodos]{todonotes}

\newcommand{\Cdot}{{\cdot}} 
\def\nn{\nonumber}

\title{Resummed spinning waveforms from  five-point amplitudes
}
\author{Andreas Brandhuber$\mbox{}^{a}$,}
\author{Graham R.~Brown$\mbox{}^{a}$,}
\author{Gang Chen$\mbox{}^{b}$,\\}
\author{\hspace{-0.2cm}Joshua Gowdy$\mbox{}^{a}$}
\author{and Gabriele Travaglini$\mbox{}^{a}$}
\affiliation{$\mbox{}^{a}$Centre for Theoretical Physics, Department of Physics and Astronomy, \\
Queen Mary University of London, Mile End Road, London E1 4NS, United Kingdom}
\affiliation{$\mbox{}^{b}$Niels Bohr International Academy,
Niels Bohr Institute, University of Copenhagen,\\
Blegdamsvej 17, DK-2100 Copenhagen \O, Denmark}
\emailAdd{a.brandhuber@qmul.ac.uk}
\emailAdd{graham.brown@qmul.ac.uk}
\emailAdd{gang.chen@nbi.ku.dk}
\emailAdd{j.k.gowdy@qmul.ac.uk}
\emailAdd{g.travaglini@qmul.ac.uk}
\begin{document}
\begin{flushright}
	QMUL-PH-23-18
\end{flushright}

\abstract{We compute the classical tree-level five-point amplitude for the two-to-two scattering of spinning celestial objects  with the emission of a graviton.
Using this five-point amplitude, we then turn to the computation of the 
leading-order time-domain  gravitational waveform. 
The method we describe is suitable for arbitrary values of classical spin of Kerr black holes and does not require any expansion in powers of the spin. In this paper we illustrate  it in  the simpler case of the scattering of one Kerr and one Schwarzschild black hole. An important ingredient of our calculation is a novel form of the Compton amplitude with spinning particles including contact terms derived from matching to black-hole perturbation theory calculations. This ensures that our waveform is valid up to  at least  fourth order in the spin.
Our method can be  applied immediately  to generate improved  waveforms once higher-order contact terms in the Compton amplitude become available.
Finally, we show the formula for the gravitational memory  to all orders in the spin, which is in agreement with our results.

}

\vspace{-2.6cm}

\maketitle

\flushbottom
 \tableofcontents
\newpage 

\section{Introduction}
Since the first direct observation of gravitational waves \cite{LIGOScientific:2016dsl,LIGOScientific:2016aoc,LIGOScientific:2016sjg,LIGOScientific:2017bnn,LIGOScientific:2017vwq}, a flurry of observations and theoretical predictions have greatly advanced the fields of black-hole physics and general relativity. Important questions regarding the intrinsic properties of black holes, the dynamics of binary 
black-hole processes, and more, can all be investigated in depth through high-precision gravitational-wave observations and theoretical calculations. 

One widely used and highly successful analytical  tool for the study of binary black-hole systems  is the Post-Newtonian (PN) expansion \cite{Damour:1985mt, Gilmore:2008gq,Damour:2001bu, Bjerrum-Bohr:2002gqz,Blanchet:2003gy, Itoh:2003fy,Foffa:2011ub,Jaranowski:2012eb, Damour:2014jta,Galley:2015kus,Damour:2015isa, Damour:2016abl, Bernard:2015njp, Bernard:2016wrg, Foffa:2012rn, Foffa:2016rgu,Porto:2017dgs,Porto:2017shd,Foffa:2019yfl,Blumlein:2020pog,Foffa:2019hrb,Blumlein:2019zku,Bini:2020wpo,Blumlein:2020pyo,Blumlein:2020znm,Bini:2020uiq,Blumlein:2021txj,Porto:2005ac,Steinhoff:2010zz,Levi:2014gsa,Levi:2015msa,Maia:2017yok,Levi:2018nxp,Levi:2020uwu} and the  effective one-body formulation \cite{Buonanno:1998gg,Damour:2016gwp,Damour:2017zjx,Vines:2017hyw,Vines:2018gqi,Damour:2019lcq}.
Recently, several varieties of modern methods, e.g. the double copy \cite{Bern:2008qj,Bern:2010ue,Bern:2019prr,Bern:2019nnu,Bern:2019crd,Bern:2021dqo,Bern:2021yeh,Bern:2022jvn}, the  Kosower-Maybee-O'Connell (KMOC) formalism  \cite{Kosower:2018adc}, heavy-mass effective theories \cite{Damgaard:2019lfh,Aoude:2020onz,Haddad:2020tvs,Brandhuber:2021kpo,Brandhuber:2021eyq,Brandhuber:2021bsf,Brandhuber:2022enp,Brandhuber:2023hhy}, the eikonal approach \cite{DiVecchia:2021bdo,DiVecchia:2022nna,DiVecchia:2022piu,DiVecchia:2023frv}, velocity cuts and the exponential representation of the $S$-matrix \cite{Bjerrum-Bohr:2021din,Damgaard:2021ipf,Damgaard:2023ttc}, worldline effective theory \cite{Goldberger:2004jt, Porto:2007qi,Goldberger:2009qd,Kalin:2020mvi, Kalin:2020fhe,Dlapa:2022lmu} and worldline quantum field theory \cite{Mogull:2020sak,Jakobsen:2021smu,Jakobsen:2023ndj}, have emerged as powerful theoretical frameworks for studying binary black-hole physics to high Post-Minkowskian (PM) order from different points of view.  
In particular they have been successfully applied to compute the conservative
part of the binary dynamics of
gravitationally interacting systems \cite{Iwasaki:1971iy,Iwasaki:1971vb,Bjerrum-Bohr:2013bxa, Bjerrum-Bohr:2014zsa,Holstein:2004dn,Neill:2013wsa, Bjerrum-Bohr:2016hpa, Bern:1994cg,Parra-Martinez:2020dzs,Cheung:2020gyp,Bjerrum-Bohr:2021din,Luna:2017dtq,Shen:2018ebu,Bautista:2019tdr,Herrmann:2021lqe,Herrmann:2021tct,Brandhuber:2019qpg, Emond:2019crr, AccettulliHuber:2019jqo,AccettulliHuber:2020oou,AccettulliHuber:2020dal,Carrillo-Gonzalez:2021mqj,Bellazzini:2021shn,Damgaard:2023ttc,Heissenberg:2023uvo,Adamo:2022ooq,Liu:2021zxr,Dlapa:2021npj,Dlapa:2021vgp,Jakobsen:2022fcj,Jakobsen:2022psy} to high orders in the PM expansion.

Research directly focused on the gravitational waveforms of binary black-hole systems in the PM expansion is  evolving rapidly. The tree-level waveforms for spinless objects were computed  in 
\cite{DEath:1976bbo,Kovacs:1977uw,Kovacs:1978eu} and reproduced in \cite{Jakobsen:2021smu,Mougiakakos:2021ckm} in the worldline picture.  The tree-level waveform was studied in \cite{Cristofoli:2021vyo,Cristofoli:2021jas} using the scattering-amplitude based KMOC formalism \cite{Kosower:2018adc,Cristofoli:2021vyo} and investigated using the eikonal approach in \cite{DiVecchia:2022nna, DiVecchia:2021bdo}.  At one loop, the study of the gravitational waveform was initiated recently in \cite{Brandhuber:2023hhy,Herderschee:2023fxh,Elkhidir:2023dco,Georgoudis:2023lgf} where the principal value contribution was obtained and shown to be consistent between KMOC and a heavy-mass effective field theory (HEFT) framework. The remaining terms beyond this principal value part were pointed out in  \cite{Caron-Huot:2023vxl} and shown to give an additional contribution to the waveform. The existence of such terms was also suggested by comparing with the Multipolar-Post-Minkowskian waveform in \cite{Bini:2023fiz}.

Gravitational waveforms 
 are influenced by various intrinsic properties of black holes. One of the most significant factors among them is their spin. An important building block for  including spin effects in waveforms is the minimal coupling between a classical spinning black hole and a graviton obtained using the  massive spinor-helicity formalism \cite{Arkani-Hamed:2017jhn}. Further important developments made use of   spinor helicity  \cite{Guevara:2017csg,Guevara:2018wpp,Chung:2018kqs,Guevara:2019fsj,Arkani-Hamed:2019ymq,Maybee:2019jus,Aoude:2020onz,Chung:2020rrz,Guevara:2020xjx,Chen:2021kxt,Chen:2022clh,Menezes:2022tcs,Chiodaroli:2021eug,Haddad:2021znf,Aoude:2022thd,Aoude:2022trd,Aoude:2023vdk}, the covariant amplitude form  ~\cite{Ochirov_2022,Cangemi:2022bew,Kosmopoulos:2021zoq,Damgaard:2019lfh,Bern:2020buy,Vines:2017hyw,Vines:2018gqi,FebresCordero:2022jts,Alessio:2022kwv,Bern:2022kto,Bern:2023ity}, gravitational solutions \cite{Bautista:2021wfy,Bautista:2022wjf,Damgaard:2022jem,Bautista:2023szu,Bianchi:2023lrg}, and the worldline picture \cite{Jakobsen:2022fcj,Jakobsen:2021zvh,Mogull:2020sak,Jakobsen:2021smu,Liu:2021zxr,Comberiati:2022ldk}. At tree level, the spin contribution  to the waveform up to quadratic order was  obtained in \cite{Jakobsen:2021lvp,Riva:2022fru} using a worldline effective theory.

In this paper, by employing the  definition of waveforms in terms of five-point amplitudes \cite{Cristofoli:2021vyo}, we compute gravitational waveforms involving spinning black holes, crucially without the need to expand in their spin.  
The building blocks entering the recursive BCFW  construction  \cite{Britto:2004ap,Britto:2005fq} of the five-point amplitude, adapted to the classical amplitude \cite{Brandhuber:2023hhy},  are the three-point and four-point Compton amplitudes with massive particles of arbitrary classical spin, which were constructed in \cite{Bjerrum-Bohr:2023jau,Bjerrum-Bohr:2023iey} using a  bootstrap technique which makes use of entire functions. After expanding in spin, this form of the Compton amplitude agrees with results obtained from black-hole perturbation theory \cite{Bautista:2022wjf,Bern:2022kto,Aoude:2022trd} for Kerr black holes up to at least fourth order in spin. However we note that this Compton amplitude can be upgraded with additional contact 
terms to match with the Teukolsky equation
\cite{Bautista:2021wfy,Bautista:2022wjf}, and the method discussed in this paper can be immediately applied to incorporate such additional terms
once these are available. 
 
 In this work we mainly focus on the time-domain waveform.
 First, we perform the Fourier transform  over the frequency; the exponential factors in the spinning amplitude then  produce a simple extra delta function when transforming to impact parameter space \cite{Jakobsen:2021lvp,Jakobsen:2021smu,Herderschee:2023fxh}. This additional delta function localises the integral further and simplifies the tree-level amplitude greatly. Finally, thanks to Cauchy's theorem, as used in  \cite{DeAngelis:2023lvf}, the remaining one-dimensional integral localises to contour integrals around  physical poles only. We stress here that our approach does not require any expansion in the spin parameters. Importantly, this allows us to preserve the (partially) resummed form of the Compton amplitude, and thus enables us to obtain a first glimpse at large-spin effects in gravitational waveforms.

The rest of the paper is organised as follows. In the next section we introduce the kinematics of the process, together with the definition of the spin variables we employ. In Section~\ref{sec:3} we introduce the three-point amplitude and the Compton amplitude with spinning particles. These are then used in Section~\ref{sec:4} to construct the five-point amplitude of four massive spinning particles with the emission of a gravitational wave, using a particular form of the BCFW recursion relation introduced in \cite{Brandhuber:2023hhy} for classical amplitudes. 
In Section~\ref{sec:timeDomainWaveform} we introduce the general method to compute the time-domain waveforms and illustrate how this computation reduces to a sum of residues on physical factorisation poles only,  in the simpler case of spinless particles.
We then present the general expression of the waveform for arbitrary spins of the two black holes. 
In Section~\ref{sec:6} we specialise to the case of a Schwarzschild and a Kerr black hole, and also  present several plots of the waveforms for increasing values of the spin of the Kerr black hole. 
In Section~\ref{sec: OrderBYOrderInSpin} we make some interesting observations by comparing the waveforms obtained using the resummed Compton amplitudes to those derived from the Compton amplitudes expanded in the spin parameter.
Section~\ref{sec-memory} presents a short derivation of the memory of the gravitational wave in the spinning case, to all orders in the spins of the Kerr black holes, which we have then used to test our analytic results. Finally, two appendices complete the paper. In 
Appendix~\ref{Appendix-A} we perform some useful simplifications of the expression of the four-point Compton amplitude, which are convenient in the derivation of the memory; and in Appendix~\ref{app: integrnadcoeffs} we list the coefficients appearing in the $q_1^2$- and $q_2^2$-channels of the classical, tree-level five-point  amplitude derived in Section~\ref{sec:timeDomainWaveform}.
 
 The interested reader can find {\it Mathematica}  notebooks with expressions for the spinning HEFT amplitudes with one  emitted graviton, and explicit time-domain waveform  results in the  system of  a Schwarzschild and a Kerr black hole
{\href{https://github.com/QMULAmplitudes/SpinningWaveformPublicData/tree/main}{{\it SpinningWaveform} GitHub repository.}}

{\bf Note added:}
While preparing this manuscript we became aware of the nice work 
\cite{DeAngelis:2023lvf}, with which our paper has  some overlap.  
We have checked  that our results agree with theirs. 


\section{Kinematics of the  scattering  and spin variables}
\label{sec:kinematics}
Here we  review the kinematics of the scattering of two heavy spinning particles of masses $m_1$ and $m_2$  and spin vectors $a_1$ and $a_2$,   with the emission of a graviton of momentum $k$:
\begin{equation}\label{eq: kinematics}
	\begin{array}{lr}

		\begin{tikzpicture}[scale=15,baseline={([yshift=-1mm]centro.base)}]
			\def\x{0}
			\def\y{0}

			\node at (0+\x,0+\y) (centro) {};
			\node at (-3pt+\x,-3pt+\y) (uno) {$p_1=\bar{p}_1 + \dfrac{q_1}{2}$};
			\node at (-3pt+\x,3pt+\y) (due) {$p_2=\bar{p}_2 + \dfrac{q_2}{2}$};
			\node at (3pt+\x,3pt+\y) (tre) {$p_2^\prime =  \bar{p}_2 - \dfrac{q_2}{2}$};
			\node at (3pt+\x,-3pt+\y) (quattro) {$\ \ p_1^\prime = \bar{p}_1 - \dfrac{q_1}{2}$};
			\node at (5.55pt+\x,\y) (cinque)
			{$k= q_1 + q_2$};

			\draw [thick,double,red] (uno) -- (centro);
			\draw [thick,double,blue] (due) -- (centro);
			\draw [thick,double,blue] (tre) -- (centro);
			\draw [thick,double,red] (quattro) -- (centro);
			\draw [vector,double] (cinque) -- (centro);

			\draw [->] (-2.8pt+\x,-2pt+\y) -- (-1.8pt+\x,-1pt+\y);
			\draw [<-] (2.8pt+\x,-2pt+\y) -- (1.8pt+\x,-1pt+\y);
			\draw [<-] (-1.8pt+\x,1pt+\y) -- (-2.8pt+\x,2pt+\y);
			\draw [->] (1.8pt+\x,1pt+\y) -- (2.8pt+\x,2pt+\y);
			\draw [->] (2.3pt+\x,-0.6pt+\y) -- (3.75pt+\x,-0.6pt+\y);

			\shade [shading=radial] (centro) circle (1.5pt);
		\end{tikzpicture}
		 & \hspace{2cm}
	\end{array}
\end{equation}
As usual we have introduced barred variables, defined as \cite{Landshoff:1969yyn,Parra-Martinez:2020dzs}
\begin{align}
	\label{barredv}
	\begin{split}
		p_1 &= \bar{p}_1 + \frac{q_1}{2}\, , \qquad p_1^\prime = \bar{p}_1 - \frac{q_1}{2}\, , \\
		p_2 &= \bar{p}_2 +  \frac{q_2}{2}\, , \qquad p_2^\prime = \bar{p}_2 - \frac{q_2}{2}\, , 
	\end{split}
\end{align}
which satisfy  \begin{align}
	\label{eq: HEFTfame}
	\bar{p}_1\Cdot q_1 =\bar{p}_2\Cdot q_2 =0 \, . 
\end{align}
We also introduce barred masses,
\begin{align}\label{eq: mbardef}
	\bar{m}_i^2 \coloneqq \bar{p}_i^2 = m_i^2 - \frac{q_i^2}{4}\, ,\qquad i=1,2\, , 
\end{align}
with the   HEFT  expansion being  organised in powers of the $\bar{m}_i$.

To parameterise the scattering process  we choose  five independent Lorentz-invariant quantities  as in \cite{Brandhuber:2023hhy}, 
\begin{equation}
	\label{fiveiv}
	\begin{split}
		y \coloneqq v_1 \Cdot v_2  \geq 1\ ,\qquad
		q_i^2 \leq 0\ ,\qquad w_i \coloneqq v_i \Cdot k  \geq 0 \, , \qquad i=1,2,
	\end{split}
\end{equation}
where the four-velocities are defined by $p_i {=} m_i v_i$, with $v_i^2{=}1$.
We also note that $y$ is the relativistic factor $\frac{1}{\sqrt{1-v_{\rm rel}^2}}$, where $v_{\rm rel}$ is the relative velocity of one of the two heavy particles in the rest frame of the other.
We will also use the   barred versions 
$\bar{w}_i \coloneqq \bar{v}_i \Cdot k$ and $\bar{y} \coloneqq \bar{v}_1 \Cdot \bar{v}_2$ of the above quantities, with $\bar{p}_i\coloneqq \bar{m}_i \bar{v}_i$ and $\bar{v_i}^2=1$.

The spin tensors for incoming and outgoing massive particles in terms of the spin vectors $s_i$ are given by
\begin{align}
    \begin{split}
        & S_i^{\mu \nu}(p_i)=-\frac{1}{m_i}\epsilon^{\mu\nu\alpha\beta}p_{i\,\alpha} s_{i\,\beta}(p_i) \, ,\qquad S_i^{\mu \nu}(p_i')=-\frac{1}{m_i}\epsilon^{\mu\nu\alpha\beta}p'_{i\,\alpha} s_{i\,\beta}(p_i')\, .
    \end{split}
\end{align}
To expand this in the heavy-mass limit we change variables from $p_i,p_i'$ to $\pb_i$ and $q_i$ as in \eqref{barredv}. We follow the method of \cite{Maybee:2019jus} and use an infinitesimal Lorentz transformation from $\pb_i$ to $\pb_i\pm \dfrac{q}{2}$ to write
\begin{align}
   s_i^\mu\big(\pb_i\pm\frac{q}{2}\big)&= (\delta_{\nu}^{\mu}\pm{\omega^{\mu}}_{\nu})s_i(\pb_i)^\nu\nn\\
    &=  \Big[ \delta_{\nu}^{\mu}\mp\frac{1}{2 \mb^2}(\pb_i^{\mu}q_{\nu}-q^{\mu}\pb_{i\,\nu})\Big]s_i(\pb_i)^\nu\\
    &= s_i(\pb_i)^\mu\mp\frac{\pb_i^\mu}{2 \mb^2}q\mdot  s_i(\pb_i) + \cO(\mb^{-2})\nn \, .
\end{align}
This is valid since $\mb_i$ (which will eventually be the classical mass) is much larger than the typical value of~$q$. This allows us to expand the spin tensors as
\begin{align}
\begin{split}
    S_i^{\mu \nu}(p_i)&=-\frac{1}{\mb_i}\epsilon^{\mu\nu\alpha\beta}\pb_{i\,\alpha} s_{i\,\beta}(\pb)-\frac{1}{2\mb_i}\epsilon^{\mu\nu\alpha\beta} q_{\alpha}s_{i\,\beta}(\pb_i)  + \cO(\mb^{-2})\, , \\
    S_i^{\mu \nu}(p'_i)&=-\frac{1}{\mb_i}\epsilon^{\mu\nu\alpha\beta}\pb_{i\,\alpha} s_{i\,\beta}(\pb_i)+\frac{1}{2\mb_i}\epsilon^{\mu\nu\alpha\beta}q_{\alpha}s_{i\,\beta}(\pb_i) + \cO(\mb^{-2})\, , 
    \end{split}
\end{align}
where, remarkably, the shifts in $s_i^\mu(p_i^{(\prime)})$ drop out to this order in the $\mb$ expansion, due to the antisymmetry of the Levi-Civita. We can also define the classical spin parameter as 
\begin{align}
    a_i^{\mu}\coloneqq \frac{s_i^\mu(\pb_i)}{\mb_i}\, , 
\end{align}
to write
\begin{align}\label{eq: spinTensorExpansion}
\begin{split}
    S_i^{\mu \nu}(p_i)&=-\epsilon^{\mu\nu\alpha\beta}\Big(\pb_{i\,\alpha}+\frac{q_{\alpha}}{2}\Big) a_{i\,\beta}+ \cO(\mb_i^{-2})\,,\\
    S_i^{\mu \nu}(p_i')&=-\epsilon^{\mu\nu\alpha\beta}\Big(\pb_{i\, \alpha}-\frac{q_{\alpha}}{2}\Big) a_{i\, \beta}+\cO(\mb_i^{-2})\,.
\end{split}
\end{align}
Finally, in the large $\mb_i$ limit the two spin tensors in \eqref{eq: spinTensorExpansion} become the same, and we define our classical spin tensors as 
\begin{equation}
\label{eq: spintensordefinition}
    S_{i}^{\mu\nu}\coloneqq-\epsilon^{\mu\nu\rho\sigma}\pb_{i\, \rho} a_{i\, \sigma}\, , 
\end{equation}
which satisfies $S^{\mu \nu}_i \pb_{i\nu}{=}0$, known as the spin supplementary condition \cite{FlemingGordon1965,Porto:2005ac}, while $a_i$ satisfies $\pb_i\Cdot a_i=0$. 
We can also invert this relation,
\begin{align}
a^{\mu}_i = -\frac{1}{2\mb^2_i}\epsilon^{\mu \nu \alpha \beta} \pb_
{i\, \nu} S_{i\, \alpha \beta}\, . 
    \end{align}
Note  that $a^\mu_i$ has mass dimension $-1$ so that $S^{\mu\nu}_i$ is dimensionless. The spin vector of a heavy particle is then \begin{align}
    s^\mu_i\coloneqq \mb_i a^\mu_i\, .
    \end{align}
    Much like $\pb^\mu$ and $\vb^\mu$, both $s^\mu$ and $a^\mu$ are well defined in the classical/large-$m$ limit. 
Finally, the gravitational coupling we use is $\kappa\coloneqq\sqrt{32\pi G}$.

\section{Classical gravitational Compton amplitude with~spin}

\label{sec:3}

\subsection{Three-point amplitude}
The three-point amplitude for two classical massive spinning particles is given by \cite{Arkani-Hamed:2017jhn,Guevara:2018wpp,Chung:2018kqs,Chen:2021kxt}
\begin{equation}\label{eq: threePointAmpTensor}
    \cM_{3}= -i \kappa\, (\pb\Cdot \varepsilon_1)^2 \exp \left(-i \frac{k_1 \Cdot S\cdot\varepsilon_1}{\pb\Cdot\varepsilon_1}\right)\,,
\end{equation}
where $p$ is the momentum of the massive particle, $k_1$ is the momentum of the graviton with polarisation $\varepsilon_1$ and $S$ is the  spin tensor of the massive particle introduced in~\eqref{eq: spintensordefinition}.  
The amplitude \eqref{eq: threePointAmpTensor} can also be written as \cite{Bern:2020buy,Johansson:2019dnu,Bjerrum-Bohr:2023jau}
\begin{equation}\label{threepointgrav}
    \cM_3= -i\kappa (\pb\Cdot\varepsilon_1)({\mathsf w}_1\mdot\varepsilon_1)\,,
\end{equation}
where 
\begin{equation}
\begin{split}
    {\mathsf w}_1^\mu&\coloneqq\cosh(k_1\Cdot a)\pb^\mu-i\frac{\sinh(k_1\Cdot a)}{k_1\Cdot a}(k_1 \Cdot S)^\mu
    %
    \, , 
    \end{split}
\end{equation}
and we have used the notation $(k_1 \mdot S)^\mu = k_{1 \nu}S^{\nu\mu}$.

\subsection{The Compton amplitude}
\label{sec:Compton}

We now move on to discuss the four-point amplitude. For convenience, in this section we will call the momenta $p,k_1,k_2, p'$ where $p, p'$ are the momenta of the massive particles, $p^2{=}(p')^2{=}m^2$ and $k_{1,2}$ are the momenta of the gravitons, with $k_{1,2}^2{=}0$. 
\begin{align}
\label{compton}
\begin{tikzpicture}[baseline={([yshift=-0.8ex]current bounding box.center)}]\tikzstyle{every node}=[font=\small]	
\begin{feynman}
    	 \vertex (a) {\( p\)};
    	 \vertex [right=2.1cm of a] (f2)[GR]{$~~$\textbf{S}$~~$};
    	 \vertex [right=2.1cm of f2] (c){$ p'$};
    	 \vertex [above=1.3cm of f2] (gm){};
    	 \vertex [left=0.8cm of gm] (g2){$ k_{1}$};
    	  \vertex [right=0.8cm of gm] (g20){$ k_{2}$};
    	  \diagram* {
(a) -- [fermion,thick] (f2) --  [fermion,thick] (c),
    	  (g2)--[photon,ultra thick,rmomentum'](f2),(g20)--[photon,ultra thick,rmomentum](f2)
    	  };
    \end{feynman}  
    \end{tikzpicture}
\end{align}

The four-point classical Compton amplitude can be divided into three pieces~\cite{Bjerrum-Bohr:2023iey}, 
\begin{align}\label{eq:M}
    \cM_4=\frac{-i\npre_{\rm dc}}{2k_1\mdot k_2}+\frac{-i\npre_{\rm r}}{4\pb\mdot k_1 \pb\mdot k_2}-i\npre_{\rm c}.
\end{align}
The first term is obtained from the double copy and corresponds to propagation  without changing the direction or magnitude  of the spin
\cite{Bjerrum-Bohr:2023jau}, 
\begin{align}
     \npre_{\rm dc}&=-\Bigg[\frac{{\mathsf w}_1\mdot F_1\mdot F_{2}\mdot {\mathsf w}_2}{k_1\mdot \pb}-
     \Big(
      i G_2\left(x_1,x_2\right) (a\mdot F_1\mdot F_{2}\mdot S\mdot k_2+ a\mdot F_{2}\mdot F_1 \mdot S\mdot k_1)\nn\\
      &+i G_1\left(x_{12}\right) \tr\left(F_1\mdot S\mdot F_2\right)+G_1(x_1) G_1(x_2) (a\mdot F_1\mdot \pb a\mdot F_2\mdot k_1- a\mdot F_1\mdot k_2 a\mdot F_2\mdot \pb)\nn\\
      &
      +k_1\mdot \pb\, G_1(x_1) G_1(x_2)  a\mdot F_1\mdot F_2\mdot a\Big)\Bigg]\Big(\frac{\pb\mdot F_1\mdot F_2\mdot \pb}{k_2\mdot \pb}\Big)\, ,
\end{align}
with 
\begin{align}
    x_i\coloneqq k_i\Cdot a\, , \quad x_{i\ldots j}\coloneqq (k_i+\cdots +k_j) \Cdot a\ , \quad F^{\mu\nu}_i=k_i^{\mu}\vareps_i^\nu-\vareps_i^\mu k_i^{\nu} \, .
\end{align} 
Note that it contains both massless and massive poles and  we already take the HEFT expansion. This term gives the minimal amplitude to fit the test particle scattering  angle in the Kerr metric.

The second term allows for a change of direction of the spin, and we refer to it as the ``spin-flip'' term \cite{Bjerrum-Bohr:2023iey},   
\begin{align}
    \npre_{\rm r}&=\Big[(\partial_{x_1}-\partial_{x_2})G_1(x_1) G_1(x_2)\Big]\nn\\
    &\times\Big[\pb\mdot k_2 \ (\pb^2 a\mdot F_1\mdot F_2\mdot a \ a\mdot F_2\mdot F_1\mdot \pb+a^2 \pb\mdot F_1\mdot F_2\mdot \pb \ a\mdot F_1\mdot F_2\mdot \pb)- (1\leftrightarrow 2)\Big]\nn\\
      &+i\Big[(\partial_{x_1}-\partial_{x_2})G_2(x_1,x_2)\Big]\nn\\
      &\times\Big[\pb\mdot k_2 \ a\mdot F_2\mdot F_1\mdot \pb \ (a\mdot F_2\mdot \pb \ a\mdot \tF_1\mdot \pb-a\mdot F_1\mdot \pb \ a\mdot \tF_2\mdot \pb)+(1\leftrightarrow 2)\Big]\, ,
\end{align}
where $\tF^{\mu\nu}\equiv \frac{1}{2}\epsilon^{\mu\nu\rho\sigma}F_{\rho\sigma}$ denotes the Hodge dual of the linearised field strength.  Note that this term only gives rise to massive poles. 
Finally, the last contribution consists of contact terms, 
\begin{align}\label{eq: comptoncontacts}
     \npre_{c}&=\Big[{(\partial_{x_1}-\partial_{x_2})^2\over 2!}{G_{1}(x_1)}{G_{1}(x_2)}\Big]\Big[a\mdot F_1\mdot \pb \ a\mdot F_2\mdot \pb\  a\mdot F_1\mdot F_2\mdot a \\
     &-{1\over 2}a^2 (a\mdot F_1 \mdot F_2\mdot \pb \ a\mdot F_2 \mdot F_1\mdot \pb- a\mdot F_1 \mdot F_2\mdot a \ \pb\mdot F_1 \mdot F_2\mdot \pb)\Big] \nn\\
     &+e_1\Big[{i(\partial_{x_1}-\partial_{x_2})^2\over 2!}{G_{2}(x_1,x_2)}\Big] \Big[ a\mdot F_1\mdot F_2\mdot a \ a\mdot F_2\mdot \pb \  a\mdot \tF_1\mdot \pb-(1\leftrightarrow 2)\Big] \nn\\ 
     &+ e_2\Big[{i(\partial_{x_1}-\partial_{x_2})^2\over 2!}{G_{2}(x_1,x_2)}\Big]\Big[\pb^2 (a\mdot F_1\mdot F_2\mdot a) (a\mdot F_2\mdot \tF_1\mdot a)-(1\leftrightarrow 2)\Big]\, \nn \\
    & +e_3\Big[{i(\partial_{x_1}-\partial_{x_2})^2\over 2!}{G_{2}(x_1,x_2)}\Big]\Big[a^2 (a\mdot F_2\mdot F_1\mdot  \pb) (a\mdot \tF_1\mdot F_2\mdot  \pb)-(1\leftrightarrow 2)\Big]\, .\nn
\end{align}
The $G$-functions appearing in the expressions above can be defined in terms of hyperbolic functions as \cite{Bjerrum-Bohr:2023jau} 
\begin{align}
    G_1(x)&:=\frac{\sinh(x)}{x}\, , \nn\\
    G_2(x_1,x_2)&:={1\over x_{2}}\Big({\sinh(x_{{12}})\over{x_{12}}}-\cosh(x_{2})\, {\sinh(x_{1})\over{x_1}}\Big),
\end{align}
and are entire functions, free of singularities. Note that $G_2(x_2, x_1)=-G_2(x_1, x_2)$.  

 The  contact terms in the first two lines of \eqref{eq: comptoncontacts} only begin contributing at quartic order in the spin and their numerical coefficients have been fixed against results at quartic order in  the spin arising from black-hole perturbation theory (BHPT) \cite{Bautista:2022wjf} or equivalently  using the  ``spin-shift symmetry'' \cite{Bern:2022kto,Aoude:2022trd}. At $\cO(a^4)$, these two methods to constrain the contact terms are in agreement. 
 
The remaining three lines in \eqref{eq: comptoncontacts} involve contact terms which contribute from quintic order in the spin. We have chosen to fix their numerical coefficients $e_1, e_2, e_3$ assuming spin-shift symmetry applied at this order \cite{Aoude:2022thd, Bern:2022kto}, setting them to be $e_1{=}-3/4,\,e_2{=}0,\,e_3{=}0$. However, we note that recent work \cite{Bautista:2022wjf} has shown that at $\cO(a^5)$ the  spin-shift symmetry is in fact broken, and instead such coefficients should be fixed by comparison to BHPT or alternatively fixed to the method of multipole moments of the Kerr BH \cite{Scheopner:2023rzp}. These two methods are in agreement at $\cO(a^5)$  \cite{Scheopner:2023rzp} but do not agree with spin-shift symmetry. Therefore, the results derived here  are only applicable to Kerr black holes up to quartic order in the spin. We  have chosen to set $e_1{=}-3/4,\,e_2{=}0,\,e_3{=}0$ simply to illustrate the general matching principle, although our method makes it easy to deal with any values of the $e_i$'s and also with further contact terms starting at $\cO(a^6)$ and beyond,  as we will discuss in~\cite{toappear}.

Finally, we note here that, as described in detail in \cite{Bautista:2021wfy, Bautista:2022wjf}, results from BHPT are valid in the physical regime $\frac{a_i}{G m_i} < 1$ but can be analytically continued to the super-extremal regime where $\frac{a_i}{G m_i} > 1$ in order to match with results formulated from amplitudes. 
Such an analytic continuation is in fact trivial up to $\cO(a^4)$. We conclude that at leading PM order and up to fourth order in spin the Compton amplitude, and hence our spin-expanded results for the waveform, do not distinguish between physical versus super-extremal Kerr.


\section{Spinning five-point amplitude}

\label{sec:4}

The crucial ingredient to compute the waveforms is the classical part of the five-point amplitude of two spinning particles with one radiated graviton.%
\footnote{In the next section we will see that actually only the residues on the physical factorisation channels are needed for computing the waveform. However, since the computation of the five-point amplitude is so simple we cannot resist to present it here.}
It can be    derived  using the HEFT BCFW recursion relation introduced in \cite{Brandhuber:2023hhy}
and 
is obtained from the following two recursive diagrams, 
\begin{align}
\label{2BCFWdiagrams}
    \begin{tikzpicture}[baseline={([yshift=-0.4ex]current bounding box.center)}]\tikzstyle{every node}=[font=\small]
		\begin{feynman}
			\vertex (p1) {\(\bar v_1,a_1\)};
			\vertex [above=1.5cm of p1](p2){$\bar v_2,a_2$};
			\vertex [right=1.5cm of p2] (u1) [HV]{H};
			\vertex [right=1.2cm of u1] (p3){};
			\vertex [right=1.5cm of p1] (b1) [dot]{};
			\vertex [right=1.2cm of b1] (p4) []{};
			\vertex [above=0.4cm of p1] (cutLt);
            \vertex [right=0.6cm of cutLt] (cutL);
			\vertex [right=1.8cm of cutL] (cutR){$k$};
			\vertex [right=0.5cm of b1] (cut1);
			\vertex [above=0.2cm of cut1] (cut1u);
			\vertex [below=0.2cm of cut1] (cut1b);
			\diagram* {
			(p2) -- [thick] (u1) -- [thick] (p3),
			(b1)--[photon,ultra thick,momentum=$q_1$](u1),  (u1)-- [photon,ultra thick,momentum] (cutR), (p1) -- [thick] (b1)-- [thick] (p4), (cutL)--[dashed, red,thick] (cutR),
			};
		\end{feynman}
	\end{tikzpicture}  &&
\begin{tikzpicture}[baseline={([yshift=-0.4ex]current bounding box.center)}]\tikzstyle{every node}=[font=\small]
		\begin{feynman}
			\vertex (p1) {\(\bar v_1,a_1\)};
			\vertex [above=1.5cm of p1](p2){$\bar v_2,a_2$};
			\vertex [right=1.5cm of p2] (u1) [dot]{};
			\vertex [right=1.2cm of u1] (p3){};
			\vertex [right=1.5cm of p1] (b1) [HV]{H};
			\vertex [right=1.2cm of b1] (p4) []{};
			\vertex [above=1.1cm of p1] (cutLt);
            \vertex [right=0.6cm of cutLt] (cutL);
			\vertex [right=1.8cm of cutL] (cutR){$k$};
			\vertex [right=0.5cm of b1] (cut1);
			\vertex [above=0.2cm of cut1] (cut1u);
			\vertex [below=0.2cm of cut1] (cut1b);
			\diagram* {
			(p2) -- [thick] (u1) -- [thick] (p3),
			(u1)--[photon,ultra thick,momentum'=$q_2$](b1), (b1)-- [photon,ultra thick,momentum'] (cutR), (p1) -- [thick] (b1)-- [thick] (p4), (cutL)--[dashed, red,thick] (cutR),
			};
		\end{feynman}
	\end{tikzpicture}
\end{align}
corresponding to the $q_1^2$ and $q_2^2$ channels, respectively. In the scalar case, these BCFW diagrams capture all of the `contact terms' in the classical amplitude (that is terms without poles in $q_1^2$ or $q_2^2$ but possibly with massive poles). In the spinning case we will follow the same procedure and, although we have no general proof that these contact terms are captured fully, we have checked that the contributions from the two BCFW diagrams satisfy the correct soft behaviour. Regardless, such contact terms without poles in $q_1^2$ or $q_2^2$ do not contribute to the tree-level waveform as we will see in 
Sections~\ref{sec:timeDomainWaveform} and~\ref{sec: OrderBYOrderInSpin}.

The contribution of each of the two diagrams is obtained by gluing a three-point amplitude with a 
four-point Compton amplitude, given in 
\eqref{eq: threePointAmpTensor} and \eqref{eq:M}, respectively. In doing so one has to sum 
 over the intermediate states of the exchanged graviton, using
\begin{align}
	\label{simplsum}
	\sum_{h}\varepsilon^{\mu_a}_{-\hat{q}} {\varepsilon}^{\nu_a}_{-\hat{q}}{\varepsilon}^{\mu_b}_{\hat{q}} \varepsilon^{\nu_b}_{\hat{q}}=\frac{1}{2}\Big[ \eta^{\mu_a \mu_b}\eta^{\nu_a \nu_b}+\eta^{\mu_a \nu_b}\eta^{\nu_a \mu_b}-\frac{2}{D-2}\eta^{\mu_a\nu_a}\eta^{\mu_b\nu_b}\Big]
	\, .
\end{align}
For convenience, we introduce a tensor current by extracting the polarisation vector from the Compton amplitude: 
\begin{align}
\mathcal{J}^{\mu\nu}_i\varepsilon_{\mu\nu}=\cM_4(-q_i,k,\bar v_i,a_i)\, .
\end{align}
Then, the amplitude in each channel is of the form
\begin{align}
\begin{split}
\label{firstBCFW}
 \cM_{q_1^2}&={1\over q_1^2} \sum_h\Big( \cosh(a_1\mdot q_1) (\bar v_1\mdot \vareps)^2  \vareps_{\mu\nu} \mathcal{J}^{\mu\nu}_1- i G(a_1\mdot q_1) \bar v_1\mdot \vareps \, q_1\mdot S_1\mdot \vareps \, \vareps_{\mu\nu} \mathcal{J}^{\mu\nu}_1 \Big)\\
 & ={1\over q_1^2} \Big(\cosh(a_1\mdot q_1)\big(\bar v_1\mdot \mathcal{J}_1\mdot \bar v_1-{1\over 2}\tr(\mathcal{J}_1)\big)- {i\over 2} G(a_1\mdot q_1) \big(q_1\mdot S_1 \mdot \mathcal{J}_1\mdot \bar v_1  -\bar v_1\mdot \mathcal{J}_1 \mdot S_1 \mdot q_1 \big) \Big)
 \, ,
 \end{split}
\end{align}
and 
\begin{align}
\begin{split}
\label{secondBCFW}
 \cM_{q_2^2}&={1\over q_2^2} \sum_h\Big( \cosh(a_2\mdot q_2) (\bar v_2\mdot \vareps)^2  \vareps_{\mu\nu} \mathcal{J}^{\mu\nu}_2- i G(a_2\mdot q_2) \bar v_2\mdot \vareps \, q_2\mdot S_2\mdot  \vareps \, \vareps_{\mu\nu} \mathcal{J}^{\mu\nu}_2 \Big)\\
 & ={1\over q_2^2} \Big(\cosh(a_2\mdot q_2)\big(\bar v_2\mdot \mathcal{J}_2\mdot \bar v_2-{1\over 2}\tr(\mathcal{J}_2)\big)- {i\over 2} G(a_2\mdot q_2) \big(q_2\mdot S_2 \mdot \mathcal{J}_2\mdot \bar v_2  -\bar v_2\mdot \mathcal{J}_2 \mdot S_2 \mdot q_2 \big) \Big)\, .
\end{split}
\end{align}
The full amplitude can be obtained directly  adding   \eqref{firstBCFW} and \eqref{secondBCFW}, 
\begin{align}
    \cM_{5,\rm HEFT}=\cM_{q_1^2}+\cM_{q_2^2}\, .
\end{align}
Both channels have the spurious pole  $1\over k\mdot q_1$, which cancels after summing the two contributions. To see this, we must use the Bianchi identity  in $D$-dimensional momentum space \cite{Feng:2020jck}
\begin{align}
	\begin{split}
		A\Cdot F_k\Cdot B \ k\Cdot C +B\Cdot F_k\Cdot C \  A\Cdot k+ C\Cdot F_k\Cdot A \ k\Cdot &B  = 0\, , 
	\end{split}
\end{align}
where $A, B, C$ can be any vector.  For example, a particular application is 
\begin{align}
    \bar v_1\mdot S_2\mdot F_k\mdot q_2=\frac{k\mdot q_2 \bar v_1\mdot F_k\mdot S_2\mdot \bar v_1-k\mdot S_2\mdot \bar v_1 \bar v_1\mdot F_k\mdot q_2}{q_2\mdot \bar v_1}.
\end{align}
The resulting expression for the amplitude  only contains the following field-strength products:
\begin{align}
   a_1\mdot F_k\mdot \bar v_1,&&a_2\mdot F_k\mdot \bar v_1,&&q_1\mdot F_k\mdot \bar v_1,&&v_1\mdot F_k\mdot \bar v_2,&&q_1\mdot S_1\mdot F_k\mdot \bar v_1,\nn\\
  q_1\mdot S_2\mdot F_k\mdot \bar v_1,&& v_1\mdot F_k\mdot S_1\mdot \bar v_2,&&\bar v_1\mdot F_k\mdot S_2\mdot \bar v_1,&&\text{tr}\left(F_k\mdot S_1\right),&&\text{tr}\left(F_k\mdot S_2\right).
\end{align}
 The complete expression for the five-point amplitude of two spinning black holes is included in the {\href{https://github.com/QMULAmplitudes/SpinningWaveformPublicData/tree/main}{GitHub repository}} associated to this paper.

In this paper we  will present waveforms in  the  simpler situation of the scattering of a Schwarzschild and a Kerr black hole, deferring the study of the waveform produced by two Kerr black holes to \cite{toappear}. Without loss of generality, we will therefore set $a_2{=}0$,  which dramatically simplifies the contribution from the $q_1^2$-channel. Then the  amplitude in each channel has a very  compact form
\begin{align}\label{eq:cutq1}
\begin{split}
 \cM_{q_1^2}
 & ={1\over q_1^2} \Big[\cosh(a_1\mdot q_1)\big(\bar v_1\mdot \left.\mathcal{J}_{2}\right|_{a_2=0}\mdot \bar v_1-\frac{1}{2} \tr(\left.\mathcal{J}_{2}\right|_{a_2=0})\big)
 \\ &
 - {i\over 2} G(a_1\mdot q_1) \big(q_1\mdot S_1 \mdot \left.\mathcal{J}_{2}\right|_{a_2=0}\mdot \bar v_1  -\bar v_1\mdot \left.\mathcal{J}_{2}\right|_{a_2=0} \mdot S_1 \mdot q_1 \big) \Big]\, , 
\end{split}
\end{align}
and 
\begin{align}
 \cM_{q_2^2}
 & ={1\over q_2^2} \Big[\bar v_2\mdot \mathcal{J}_1\mdot \bar v_2-\frac{1}{2}\tr(\mathcal{J}_1) \Big]\, .
\end{align}


\section{The time-domain waveform}\label{sec:timeDomainWaveform}

\subsection{Waveforms from amplitudes}
We begin by briefly reviewing the emergence of waveforms in black-hole scattering. We consider the  classical gravitational field 
produced by the scattering of two black holes which are modelled by two massive  spinning  particles
using the KMOC approach
\cite{Kosower:2018adc,Cristofoli:2021vyo}. The corresponding  initial two-particle state has the form 
\begin{align}
	\label{psi}
	|\psi\rangle_{\rm in} \coloneqq \int\!\prod_{j=1}^2d\Phi (p_j)  e^{i p_1 \Cdot b} \phi(p_1) \phi(p_2) |p_1, a_1,  p_2, a_2\rangle_{\rm in}
	\ . 
\end{align}
Following  \cite{Kosower:2018adc,Cristofoli:2021vyo, Brandhuber:2023hhy,Herderschee:2023fxh, Elkhidir:2023dco}, one finds that 
\begin{align}
	\label{KMOCsubfinalforhmunu}
	\begin{split}
		\langle h^{\rm out}_{\mu \nu} (x) \rangle_{\rm \psi}  \!=\!     \kappa \,  \int\!\prod_{j=1}^2d\Phi (\pb_j)  \, |\phi(\pb_1)|^2 |\phi (\pb_2)|^2
		\Big[
	\sum_h \int\!d\Phi(k) e^{-i k \Cdot x} \,
		\varepsilon^{(h)\ast}_{\mu\nu} (\hat{k}) 
		\, 
		\big[ i \,  {W} \big] \, + \, {\rm h.c.}\Big]
		, 
	\end{split}
\end{align}
where $k{\coloneqq} \omega \hat{k}$. 
Here ${W}{=}{W}(\vec{b}, k; h)$  is the spectral waveform for the emission of a graviton of momentum $k$ and helicity $h$, which at leading order in the PM expansion~is%
\footnote{The factor of $-i$ cancels the $i$  from our definition of amplitudes as matrix elements of~$i\, T$.  Furthermore, we have defined the physical impact parameter $b\coloneqq b_1 - b_2$, where the $b_i$ are taken to be orthogonal to $p_i$, and finally we have set  $b_2=0$. } 
\begin{align}
\label{KMOCsubfinalbis}
{W}(b, k^h ) \coloneqq  -i \int\!d\mu^{(D)}\  e^{iq_1\Cdot b} \ \mathcal{M}_{5,\rm HEFT}(q_1, q_2, a_1, a_2; h) \, , 
\end{align}
where we have introduced the $D$-dimensional measure (for regularisation purposes) 
\begin{align}
\label{measureinnn}
	d\mu^{(D)} \coloneqq \frac{d^Dq_1}{(2\pi)^{D{-}1}} \frac{d^Dq_2}{(2\pi)^{D{-}1}}
	\, (2\pi)^D  \delta^{(D)} (q_1 + q_2 - k) \delta(2  {\pb}_1\Cdot q_1 ) \delta(2  {\pb}_2\Cdot q_2 )\, ,
\end{align}
with  $q_{1,2}{=} p_{1,2} {-} p_{1,2}^\prime$ being the momentum transfers, and  $D{=}4{-}2\epsilon$. Here we are ignoring zero-modes in the amplitude which only have support when the graviton energy $\omega$ is zero.

In the far-field limit, corresponding to large observer distance $r{\coloneqq }|\vec{x}|$  and time $t$ with fixed retarded time $u{\coloneqq} t{-}r$, \eqref{KMOCsubfinalforhmunu} can be   simplified to%
\footnote{Henceforth, we  omit an overall factor of $\int\!\prod_{j=1}^2d\Phi (p_j)  \, |\phi(p_1)|^2 |\phi (p_2)|^2$.} 
\begin{align}
	\label{KMOCsubfinalforhmunuBIS}
	\begin{split}
		\langle h^{\rm out}_{\mu \nu} (x) \rangle_{\rm \psi}  \!=\!     \frac{\kappa}{4\pi r} \,  
		\Big[
	\sum_h \varepsilon^{(h)\ast}_{\mu\nu} (\hat{k})  \int_0^{+\infty}\!\frac{d\omega}{2\pi} e^{-i \omega u } \,	 
		 {W}(b, k^h )    \, + \, {\rm h.c.}\Big]_{k= \omega (1, \mathbf{\hat{x}})}
		. 
	\end{split}
\end{align}
Alternatively, extending the $\omega$ integration from $-\infty$ to $+\infty$, 
\begin{align}
	\label{KMOCsubfinalforhmunuTER}
	\begin{split}
		&\langle h^{\rm out}_{\mu \nu} (x) \rangle_{\rm \psi}  =    \frac{\kappa}{4\pi r} \,   
	\sum_h   \int_{-\infty}^{+\infty}\!\frac{d\omega}{2\pi} e^{-i \omega u } \\ &
 \Big[ 	 
		 \varepsilon^{(h)\ast}_{\mu\nu}(\hat{k}) \, \theta(\omega)\,  \left.{W}\big(b, k^h\big)  \right|_{k= \omega (1, \mathbf{\hat{x}})}  
 +  \, \varepsilon^{(h)}_{\mu\nu}(\hat{k})\theta(-\omega)\,  \left.{W}^\ast\big(b, k^h \big) \right|_{k= -\omega (1, \mathbf{\hat{x}})}\Big]\, 
		.
	\end{split}
\end{align}
We now define%
\footnote{We comment that in our normalisations, the combination  $\langle h_+ {-} i h_\times\rangle$ is proportional to the strain $h(x)$, specifically   $h(x) {=} -(1/2) \langle h_+ {-} i h_\times\rangle$, where the strain  is related to the Newman-Penrose scalar $\Psi_4$ as $\Psi_4 {=} d^2h/ du^2$.}
\begin{align}
    \begin{split}
    \langle h_{+}\pm i h_{\times}\rangle &\coloneqq \langle h_{\mu \nu}^{\rm out}\rangle  \vareps^{\mu \nu}_{(\pm \pm)}
    \coloneqq\frac{1}{4\pi r }(h_{+}^\infty\pm i h_{\times}^\infty) \, .
    \end{split}
\end{align}
Using the properties of the positive/negative helicity polarisation vectors $\vareps_{\mu}^{(\pm) \ast} {=} \vareps_{\mu}^{(\mp) }$, $\vareps_{\mu}^{(\pm) \ast} \vareps^{\mu (\pm) }=-1$, $\vareps_{\mu}^{(\pm) \ast} \vareps^{\mu (\mp) }=0$, we get
\begin{align}
	\label{KMOCsubfinalforhmunuquater}
	\begin{split}
		& h^{\infty}_{+} \pm i  h^{\infty}_{\times}  =    
  \kappa    
	   \int_{-\infty}^{+\infty}\!\frac{d\omega}{2\pi} e^{-i \omega u } 
 \Big[ 	 
   \theta(\omega)\,  \left.{W}\big(b, k^\pm\big)  \right|_{k= \omega (1, \mathbf{\hat{x}})}  
 +  \, 
 \theta(-\omega)\,  \left. {W}^\ast\big(b,k^\mp\big) \right|_{k= -\omega (1, \mathbf{\hat{x}})}\Big]
		.
	\end{split}
\end{align}
We can now combine the two terms in \eqref{KMOCsubfinalforhmunuquater}. In order to do so, we  first  note that the five-point spinning amplitude has the form
\begin{align}
\label{decoddeven}
    -i \cM_{5, \rm HEFT} = \varepsilon_{\mu \nu}(k) m^{\mu\nu}\, , \qquad \text{with}\qquad
    m^{\mu \nu} = m^{\mu \nu}_{\rm even} +i m^{\mu \nu}_{\rm odd}\, , 
\end{align}
where $m^{\mu \nu}_{\rm even}$ and $m^{\mu \nu}_{\rm odd}$ are real, and contain even and odd powers of the spin, respectively. 
Then we observe that we can  separate out the $\omega$ dependence of the amplitude: we perform a rescaling of $q_{1,2}$ and define
\begin{equation}\label{eq:omegarescaling}
    q_{1,2}\coloneqq\omega\hat{q}_{1,2}, \qquad k\coloneqq\omega \hat{k}, \qquad w_{1,2}\coloneqq\omega\hat{w}_{1,2}\, , 
\end{equation}
where the $w_i$ variables were defined in \eqref{fiveiv}.
Then we have 
\begin{align}
\label{omegapower}
  \left.  \cM_{5, \rm HEFT}(q_1, q_2, k^h, a_1, a_2)  \right|_{S^n} = \frac{\omega^n}{\omega^2}\left.\cM_{5, \rm HEFT}(\hat{q}_1, \hat{q}_2, \hat{k}^h, a_1, a_2)\right|_{S^n}\, ,  
\end{align}
where $\left. \right|_{S^n}$ denotes the term containing $n$ powers of the spin in the HEFT amplitude.
Note that  $\cM_{5, \rm HEFT}(\hat{q}_1, \hat{q}_2, \hat{k}^h, a)$ is $\omega$-independent. 
Combining \eqref{decoddeven} and \eqref{omegapower} we find that 
\begin{align}
\left.    W^\ast (b, k^h)\right|_{k=-\omega(1, \hat{\mathbf{x}})} = 
\left.    W(b, k^{-h})\right|_{k=\omega(1, \hat{\mathbf{x}})}\, , 
\end{align}
and we can thus rewrite
\begin{align}
\label{combinedres}
   h^{\infty}_{+} \pm i  h^{\infty}_{\times} = 
\kappa \int_{-\infty}^{+\infty}\!
\frac{d\omega}{2\pi}e^{- i \omega u} 
 \left. 
W(b, k^{\pm})\right|_{k=\omega(1, \hat{\mathbf{x}})}
 \, . 
\end{align}
For convenience, in the following we will call this 
quantity
\begin{align}
\begin{split}
\label{KMOCsubfinalbis2}
 h^\infty(u) &\coloneqq 
 \kappa \int_{-\infty}^{+\infty}\!
\frac{d\omega}{2\pi}e^{- i \omega u} 
 \left. 
W(b, k)\right|_{k=\omega(1, \hat{\mathbf{x}})}\\ 
&= -i\kappa \int_{-\infty}^{+\infty}\!
\frac{d\omega}{2\pi}e^{- i \omega u} 
   \int\!  \frac{d^4q_1}{(2\pi)^{2}}  \delta(2  {\pb}_1\Cdot q_1 ) \delta(2  {\pb}_2\Cdot (k-q_1) )\  e^{iq_1\Cdot b} \ \mathcal{M}_{5,\rm HEFT} \, ,
\end{split}
\end{align}
leaving the dependence on the helicity understood, and where in all formulae  
$k{=}\omega(1, \hat{\mathbf{x}})$.

The above no longer appears manifestly real but in fact it is  (when expressed in a basis of real polarisation tensors) thanks to the properties of $-i \cM_{5, \rm HEFT}$ in \eqref{decoddeven} and \eqref{omegapower}. That is, a real term in the amplitude has an even power of the spin and hence after the re-scaling \eqref{eq:omegarescaling} is an even function of $\omega$; its Fourier transform is thus real. On the other hand, terms containing a factor of $i$ will feature an odd power of the spin and so are  odd functions of $\omega$; their Fourier transform is thus imaginary and this cancels the additional factor of $i$, with the final result   being real.
 
\subsection{A scalar warm-up}
Here we detail the computation of the scalar tree-level waveform, as a warm-up to the spinning case. Many of the simplifications we discuss here apply to the spinning waveform as well, in particular the intriguing fact that the computation boils down to a simple application of Cauchy's theorem. We begin with the expression for the waveform \eqref{KMOCsubfinalbis2}  derived in the previous section (here and for the rest of the paper we will drop the explicit bars on all of the variables to reduce clutter)
\begin{align}
\begin{split}
h^\infty(u)= -i\kappa \int_{-\infty}^{+\infty}\!
\frac{d\omega}{2\pi}e^{- i \omega u} 
   \int\!  \frac{d^4q_1}{(2\pi)^{2}}  \delta(2  {p}_1\Cdot q_1 ) \delta(2  {p}_2\Cdot (k-q_1) )\  e^{iq_1\Cdot b} \ \mathcal{M}_{5,\rm HEFT} \, .
\end{split}
\end{align}
First, we rescale the momentum transfers by $\omega$, as discussed above, introducing hatted momenta \eqref{eq:omegarescaling}.
The classical scalar amplitude then scales universally like $\omega^{-2}$, which cancels the power of $\omega^2$ coming from the change of variables, to give
\begin{align}
\begin{split}
h^\infty(u){=} {-i\kappa} \int_{-\infty}^{+\infty}\!
\frac{d\omega}{2\pi}e^{- i \omega u} 
   \int\!  \frac{d^4\hat{q}_1}{(2\pi)^{2}}  \delta(2  {p}_1\Cdot \hat{q}_1 ) \delta(2  {p}_2\Cdot (\hat{k}-\hat{q}_1) )\  e^{i\omega\hat{q}_1\Cdot b} \ \mathcal{M}_{5,\rm HEFT}(\hat{q}_1,\hat{q}_2,\hat{k}^h) \, .
\end{split}
\end{align}
In addition, it is useful to rescale the energy and retarded time by $\sqrt{-b^2}$, as  $\omega{\rightarrow}\omega/\sqrt{-b^2}$ and $u{\rightarrow}\sqrt{-b^2}u$. Effectively this means we are measuring the retarded time $u$ in units of $\sqrt{-b^2}$. With this choice, the tree-level waveform becomes
\begin{equation}
    h^\infty(u){=} \frac{-i\kappa}{\sqrt{-b^2}} \int_{-\infty}^{+\infty}\!
\frac{d\omega}{2\pi}e^{- i \omega u} 
   \int\!  \frac{d^4\hat{q}_1}{(2\pi)^{2}}  \delta(2  {p}_1\Cdot \hat{q}_1 ) \delta(2  {p}_2\Cdot (\hat{k}-\hat{q}_1) )\  e^{i\omega \frac{\hat{q}_1\Cdot b}{\sqrt{-b^2}}} \ \mathcal{M}_{5,\rm HEFT}(\hat{q}_1,\hat{q}_2,\hat{k}^h) \, .
\end{equation}
In fact, we are free to set $\sqrt{-b^2}=1$ in the expression above (and in all subsequent expressions) since $b^\mu$ only appears 
in the exponent through $b^\mu/\sqrt{-b^2}$. To restore $\sqrt{-b^2}$ we simply count the mass dimension of the expression, obtaining the $1/\sqrt{-b^2}$ factor above. Similarly, one can recover the original definition of the retarded time $u$ by counting mass dimension.

Next, it is useful to split the amplitude into the two terms coming from the BCFW diagrams \eqref{2BCFWdiagrams}. This gives us two contributions to the waveform, which we call $h_{q_1^2}^{\infty}(u)$ and $h_{q_2^2}^{\infty}(u)$,

\begin{equation}\label{eq: scalarWaveformCuts}
    h_{q_1^2,q_2^2}^{\infty}(u)=  {-i\kappa} \int_{-\infty}^{+\infty}\!
\frac{d\omega}{2\pi}e^{- i \omega u} 
   \int\!  \frac{d^4\hat{q}_1}{(2\pi)^{2}}  \delta(2  {p}_1\Cdot \hat{q}_1 ) \delta(2  {p}_2\Cdot (\hat{k}-\hat{q}_1) )\  e^{i\omega\hat{q}_1\Cdot b} \ \mathcal{M}_{q_1^2,q_2^2}(\hat{q}_1,\hat{q}_2,\hat{k}) \, .
\end{equation}
The two contributions $\mathcal{M}_{q_1^2}$ and $\mathcal{M}_{q_2^2}$ are related by the replacements $v_1{\leftrightarrow}v_2, q_1{\leftrightarrow} q_2$, which allows us to obtain the waveform contribution in  the $q_2^2$-channel from the $q_1^2$-channel.   To do this we perform the following replacements 
\begin{equation}\label{eq: scalar2from1}
    h_{q_1^2}^{\infty}(b\Cdot \hat{k} -u)\xrightarrow[]{v_1\leftrightarrow v_2} h_{q_2^2}^{\infty}(u)\,,
\end{equation}
which can be seen immediately using the definition \eqref{eq: scalarWaveformCuts}. The asymmetric shift in the proper time $u$ is due to our asymmetric choice of impact parameter in \eqref{psi}.

To compute the first cut, we decompose $\hat{q}_1$ onto a basis of four-vectors \cite{Cristofoli:2021vyo}
\begin{align}\label{eq: ScalarBasis}
    \hat{q}_1^\mu= z_1 v_1^\mu + z_2 v_2^\mu+ z_v v_\perp^\mu +z_b b^\mu\,,
\end{align}
where 
\begin{equation}
v_{\perp}\coloneqq\epsilon(v_1 \, v_2\,  b \, \bullet)\, ,
\end{equation}
and then change integration variables from $q_1$ to $z_1, z_2, z_v, z_b$.
In this parameterisation, we can use the two delta functions in \eqref{eq: scalarWaveformCuts} to localise the variables $z_1$ and  $z_2$~to 
\begin{equation}
    z_1= \frac{(v_2\Cdot \hat{k})(v_1\Cdot v_2)}{(v_1\Cdot v_2)^2-1}= \frac{\hat{w}_2 y}{y^2-1},\quad z_2= -\frac{v_2\Cdot \hat{k}}{(v_1\Cdot v_2)^2-1}= -\frac{\hat{w}_2 }{y^2-1}
    \, .
\end{equation}
The remaining integrals are then over $z_v,z_b$ and $\omega$,  
\begin{equation}
    h_{q_1^2}^{\infty}(u)= \frac{-i\kappa}{(4\pi)^2 m_1 m_2} \int_{-\infty}^{+\infty}\!
\frac{d\omega}{2\pi} dz_v dz_b e^{- i \omega (u+z_b)} \mathcal{M}_{q_1^2}\Big|_{z_1=\frac{\hat{w}_2 y}{y^2-1},\,z_2=-\frac{\hat{w}_2 }{y^2-1}}\,.
\end{equation}
The integral over $\omega$ also gives a delta function which we can immediately use to localise the $z_b$ integral, 
\begin{equation}\label{eq: scalarWaveformzvOnly}
    h_{q_1^2}^{\infty}(u)= \frac{-i\kappa}{(4\pi)^2 m_1 m_2} \int_{-\infty}^{+\infty}\!
dz_v  \, \mathcal{M}_{q_1^2}\Big|_{z_1=\frac{\hat{w}_2 y}{y^2-1},\,z_2=-\frac{\hat{w}_2 }{y^2-1},\, z_b=-u}\,.
\end{equation}
 To compute the final integral in $z_v$ we use  Cauchy's residue theorem, as done in  \cite{DeAngelis:2023lvf},
 hence we need to examine the pole structure of the $q_1^2$-cut. The integrand contains three types of poles in $z_v$ which arise from certain denominator structures in the tree-level amplitude. These are 
 \begin{align}\label{eq: scalarPhysicalPole}
       &\text{Physical pole:}\quad\frac{1}{q_1^2} \sim \frac{1}{(z_v-iA)(z_v+iA)},\\
       &\text{Spurious pole:}\quad \frac{1}{q_1^2\,q_1\Cdot k}\sim \frac{1}{(z_v-iA)(z_v+iA)(z_v-B)}\,,\label{eq: scalarSpuriousPole}\\
       &\text{Pole at infinity:}\, \begin{cases}
           \frac{z_v}{q_1^2}\sim \frac{z_v}{(z_v-iA)(z_v+iA)}\xrightarrow[z_v\rightarrow\infty]{}\frac{1}{z_v}\,,\\
           \frac{z_v^2}{q_1^2\,q_1\Cdot k}\sim \frac{z_v^2}{(z_v-iA)(z_v+iA)(z_v-B)}\xrightarrow[z_v\rightarrow\infty]{}\frac{1}{z_v}\,,\label{eq: scalarInfPole}
       \end{cases}
 \end{align}
 where $A$ and $B$ are real functions of the external kinematics. 
To compute the $z_v$ integral we will close the integration contour in the upper half plane to capture the pole at $z_v=iA$ and regulate the pole at infinity with a principal value prescription. This is equivalent to taking the integration limits $z_v{\rightarrow}-\infty$ and $z_v{\rightarrow}+\infty$ in a symmetric fashion, and implies that the pole at infinity receives an extra factor of $\frac{1}{2}$. The spurious pole at $z_v=B$ (coming from the factor $q_1\Cdot k$) lies on the integration contour, however we know that this pole cancels when we combine the two cuts in $q_1^2$ and $q_2^2$. Hence we are free to ignore the residue on this spurious pole since it would cancel at the end of the computation (as we have checked explicitly). 

In fact, we can further simplify the integral \eqref{eq: scalarWaveformzvOnly} using the following observations. First, the integral of one of the terms with a pole at infinity in \eqref{eq: scalarInfPole} is actually zero, 
\begin{equation}\label{eq: zeroIntegral}
    \int_{-\infty}^{+\infty}\!
dz_v  \frac{z_v}{(z_v-iA)(z_v+iA)}=0\,.
\end{equation}
This can be seen from the fact that the integrand is odd in $z_v$, or that the residue at $z_v=iA$ cancels with half the residue at infinity (recalling the principal value prescription mentioned earlier). The second term with a pole at infinity in \eqref{eq: scalarInfPole} can also be simplified as
\begin{align}\label{eq: infPoleSimplify}
    \frac{z_v^2}{(z_v{-}iA)(z_v{+}iA)(z_v{-}B)}&= \frac{\big((z_v{-}B){+}B\big)^2}{(z_v-iA)(z_v+iA)(z_v-B)}\nn\\
    &=\frac{B}{(z_v{-}iA)(z_v{+}iA)} +\frac{B^2}{(z_v{-}iA)(z_v{+}iA)(z_v{-}B)}+\cdots
\end{align}
where $+\cdots$ are terms which vanish after integration due to \eqref{eq: zeroIntegral}. The remaining terms above are in the form of \eqref{eq: scalarPhysicalPole} and \eqref{eq: scalarSpuriousPole}. Thus, after simplifications the only terms relevant to the waveform integral \eqref{eq: scalarWaveformzvOnly} are
\begin{equation}
    \frac{1}{q_1^2} \sim \frac{1}{(z_v-iA)(z_v+iA)}\,,\qquad \frac{1}{q_1^2q_1\Cdot k}\sim \frac{1}{(z_v-iA)(z_v+iA)(z_v-B)}\,,
\end{equation}
for which we only compute the residue on the physical pole $z_v=iA$. The computation for the second cut $\cM_{q_2^2}$ proceeds in an identical way, or alternatively we can obtain the second cut using the replacements \eqref{eq: scalar2from1}.

We have thus learned that the computation of the waveform can be efficiently reduced to the evaluation of residues on physical poles. The same general principle will be used in the spinning case.  The final expression for the scalar waveform is simply the sum of $h_{q_1^2}^\infty$ and $h_{q_2^2}^\infty$, and is included in the {\href{https://github.com/QMULAmplitudes/SpinningWaveformPublicData/tree/main}{GitHub repository}}. 

\begin{figure}
    \centering
\begin{tikzpicture}
	\begin{axis}[width=0.6\textwidth,
			title={Scalar waveforms},
			xmin=-4., xmax=4,
			ymin=-0.01, ymax=0.01,
			axis lines=left,
			compat=newest,
			xlabel=$u$, 
			ylabel= ${h}_{+}^{\infty}\,\,$, ylabel style={rotate=-90,at={(0,1)},anchor=south},
			legend pos=south east,
			ymajorgrids=true,
			grid style=dashed,
            every axis plot/.append style={thick},
		]

   \addplot[
			color=orange,
			mark=none,
		]
		coordinates {(-4.,-0.00325304)(-3.9,-0.00323509)(-3.8,-0.00321544)(-3.7,-0.00319389)(-3.6,-0.00317021)(-3.5,-0.00314416)(-3.4,-0.00311545)(-3.3,-0.00308375)(-3.2,-0.0030487)(-3.1,-0.0030099)(-3.,-0.00296687)(-2.9,-0.0029191)(-2.8,-0.00286601)(-2.7,-0.00280694)(-2.6,-0.00274117)(-2.5,-0.00266789)(-2.4,-0.00258624)(-2.3,-0.00249523)(-2.2,-0.00239386)(-2.1,-0.00228101)(-2.,-0.00215554)(-1.9,-0.00201628)(-1.8,-0.00186207)(-1.7,-0.00169181)(-1.6,-0.00150452)(-1.5,-0.00129941)(-1.4,-0.00107596)(-1.3,-0.00083402)(-1.2,-0.000573897)(-1.1,-0.000296456)(-1.,-3.18963*10^-6)(-0.9,0.000303732)(-0.8,0.000621449)(-0.7,0.000946442)(-0.6,0.00127463)(-0.5,0.0016015)(-0.4,0.00192233)(-0.3,0.00223239)(-0.2,0.0025272)(-0.1,0.0028028)(0.,0.00305594)(0.1,0.00328426)(0.2,0.00348634)(0.3,0.00366172)(0.4,0.00381085)(0.5,0.00393491)(0.6,0.00403569)(0.7,0.00411535)(0.8,0.00417629)(0.9,0.00422097)(1.,0.0042518)(1.1,0.00427104)(1.2,0.00428076)(1.3,0.00428277)(1.4,0.00427867)(1.5,0.0042698)(1.6,0.00425731)(1.7,0.00424214)(1.8,0.00422505)(1.9,0.00420666)(2.,0.00418747)(2.1,0.00416786)(2.2,0.00414815)(2.3,0.00412856)(2.4,0.00410928)(2.5,0.00409042)(2.6,0.0040721)(2.7,0.00405436)(2.8,0.00403725)(2.9,0.00402079)(3.,0.004005)(3.1,0.00398987)(3.2,0.0039754)(3.3,0.00396157)(3.4,0.00394837)(3.5,0.00393577)(3.6,0.00392374)(3.7,0.00391228)(3.8,0.00390134)(3.9,0.00389091)(4.,0.00388097)};
  
    \addplot[
			color=blue,
			mark=none,
		]
		coordinates {(-4.,-0.00511064)(-3.9,-0.0051026)(-3.8,-0.00509363)(-3.7,-0.0050836)(-3.6,-0.00507237)(-3.5,-0.00505975)(-3.4,-0.00504553)(-3.3,-0.0050295)(-3.2,-0.00501135)(-3.1,-0.00499076)(-3.,-0.00496735)(-2.9,-0.00494065)(-2.8,-0.00491012)(-2.7,-0.00487513)(-2.6,-0.00483489)(-2.5,-0.00478853)(-2.4,-0.00473496)(-2.3,-0.00467293)(-2.2,-0.00460095)(-2.1,-0.00451728)(-2.,-0.00441989)(-1.9,-0.00430644)(-1.8,-0.00417428)(-1.7,-0.00402043)(-1.6,-0.00384164)(-1.5,-0.0036345)(-1.4,-0.00339555)(-1.3,-0.00312157)(-1.2,-0.00280982)(-1.1,-0.00245849)(-1.,-0.00206704)(-0.9,-0.00163664)(-0.8,-0.00117044)(-0.7,-0.000673711)(-0.6,-0.000153793)(-0.5,0.000380156)(-0.4,0.000917602)(-0.3,0.00144724)(-0.2,0.00195772)(-0.1,0.0024385)(0.,0.00288066)(0.1,0.00327766)(0.2,0.00362583)(0.3,0.00392438)(0.4,0.00417515)(0.5,0.00438193)(0.6,0.00454974)(0.7,0.00468415)(0.8,0.00479064)(0.9,0.00487432)(1.,0.00493964)(1.1,0.00499039)(1.2,0.00502969)(1.3,0.00506006)(1.4,0.00508348)(1.5,0.00510153)(1.6,0.00511544)(1.7,0.00512614)(1.8,0.00513439)(1.9,0.00514073)(2.,0.00514561)(2.1,0.00514937)(2.2,0.00515226)(2.3,0.00515448)(2.4,0.00515619)(2.5,0.00515751)(2.6,0.00515852)(2.7,0.0051593)(2.8,0.0051599)(2.9,0.00516037)(3.,0.00516074)(3.1,0.00516104)(3.2,0.00516127)(3.3,0.00516147)(3.4,0.00516164)(3.5,0.00516178)(3.6,0.00516192)(3.7,0.00516204)(3.8,0.00516216)(3.9,0.00516228)(4.,0.00516239)};
  \addplot[
			color=red,
			mark=none,
		]
		coordinates {(-4.,-0.00866514)(-3.9,-0.00866419)(-3.8,-0.00866279)(-3.7,-0.00866085)(-3.6,-0.00865828)(-3.5,-0.00865496)(-3.4,-0.00865076)(-3.3,-0.00864553)(-3.2,-0.00863906)(-3.1,-0.00863113)(-3.,-0.00862145)(-2.9,-0.00860968)(-2.8,-0.0085954)(-2.7,-0.00857811)(-2.6,-0.00855716)(-2.5,-0.00853178)(-2.4,-0.008501)(-2.3,-0.00846362)(-2.2,-0.00841813)(-2.1,-0.00836265)(-2.,-0.0082948)(-1.9,-0.00821163)(-1.8,-0.00810941)(-1.7,-0.00798355)(-1.6,-0.00782838)(-1.5,-0.00763707)(-1.4,-0.00740162)(-1.3,-0.00711306)(-1.2,-0.00676199)(-1.1,-0.00633962)(-1.,-0.00583921)(-0.9,-0.00525783)(-0.8,-0.00459762)(-0.7,-0.00386607)(-0.6,-0.0030748)(-0.5,-0.0022374)(-0.4,-0.00136772)(-0.3,-0.000479602)(-0.2,0.000411815)(-0.1,0.00128869)(0.,0.00213054)(0.1,0.00291664)(0.2,0.00362986)(0.3,0.00425984)(0.4,0.00480405)(0.5,0.00526655)(0.6,0.00565562)(0.7,0.00598135)(0.8,0.00625395)(0.9,0.00648272)(1.,0.00667565)(1.1,0.00683937)(1.2,0.00697927)(1.3,0.00709966)(1.4,0.00720402)(1.5,0.00729509)(1.6,0.00737511)(1.7,0.00744584)(1.8,0.00750872)(1.9,0.00756493)(2.,0.00761544)(2.1,0.00766102)(2.2,0.00770235)(2.3,0.00773996)(2.4,0.00777432)(2.5,0.00780582)(2.6,0.00783479)(2.7,0.00786152)(2.8,0.00788624)(2.9,0.00790917)(3.,0.00793049)(3.1,0.00795036)(3.2,0.00796891)(3.3,0.00798628)(3.4,0.00800257)(3.5,0.00801787)(3.6,0.00803227)(3.7,0.00804584)(3.8,0.00805865)(3.9,0.00807077)(4.,0.00808224)};

\legend{{\rm $y$=9/8}, {\rm $y$=5/4}, {\rm $y$=3/2}, {\rm $y$=7/4}}
	\end{axis}
\end{tikzpicture}
 \caption{Scalar waveforms $h^{\infty}_+$ at various values of $y$.}
    \label{fig:scalarwaveform}
\end{figure}
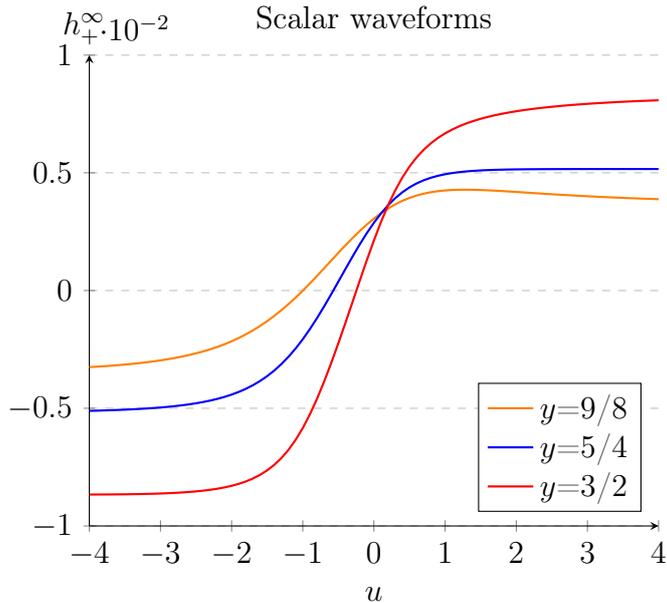

We can choose a frame such that the kinematics are given by
\begin{align}
    v_1&=(1,0,0,0),  ~~~~~~~~~~~~~~~~~~~~~~~~~~~~~~~~~~~~~~~~~~~~~~~~~~~~~~~~~v_2=(y,\sqrt{y^2-1},0,0)\nn\\
    \hat{k}&=(1,\sin \theta  \cos \phi ,\sin\theta  \sin \phi ,\cos \theta ), ~~~~~~~~~~~~~~~~~~~~~~~~~~~~~~\,v_\perp= (0,0,\sqrt{y^2-1},0) \nn\\
   \vareps^{(+)}&=\frac{1}{\sqrt{2}}\big(0,\cos \theta  \cos \phi -i \sin \phi ,\cos \theta  \sin \phi +i \cos\phi ,-\sin \theta  \big), ~~~b=(0, 0, 0, 1)\, ,
\end{align}
and then in Figure~\ref{fig:scalarwaveform} we present the scalar waveform at fixed angles $\theta={\pi \over 4}$ and $\phi={\pi \over 4}$ for various values of $y$.


\subsection{General expression of the time-domain waveform for arbitrary spins}\label{sec: spinningWFMethodGen}

We now turn to the spinning case. The first observation to make is that, in principle, the   Fourier transform to impact parameter space in \eqref{KMOCsubfinalbis2}   is ill-defined due to the  
 large-$q_1$ behaviour of the integrand giving rise to an ultraviolet (UV) divergence.
An elegant way to regularise this is to leave
the hyperbolic and exponential functions in the Compton amplitudes unexpanded (in the spin vectors),
introduce a new  spin parameter as
\begin{align}\label{eq: spinAnalyticCont}
    \tilde{a}_{1,2}\coloneqq i a_{1,2}\, , 
\end{align}
and temporarily take $\tilde{a}_{1,2}$ to be real. Assuming that the final spinning waveform has an expansion around $a_{1,2}\rightarrow 0$, this analytic continuation should not change the expansion coefficients. In support of this approach we mention that  the $a_1\rightarrow 0$ limit of our waveform gives the correct scalar result, and for $a_1\neq 0$ has the correct gravitational memory (computed in Section~\ref{sec-memory}); and finally, our results also agree with the recently derived waveform of \cite{DeAngelis:2023lvf}, obtained by expanding in spin and then integrating,   up to and including $\cO(a_1^4)$. Indeed, one can expand the amplitude in the spin parameters before integration, and the amplitude's degree of divergence would grow with each additional order in the spin. However, as we see in 
Section~\ref{sec: OrderBYOrderInSpin}, these divergences can be ignored since they only contribute to contact terms in $q_1^2$ and $q_2^2$, and both methods (the analytic continuation and expanding in spin before integration)  give the same result.

Proceeding now with the analytic continuation in the spin \eqref{eq: spinAnalyticCont}, we observe that 
in the large-$q_1$ limit, 
i.e.~$q_1{\rightarrow} \lambda \, q_1$ with $\lambda{\rightarrow} \infty$, the scaling behaviour of the amplitude is now   $\cO(\lambda^{-1})$ as $\lambda\to \infty$.
Pleasingly, this is precisely the same behaviour as that of the scalar amplitude. This logarithmic divergence will appear, identically to the scalar case, as a pole at infinity which we can again regulate with a principal value prescription.
The waveform is therefore well-defined once we tame this logarithmic divergence, 
\begin{align}
\begin{split}
\label{KMOCsubfinalbis3}
 h^\infty(u) 
  =-i\kappa &\int_{-\infty}^{+\infty}\!
\frac{d\omega}{2\pi}e^{- i \omega u} 
  \int \frac{d^4\hat{q}_1  }{(2\pi)^{2}}\delta(2  {p}_1\Cdot \hat{q}_1 ) \delta\big(2  {p}_2\Cdot (\hat{k}-\hat{q}_1) \big)e^{i\omega \hat{q}_1\Cdot b} \\
  &\omega^2\big(\mathcal{M}_{q_1^2}(\omega\hat{q}_1,\omega\hat{q}_2,\omega\hat{k})+\mathcal{M}_{q_2^2}(\omega\hat{q}_1,\omega\hat{q}_2,\omega\hat{k})\big)\, ,
\end{split}
\end{align}
where the hatted variables were defined in \eqref{eq:omegarescaling}. The factor of  $\omega^2$ comes from the re-scaled measure, while the amplitude itself depends on $\omega$ in a manner which we now describe.
Writing the hyperbolic functions within the expression of the Compton amplitude  \eqref{2BCFWdiagrams} in terms of exponential functions, we find that the tree-level amplitude can be rewritten as a linear combination of at most  eight exponential factors, with a very simple frequency dependence. Specifically, we find that   only three different powers of the frequency $\omega$ can appear  for arbitrary classical spins,  
\begin{align}  
\begin{split}\cM_{q_1^2}&=\sum_{\cpm_1,\cpm_2,\cpm_3{=}\pm1}{e^{\cpm_1 i\omega\ta_1\mdot \hat{q}_1+\cpm_2 i\omega\ta_2\mdot \hat{q}_1+\cpm_3 i\omega\ta_2\mdot \hat{k}}  }{1\over\omega^2}\Big(\sum_{i=0}^{2}\cM^{(i)}_{q_1^2}(\cpm_1,\cpm_2,\cpm_3)\omega^i\Big)\, , \\    \cM_{q_2^2}&=\sum_{\cpm_1,\cpm_2,\cpm_3{=}\pm1}e^{\cpm_1 i\omega\ta_2\mdot \hat{q}_2+\cpm_2 i\omega\ta_1\mdot \hat{q}_2+\cpm_3 i\omega\ta_1\mdot \hat{k}} {1\over\omega^2}\Big(\sum_{i=0}^{2}\cM^{(i)}_{q_2^2}(\cpm_1,\cpm_2,\cpm_3 )\omega^i\Big)\, , 
\end{split}
\end{align} 
where the sum is extended to all values of  $\rho_i\!\in\!\!\{{-}1,1\}$, $i{=}1,2,3$.
Note that the $\cM^{(i)}_{q_2^2}(\cpm_1,\cpm_2,\cpm_3)$ now do not depend on $\omega$ and thus are functions of the hatted variables $\hat{q}_1,\,\hat{q}_2,\, \hat{k}$.
Hence, the  waveform integral  has a simple general structure.  In the remainder of this section we will focus on $\cM^{(i)}_{q_1^2}$,  and  the case of  $\cM^{(i)}_{q_2^2}$ is similar. 
Similarly to the scalar case, the four-dimensional integration is immediately reduced to a two-dimensional one using the $\delta$-functions in \eqref{KMOCsubfinalbis3}.
Furthermore, for each exponential factor, the Fourier transform to the time domain generates a third delta function, which  constrains   the integration  over $\hat{q}_1$    to the hyperplane defined by%
\begin{align}
   b\mdot \hat{q}_1+ \cpm_1 \ta_1\mdot \hat{q}_1+\cpm_2 \ta_2\mdot \hat{q}_1+\cpm_3 \ta_2\mdot \hat{k}-u=0
\, . 
\end{align}
Following similar manipulations to \eqref{eq: infPoleSimplify} in the scalar case, the master integrands are of the form
\begin{align}\label{eq:masterIntegral}
  {1\over (\hat{q}_1\mdot X + Y\mdot  Z) \hat{q}_1^2},&&{1 \over  \hat{q}_1^2},&&{\hat{q}_1\mdot W \over  \hat{q}_1^2},
\end{align}
where $W$ can be  chosen to be orthogonal to the localising hyperplane
and $\hat{q}_1\mdot X + Y\mdot  Z$ denotes a generic \textit{spurious} pole linear in $\hat{q}_1$ and featuring external vectors $X, Y$ and $Z$ which may be the spins $a_i$ or $k$. 
The first two master integrals are UV convergent, while the last one is logarithmically divergent. However, the last master integral is an odd function of $q_1$, and hence vanishes when integrated on a symmetric domain, identically to \eqref{eq: zeroIntegral} in the scalar case.  This corresponds to a principal value (PV) regularisation of the divergent integral, or equivalently a PV regularisation of the pole at infinity. 
With this regularisation,  the residue of the pole at $\hat{q}_1^2 =0$ of the third term in \eqref{eq:masterIntegral} cancels the residue of the pole at infinity. 
Therefore, we can drop the last master integral altogether.

Now that the pole at infinity has been removed,  we can perform the integration of the remaining terms using  Cauchy's theorem on the finite poles. 
 There is only one physical pole in this channel, namely $ \hat{q}_1^2 {=}0$. The residues on the spurious poles $\hat{k} \mdot \hat{q}_1$ in the integrand can be discarded since they cancel when combining with the  $\hat{q}_2^2$-channel,   a fact we have confirmed by explicit calculations.  
The residues of the spin-dependent spurious poles in the three-point and Compton amplitudes (coming from the entire functions $G_i$) cancel when performing an expansion in the spins $a_1$ and $a_2$, and so they can also be ignored. A similar statement holds for these poles in the final integrated waveform.

In summary, the closed-form expression  of the  time-domain waveform with arbitrary spin at tree level is then
\begin{align}
\label{eq:WFclosedForm}
\begin{split}
  h^\infty(u)=&  -i\kappa \sum_{\cpm_1,\cpm_2,\cpm_3{=}\pm1}\sum_{j=0}^{2}(i\partial_u)^{j} \Big[
\ointctrclockwise_{(\hat{q}_1^2)^+=0} \frac{d^4\hat{q}_1}{(2\pi)^{2}} \delta(2  {v}_1\Cdot \hat{q}_1 ) \delta(2  {v}_2\Cdot (\hat{k}-\hat{q}_1) )\\
  &\delta(b\mdot \hat{q}_1+ \cpm_1 \ta_1\mdot \hat{q}_1+\cpm_2 \ta_2\mdot \hat{q}_1+\cpm_3 \ta_2\mdot \hat{k}-u)\, \cM^{(i)}_{q_1^2,{\rm fin}}(\cpm_1,\cpm_2,\cpm_3)\\
+& \ointctrclockwise_{(\hat{q}_2^2)^+=0} \frac{d^4\hat{q}_1}{(2\pi)^{2}} \delta(2  {v}_1\Cdot \hat{q}_1 ) \delta(2  {v}_2\Cdot (\hat{k}-\hat{q}_1) )\\
&\delta(b\mdot \hat{q}_1+ \cpm_1 \ta_2\mdot \hat{q}_2+\cpm_2 \ta_1\mdot \hat{q}_2+\cpm_3 \ta_1\mdot \hat{k}-u)\,\cM^{(i)}_{q_2^2,{\rm fin}}(\cpm_1,\cpm_2,\cpm_3)\Big]\, , 
\end{split}
\end{align}
where $\cM^{(i)}_{q_j^2, {\rm fin}}$ denotes the UV-convergent part of the amplitude coming from the first two master integrals in \eqref{eq:masterIntegral}.  We denote as $(\hat{q}_{1,2}^2)^+ =0$  the physical poles in the upper half plane.

\section{The waveform from the scattering of a Schwarzschild and a Kerr black hole}

\label{sec:6}

In this paper, we will focus on the 
case where the first black hole is spinning while the second   is spinless, that is $a_2 {=}0$.
Furthermore, in order to show explicitly powers of the frequency $\omega$, in this section  we rescale $k$, $q$ and the $w_i$ by $\omega$, as in \eqref{eq:omegarescaling}, dropping the hats on these rescaled quantities in order not to clutter formulae. 

\subsection{The $q_1^2$-channel}
\label{sec:q1^2channel}
For the contribution to the amplitude in the $q_1^2$-channel, the waveform integrand is obtained from gluing a three-point spinning amplitude with a four-point spinless amplitude. The amplitude in this channel is very simple thanks to our restriction $a_2=0$, and from \eqref{eq:cutq1} we obtain, up to overall constant pre-factors
\begin{align}
\begin{split}
 \omega ^2 e^{i \omega  b\cdot q_1-i  \omega u}\cM_{q_1^2}
 &=\omega ^2 e^{i \omega  b\cdot q_1-i  \omega u}\Big(\frac{c_1 \cosh \left(\omega  a_1\mdot q_1\right)}{\omega ^2 q_1\mdot q_1 k\mdot q_1} + \frac{c_2 \left(q_1\mdot F_k\mdot v_1\right){}^2 \cosh \left(\omega  a_1\mdot q_1\right)}{\omega ^2 q_1\mdot q_1 k\mdot q_1}\\
 & +
 \frac{c_3 \left(q_1\mdot F_k\mdot v_1\right){}^2 q_1\mdot S_1\mdot v_2 G_1\left(\omega  a_1\mdot q_1\right)}{\omega  q_1\mdot q_1 k\mdot q_1}+\frac{c_4 q_1\mdot F_k\mdot v_1 k\mdot S_1\mdot q_1 G_1\left(\omega  a_1\mdot q_1\right)}{\omega  q_1\mdot q_1 k\mdot q_1}\\
 &+\frac{c_5 G_1\left(\omega  a_1\mdot q_1\right) q_1\mdot S_1\mdot F_k\mdot v_1}{\omega  q_1\mdot q_1}+\frac{c_6 q_1\mdot F_k\mdot v_1 \cosh \left(\omega  a_1\mdot q_1\right)}{\omega ^2 q_1\mdot q_1 k\mdot q_1}\\
   &+\frac{c_7 q_1\mdot F_k\mdot v_1 q_1\mdot S_1\mdot v_2 G_1\left(\omega  a_1\mdot q_1\right)}{\omega  q_1\mdot q_1 k\mdot q_1}
   +\frac{c_8 k\mdot S_1\mdot q_1 G_1\left(\omega  a_1\mdot q_1\right)}{\omega  q_1\mdot q_1 k\mdot q_1}
   \\ & 
   +\frac{c_9 k\mdot q_1 \cosh \left(\omega  a_1\mdot q_1\right)}{\omega ^2 q_1\mdot q_1}+\frac{c_{10} q_1\mdot F_k\mdot v_1 \cosh \left(\omega  a_1\mdot q_1\right)}{\omega ^2 q_1\mdot q_1}
   \\
   &+\frac{c_{11} k\mdot q_1 q_1\mdot S_1\mdot v_2 G_1\left(\omega  a_1\mdot q_1\right)}{\omega  q_1\mdot q_1}+\frac{c_{12} q_1\mdot F_k\mdot v_1 q_1\mdot S_1\mdot v_2 G_1\left(\omega  a_1\mdot q_1\right)}{\omega  q_1\mdot q_1}\\
   &+\frac{c_{13} k\mdot S_1\mdot q_1 G_1\left(\omega  a_1\mdot q_1\right)}{\omega  q_1\mdot q_1} +\frac{c_{14} \cosh \left(\omega  a_1\mdot q_1\right)}{\omega ^2 q_1\mdot q_1}
   \\ & +\frac{c_{15} q_1\mdot S_1\mdot v_2 G_1\left(\omega  a_1\mdot q_1\right)}{\omega  q_1\mdot q_1}
   +\frac{c_{16} q_1\mdot F_k\mdot v_1 G_1\left(\omega  a_1\mdot q_1\right) q_1\mdot S_1\mdot F_k\mdot v_1}{\omega  q_1\mdot q_1 k\mdot q_1}\\
   &+\frac{c_{17} G_1\left(\omega  a_1\mdot q_1\right) q_1\mdot S_1\mdot F_k\mdot v_1}{\omega  q_1\mdot q_1 k\mdot q_1}
   \Big)\, ,
\end{split}
\end{align}
where the coefficients $c_i$ are independent of $q_1$  and $\omega$,and hence can be factored out in the waveform integration, their precise form is given in Appendix~\ref{app: integrnadcoeffs}. For this channel the amplitude scales with $\omega$ as $\omega^0$ with the remaining $\omega$ dependence exponentiating. 
In this channel, there are only two sectors from the exponential   factors:
\begin{align}
 {\rm (I)}:  \quad e^{-i \omega  \left(-\tilde{a}_1\mdot q_1-b\mdot q_1+u\right)}, && {\rm (II)}: \quad e^{-i \omega  \left(\tilde{a}_1\mdot q_1-b\mdot q_1+u\right)}.
\end{align}
Again we have the parameterisation of $q_1$ on the four-dimensional vector basis given by the vectors 
\begin{align}\label{eq:Basis}
    v_1, v_2, b, v_\perp.
\end{align}
As in the scalar case, we temporarily set $b^2=-1$ which means regarding the spins $a_i$ and retarded time $u$ as dimensionless and measured in units of $\sqrt{{-}b^2}$. The overall dependence on $b$ can then be reinstated by counting of mass dimension and gives simply a prefactor of $\frac{1}{\sqrt{{-}b^2}}$.

However, the parameterisation \eqref{eq:Basis} is not well suited  to the particular sectors and does not cleanly identify the 
UV-divergent term in \eqref{eq:masterIntegral}. It is more convenient to introduce a 
sector-dependent basis as 
\begin{align}\label{eq:secBasis}
    v_1, v_2, \tilde b_j, \tilde v_j,
\end{align}
where 
in each sector we introduce an effective impact parameter
\begin{align}
    \begin{cases}
        \tilde b_{(\rm I)}\coloneqq-b-\tilde{a}_1&\\
        \tilde b_{(\rm II)}\coloneqq-b+\tilde{a}_1&
    \end{cases},
\end{align}
and correspondingly
\begin{align}
    \tilde v_j\coloneqq\eps(v_1 v_2 \tilde b_j  \bullet),\, \quad j=\rm I,II\,.
\end{align}
We then parameterise 
 $q_1$ as 
\begin{align}
	\label{qparam}
	q_1= z_1 v_1 + z_2 v_2 + z_v \tilde v_j + z_b \tilde{b}_j,\quad j=\rm I,II\,,
\end{align}
in terms of the basis vectors defined above. 
The divergent part in  \eqref{eq:masterIntegral} is  then of the form 
\begin{align}
    {z_v\over c-z_v^2},
\end{align}
which vanishes once we perform the integration as in the scalar case;  hence we drop such terms.

\textbf{Examples with constrained spin: } In this paper we present results for the case where the Kerr black hole spin $a_1$ satisfies the additional constraint
\begin{align}
    \tilde a_1\mdot v_2=0
\, . 
\end{align} In this case $b, \tilde a_1$ are both constrained to the hyperplane  orthogonal to $v_1$ and  $v_2$ as $\tilde a_1{\mdot} v_i{=}b{\mdot} v_i{=}0$. The $q_1$ variable is also constrained to another  parallel hyperplane defined by  $q_1\mdot v_1{=}0, q_1\mdot v_2{=}k{\mdot} v_2{=}w_2$. 
Then the extra $\delta$-function after the time-domain Fourier transform is, in the two sectors, 
\begin{align}\label{eq: sectors}
 \quad \delta \left(-\tilde{a}_1\mdot q_1-b\mdot q_1+u\right)=\delta(\tilde{b}_{(\rm I )}\mdot q_1+u), && \quad  \delta \left(\tilde{a}_1\mdot q_1-b\mdot q_1+u\right)=\delta(\tilde{b}_{(\rm II )}\mdot q_1+u)
\end{align}
and the $q_1$ integral  localises to the line as shown in the following figure, 
\begin{align}
\label{vectors}
    \begin{tikzpicture}[baseline={([yshift=-0.0ex]current bounding box.center)},extended line/.style={shorten >=-#1,shorten <=-#1},
  extended line/.default=1cm]\tikzstyle{every node}=[font=\small]
		\draw[thick,->] (0,0) -- (4.5,0);
  \draw  (0.4,2.5) node {$v_\perp$};
\draw[thick,->] (0,-2.5) -- (0,2.5);
\draw[ultra thick,blue,->] (0,0) -- (3.5,0);
\draw  (3.5,-0.3) node {$-b$};
\draw[ultra thick,blue,->] (0,0) -- (2,-2) ;
\draw (2,-2) node {~~~~~~~~~$-b-\tilde{a}$};
\draw[ultra thick,blue,->] (0,0) -- (5,2);
\draw (5,2) node {~~~~~~~~~$-b+\tilde{a}$};
\draw [ extended line, ultra thick,red] ($(5,2)!(1.5,0)!(0,0)$) -- (1.5,0);
\draw [ extended line, ultra thick,red] ($(2,-2)!(1.9,0)!(0,0)$) -- (1.9,0);
\fill [black] ($(5,2)!(1.5,0)!(0,0)$) circle [radius=2pt] node {$~~~~~{-}u$};
\fill [black] ($(2,-2)!(1.9,0)!(0,0)$) circle [radius=2pt] node {$~~~~~{-}u$};
	\end{tikzpicture} &\xrightarrow[ \text{vector  basis} ]{\text{changing}} & 
  \begin{tikzpicture}[baseline={([yshift=-0.0ex]current bounding box.center)},extended line/.style={shorten >=-#1,shorten <=-#1},
  extended line/.default=1cm]\tikzstyle{every node}=[font=\small]
		\draw[thick,->] (0,0) -- (4.5,0);
  \draw  (0.4,4.5) node {$\tilde v$};
\draw[thick,->] (0,0) -- (0,4.5);
\draw[ultra thick,blue,->] (0,0) -- (3.5,0);
\draw  (3.5,-0.3) node {$\tilde b$};
\draw [extended line, ultra thick,red] ($(3.5,0)!(1.5,0)!(0,0)$) -- (1.5,0) ;
\fill [black] ($(3.5,0)!(1.5,0)!(0,0)$) circle [radius=2pt];
\draw  (1.7,-0.3) node {$-u$};
	\end{tikzpicture} 
 \end{align}
The plane depicted here is the one orthogonal to $v_1$ and $v_2$ which corresponds to the integrals over $z_v$ and $z_b$ in each sector \eqref{qparam}. 
The variable $\tilde b\mdot q_1= \tilde{b}^2 z_b$ is localised to $-u$  using \eqref{eq: sectors} and the integral over $z_v$ is taken along the red line orthogonal to the basis vector $\tilde b$. In the following we use $\tilde b_{\rm (I)},\tilde b_{\rm (II)}$ to denote the shifted impact parameters in the two sectors, and $\tilde v_{(\rm I)},\tilde v_{(\rm II)}$ to denote the corresponding orthogonal directions. We also note that when we replace back $\tilde a_1 = i a_1$  in terms of the physical spin the quantities $\tilde b_{\rm (I)},\tilde b_{\rm (II)}$ and $\tilde v_{(\rm I)},\tilde v_{(\rm II)}$ are complex conjugates of each other.

We now go into some explicit examples. First, consider the term
\begin{equation}
    \frac{c_{14} \cosh \left(\omega  a_1\mdot q_1\right) e^{i \omega  b\mdot q_1-i  \omega u}}{q_1\mdot q_1}.
\end{equation}
Then according to \eqref{eq:WFclosedForm}, we need to sum over the two sectors and get
\begin{align}
    \frac{c_{14}}{4 \sqrt{w_2^2 \tilde{b}_{\rm (I)}\mdot \tilde{b}_{\rm (I)}-u^2 \left(y^2-1\right)}}+\frac{c_{14}}{4 \sqrt{w_2^2 \tilde{b}_{\rm (II)}\mdot \tilde{b}_{\rm (II)}-u^2 \left(y^2-1\right)}},
\end{align}
which is a real result since $\tilde b_{\rm (I)},\tilde b_{\rm (II)}$ are a complex conjugate pair. This is a general feature of the integrals encountered in the following calculation, namely when replacing $\tilde a_1 = i a_1$ the sector variables are complex but appear in combinations such that the resulting waveform is real (for a basis of real polarisations).
A second example is 
\begin{align}
   \frac{c_{13} \omega  k\mdot S_1\mdot q_1 e^{i \omega  b\mdot q_1-i  \omega  u} G_1\left(-i \omega  \tilde{a}_1\mdot q_1\right)}{q_1\mdot q_1}.
\end{align}
The contribution to the waveform is 
\begin{align}
  {i c_{13} \over 4} \Bigg(\frac{ -k\mdot S_1\mdot \tilde{v}_{\rm (I)} \sqrt{w_2^2 \tilde{b}_{\rm (I)}\mdot \tilde{b}_{\rm (I)}+u^2 \left(1-y^2\right)}+u \left(y^2-1\right) \tilde{b}_{\rm (I)}\mdot S_1\mdot k-w_2 \tilde{b}_{\rm (I)}\mdot \tilde{b}_{\rm (I)} k\mdot S_1\mdot v_2}{ u \left(y^2-1\right) \tilde{a}_1\mdot \tilde{b}_{\rm (I)} \sqrt{w_2^2 \tilde{b}_{\rm (I)}\mdot \tilde{b}_{\rm (I)}-u^2 \left(y^2-1\right)}-\tilde{a}_1\mdot \tilde{v}_{\rm (I)} \left(u^2 \left(y^2-1\right)-w_2^2 \tilde{b}_{\rm (I)}\mdot \tilde{b}_{\rm (I)}\right)}\nn\\
   -\frac{-k\mdot S_1\mdot \tilde{v}_{\rm (II)} \sqrt{w_2^2 \tilde{b}_{\rm (II)}\mdot \tilde{b}_{\rm (II)}+u^2 \left(1-y^2\right)}+u \left(y^2-1\right) \tilde{b}_{\rm (II)}\mdot S_1\mdot k-w_2 \tilde{b}_{\rm (II)}\mdot \tilde{b}_{\rm (II)} k\mdot S_1\mdot v_2}{u \left(y^2-1\right) \tilde{a}_1\mdot \tilde{b}_{\rm (II)} \sqrt{w_2^2 \tilde{b}_{\rm (II)}\mdot \tilde{b}_{\rm (II)}-u^2 \left(y^2-1\right)}-\tilde{a}_1\mdot \tilde{v}_{\rm (II)} \left(u^2 \left(y^2-1\right)-w_2^2 \tilde{b}_{\rm (II)}\mdot \tilde{b}_{\rm (II)}\right)}\Bigg).
\end{align}
In this form, the poles which depend on the spin vector $a_1$ are due to the spurious pole in the $G_1$ function. As with the $G_1$ function itself, this pole explicitly cancels once we expand for $|a_1|\ll 1$ giving
\begin{align}
  &-\frac{c_{13} w_2 \left(u w_0 \left(y^2-1\right) \tilde{a}_1\mdot b+\left(-u w_3 y^2+u w_3+w_1 w_2 y-w_2^2\right) \tilde{a}_1\mdot v_\perp\right)}{2 \left(y^2-1\right) \left(-u^2 \left(y^2-1\right)-w_2^2\right){}^{3/2}}\nn\\
  &+  \frac{c_{13} w_2^3\left(u w_0 \left(y^2-1\right) \tilde{a}_1\mdot b+\left(-u w_3 y^2+u w_3+w_1 w_2 y-w_2^2\right) \tilde{a}_1\mdot v_\perp\right)}{4 \left(y^2-1\right)^2 \left(-u^2 \left(y^2-1\right)-w_2^2\right){}^{7/2}}  \nn\\
  &\times\left(\left(y^2-1\right) \left(u^2 \left(y^2-1\right)-4 w_2^2\right) \left(\tilde{a}_1\mdot b\right){}^2+\left(u^2 \left(y^2-1\right)+w_2^2\right) \left(\tilde{a}_1\mdot v_\perp\right)^2\right)+\cdots,
\end{align}
where $w_1:=k\mdot v_1,w_2:=k\mdot v_2,w_3:=k\mdot b,w_0:=k\mdot v_\perp$. 

A third example is 
\begin{align}
    \frac{c_3 \omega  \left(q_1\mdot F_k\mdot v_1\right){}^2 q_1\mdot S_1\mdot v_2 e^{i \omega  b\mdot q_1-i u \omega } G_1\left(-i \omega  \tilde{a}_1\mdot q_1\right)}{q_1\mdot q_1 k\mdot q_1}.
\end{align}
In this case, there is a trivial log-divergent term which we remove using the method described in Section~\ref{sec: spinningWFMethodGen}. Thus the integral gives 
\begin{align}
    &\sum_{j=\rm I}^{\rm II}{(-1)^{j-1}\over 4(y^2-1)}\Bigg[\frac{i c_3  \left(\tilde{v}_j\mdot S_1\mdot v_2 \sqrt{w_2^2 \tilde{b}_j\mdot \tilde{b}_j-u^2 \left(y^2-1\right)}+u \left(y^2-1\right) \tilde{b}_j\mdot S_1\mdot v_2\right)}{ k\mdot \tilde{v}_j \sqrt{w_2^2 \tilde{b}_j\mdot \tilde{b}_j+u^2 \left(1-y^2\right)}+u \left(y^2-1\right) \tilde{b}_j\mdot k+w_2 \left(w_2-w_1 y\right) \tilde{b}_j\mdot \tilde{b}_j}\nn\\
    &{\left(\tilde{v}_j\mdot F_k\mdot v_1 \sqrt{w_2^2 \tilde{b}_j\mdot \tilde{b}_j+u^2 \left(1-y^2\right)}+u \left(y^2-1\right) \tilde{b}_j\mdot F_k\mdot v_1-w_2 \tilde{b}_j\mdot \tilde{b}_j v_1\mdot F_k\mdot v_2\right)^2\over  \tilde{b}_j\mdot \tilde{b}_j \sqrt{w_2^2 \tilde{b}_j\mdot \tilde{b}_j-u^2 \left(y^2-1\right)}\left(\tilde{a}_1\mdot \tilde{v}_j \sqrt{w_2^2 \tilde{b}_j\mdot \tilde{b}_j-u^2 \left(y^2-1\right)}+u \left(y^2-1\right) \tilde{a}_1\mdot \tilde{b}_j\right) }\nn\\
    &-\frac{i c_3 \left(\tilde{v}_j\mdot F_k\mdot v_1\right){}^2 \tilde{v}_j\mdot S_1\mdot v_2}{ \tilde{b}_j\mdot \tilde{b}_j \tilde{a}_1\mdot \tilde{v}_j k\mdot \tilde{v}_j}\Bigg].
\end{align}
The last term is the removed log-divergent term  that can be removed trivially.  
One can also directly check that the spurious poles $1\over \tilde a_1\mdot \tilde v_j$ and $1\over \left(\tilde{a}_1\mdot \tilde{v}_j \sqrt{w_2^2 \tilde{b}_j\mdot \tilde{b}_j-u^2 \left(y^2-1\right)}+u \left(y^2-1\right) \tilde{a}_1\mdot \tilde{b}_j\right) $ cancel among the sectors.  Again, by expanding for $|{a}_1|\ll 1$ we see that the spurious poles cancel,
\begin{align}
&\frac{-i c_3 w_2^2 \left(v_1\mdot F_k\mdot v_2\right){}^2 b\mdot S_1\mdot v_2}{2 \left(y^2-1\right) \left(u^2 \left(1-y^2\right)-w_2^2\right){}^{3/2}} \frac{1}{\big(w_1 w_2 y-w_2^2-w_0 \sqrt{u^2 \left(1-y^2\right)-w_2^2}-u w_3 \left(y^2-1\right)\big)^2}\nn\\
&
\times\Big({-w_0 (u^2 (y^2-1)-w_2^2) \sqrt{u^2 \left(1-y^2\right)-w_2^2}- u^3 \left(y^2-1\right)^2 w_3 -w_1 w_2^3 y+w_2^4}\Big)\nn\\
   &+\cdots\,.
\end{align} 
The spin-independent spurious pole is still present and will only cancel after  summing with the corresponding terms in the $q_2^2$-channel.

\subsection{The $q_2^2$-channel}
\label{sec:q2^2channel}
For the second graph in \eqref{2BCFWdiagrams}, the physical propagator is 
\begin{align}
    {1\over q_2^2}={1\over (k-q_1)^2}.
\end{align}
It is convenient to shift the integration variable as $q_1\rightarrow q_1+k$ and the physical propagator becomes simply ${1\over q_1^2}$, the same as in the $q_1^2$-channel. The spurious pole $1\over k\mdot q_1$ is invariant under the shift due to the on-shell condition of the external graviton, while the spin dependent spurious poles become 
\begin{align}
    {1\over a\mdot q_1}, && {1\over a\mdot k}.
\end{align}
The delta functions coming from the definition of the waveform \eqref{KMOCsubfinalbis2} are shifted correspondingly as 
\begin{align}
    \delta(2v_1\mdot q_1+2w_1), && \delta(2v_2\mdot q_1).
\end{align}
Applying the residue theorem to evaluate the integrals is then the exact same process as the $q_1^2$-channel with the following integrand
\begin{align}
&( \omega ^2 e^{i \omega  \left(b\mdot k+b\mdot q_1\right)-i \omega u})\cM_{q_2^2}
=( \omega ^2 e^{i \omega  \left(b\mdot k+b\mdot q_1\right)-i \omega u})\times \nn\\
 &\Bigg[\frac{\cosh \left(\omega  a_1\mdot (k+ q_1)\right) \left(2 w_1^2 \left(v_1\mdot F_k\mdot v_2\right)^2-4 w_1 y v_1\mdot F_k\mdot v_2 q_1\mdot F_k\mdot v_1+\left(2 y^2-1\right) \left(q_1\mdot F_k\mdot v_1\right){}^2\right)}{4 w_1^2 \omega ^2 q_1\mdot q_1 k\mdot q_1}\nn\\
 &+\frac{i G_1 \left(\omega  a_1\mdot (k+ q_1)\right) }{2 w_1^2 \omega  q_1\mdot q_1 k\mdot q_1 } \Big(w_1 v_1\mdot F_k\mdot v_2-y q_1\mdot F_k\mdot v_1\Big)\Big(-w_2 q_1\mdot S_1\mdot F_k\mdot v_1\nn\\
 &+k\mdot S_1\mdot v_2 q_1\mdot F_k\mdot v_1+q_1\mdot S_1\mdot v_2 q_1\mdot F_k\mdot v_1+v_1\mdot F_k\mdot v_2 k\mdot S_1\mdot q_1-w_2 k\mdot S_1\mdot F_k\mdot v_1\Big)\nn\\
 &+\frac{G_2\left(\omega  a_1\mdot q_1,\omega  a_1\mdot k\right) \left(c_{33} q_1\mdot S_1\mdot F_k\mdot v_1+\left(c_{25} q_1\mdot S_1\mdot v_2+c_{48}\right) q_1\mdot F_k\mdot v_1+c_{41} q_1\mdot S_1\mdot v_2+c_{19}\right)}{q_1\mdot q_1}\nn\\
 &+\frac{G_1 \left(\omega  a_1\mdot (k+ q_1)\right) \left(c_{29} q_1\mdot F_k\mdot v_1+c_{43}\right)}{\omega  q_1\mdot q_1}+{G_1\left(\omega  a_1\mdot k\right)G_1\left(\omega  a_1\mdot q_1\right)\over  q_1\mdot q_1}\Big(c_2+c_{68} q_1\mdot F_k\mdot v_1\nn\\
 & +a_1\mdot q_1 \left(c_{63} q_1\mdot F_k\mdot v_1+c_{70}\right)+c_{62} \left(q_1\mdot F_k\mdot v_1\right){}^2+k\mdot q_1 \left(c_7 q_1\mdot F_k\mdot v_1+c_{11}\right)\Big)\nn\\
 &+{G_1\left(\omega  a_1\mdot k\right)\cosh \left(\omega  a_1\mdot q_1\right)\over \omega q_1\mdot q_1} \left(c_{12} q_1\mdot F_k\mdot v_1+c_{51}\right)+G''_e\left(\omega  a_1\mdot q_1,\omega  a_1\mdot k\right) \omega ^2\Big(\frac{c_4  k\mdot q_1}{q_1\mdot q_1}\nn\\
 &+\frac{c_{16}  a_1\mdot q_1 k\mdot q_1}{q_1\mdot q_1}+\frac{c_{58}  a_1\mdot q_1}{q_1\mdot q_1}+\frac{c_{74}  \left(a_1\mdot q_1\right){}^2}{q_1\mdot q_1}+\frac{c_{57}  q_1\mdot F_k\mdot v_1}{q_1\mdot q_1}+\frac{c_{72}  a_1\mdot q_1 q_1\mdot F_k\mdot v_1}{q_1\mdot q_1}\Big)\nn\\
 &+ G'_e\left(\omega  a_1\mdot q_1,\omega  a_1\mdot k\right)\omega \Big(\frac{c_{66}   a_1\mdot q_1 q_1\mdot F_k\mdot v_1}{q_1\mdot q_1}+\frac{c_{64}   a_1\mdot q_1 k\mdot q_1}{q_1\mdot q_1}+\frac{c_{69}   \left(a_1\mdot q_1\right)^2}{q_1\mdot q_1}+\frac{c_{82}   a_1\mdot q_1}{q_1\mdot q_1}\nn\\
 &+\frac{c_{10}  k\mdot q_1 q_1\mdot F_k\mdot v_1}{q_1\mdot q_1}+\frac{c_{14}  \left(q_1\mdot F_k\mdot v_1\right)^2}{q_1\mdot q_1}+\frac{c_{71} q_1\mdot F_k\mdot v_1}{q_1\mdot q_1}+\frac{c_{22}   \left(k\mdot q_1\right)^2}{q_1\mdot q_1}+\frac{c_{67}   k\mdot q_1}{q_1\mdot q_1}+\frac{c_{59}  }{q_1\mdot q_1}\Big)\nn\\
 &+G^{''}_o\left(\omega  a_1\mdot q_1,\omega  a_1\mdot k\right)\omega ^2\Big(\frac{c_{44}  a_1\mdot q_1 q_1\mdot S_1\mdot F_k\mdot v_1}{q_1\mdot q_1}+\frac{c_{52}  a_1\mdot q_1 q_1\mdot F_k\mdot v_1}{q_1\mdot q_1}+\frac{c_{46}  a_1\mdot q_1 k\mdot S_1\mdot q_1}{q_1\mdot q_1}\nn\\
 &+\frac{c_{53}  a_1\mdot q_1 k\mdot q_1}{q_1\mdot q_1}+\frac{c_{76}  a_1\mdot q_1 q_1\mdot S_1\mdot v_2}{q_1\mdot q_1}+\frac{c_{61} a_1\mdot q_1}{q_1\mdot q_1}+\frac{c_{81}  \left(a_1\mdot q_1\right){}^2}{q_1\mdot q_1}+\frac{c_{45}  q_1\mdot S_1\mdot v_2 q_1\mdot F_k\mdot v_1}{q_1\mdot q_1}\nn\\
 &+\frac{c_5  q_1\mdot F_k\mdot v_1}{q_1\mdot q_1}+\frac{c_{47}  k\mdot q_1 q_1\mdot S_1\mdot v_2}{q_1\mdot q_1}+\frac{c_6  k\mdot q_1}{q_1\mdot q_1}\Big)+G^{'}_o\left(\omega  a_1\mdot q_1,\omega  a_1\mdot k\right)\omega\Big(\frac{c_{32}   a_1\mdot q_1 q_1\mdot S_1\mdot F_k\mdot v_1}{q_1\mdot q_1}\nn\\
 &+\frac{c_{20}   a_1\mdot q_1 q_1\mdot F_k\mdot v_1}{q_1\mdot q_1}+\frac{c_{13}   a_1\mdot q_1 k\mdot q_1}{q_1\mdot q_1}+\frac{c_{40}  a_1\mdot q_1 q_1\mdot S_1\mdot v_2}{q_1\mdot q_1}+\frac{c_{50}   \left(a_1\mdot q_1\right){}^2}{q_1\mdot q_1}+\frac{c_{80}  a_1\mdot q_1}{q_1\mdot q_1}\nn\\
 &+\frac{c_{15}   q_1\mdot S_1\mdot v_2 q_1\mdot F_k\mdot v_1}{q_1\mdot q_1}+\frac{c_{34}  q_1\mdot S_1\mdot F_k\mdot v_1}{q_1\mdot q_1}+\frac{c_{21}  q_1\mdot F_k\mdot v_1}{q_1\mdot q_1}+\frac{c_{26}  k\mdot q_1 q_1\mdot S_1\mdot v_2}{q_1\mdot q_1}+\frac{c_{35}  k\mdot S_1\mdot q_1}{q_1\mdot q_1}\nn\\
 &+\frac{c_{49}   k\mdot q_1}{q_1\mdot q_1}+\frac{c_{73}  q_1\mdot S_1\mdot v_2}{q_1\mdot q_1}+\frac{c_{60} }{q_1\mdot q_1}\Big)\Bigg],
\end{align}
where 
\begin{align}
G_o^{'}(x_1,x_2)&:=(\partial_{x_1}-\partial_{x_2})G_2(x_1,x_2), &G_e^{'}(x_1,x_2)&:=(\partial_{x_1}-\partial_{x_2})G_1(x_1)G_1(x_2)\nn\\
    G_o^{''}(x_1,x_2)&:={(\partial_{x_1}-\partial_{x_2})^2\over 2}G_2(x_1,x_2), &G_e^{''}(x_1,x_2)&:={(\partial_{x_1}-\partial_{x_2})^2\over 2}G_1(x_1)G_1(x_2)\, .
\end{align} The coefficients are listed in Appendix \ref{app: integrnadcoeffs}. The integrand is composed of four parts:
\begin{itemize}
    \item terms including the functions $G_1$ and $\cosh$ and with spurious pole $1\over k\mdot q_1$: In this part, the entire function are $G_1(\omega a_1\mdot (k+q_1)), \cosh(\omega a_1\mdot (k+q_1))$.  All the terms are of $\cO(\omega^0)$. It is easy to see that the spurious pole is cancelled when adding  the corresponding terms in the $q_1^2$-channel. 
    \item terms with the functions $\cosh, G_1, G_2$ and without the spurious pole $1\over k\mdot q_1$: They are all of $\cO(\omega^0)$.
     \item terms with the functions $G'_o,G'_e$: They are of $\cO(\omega^0)$ and  $\cO(\omega^1)$. All of them do not contain the spurious pole $1\over k\mdot q_1$
      \item terms with the functions $G^{''}_o,G^{''}_e$: They are of $\cO(\omega^0)$,  $\cO(\omega^1)$ and $\cO(\omega^2)$. All of them do not contain the spurious pole $1\over k\mdot q_1$. They also do not contain the physical massive pole ${1\over k\mdot v_1}={1\over w_1}$. 
\end{itemize}

Unlike in the $q_1^2$-channel, here we have more general entire functions coming from the Compton amplitude for the particle with spin $a_1$ and consequently we now have four sectors with different exponential factors
\begin{align}
     {\rm (I)}: \quad &\exp \left(-i \omega  \left(-\tilde{a}_1\mdot k-\tilde{a}_1\mdot q_1-b\mdot k-b\mdot q_1+u\right)\right), \nn\\
     {\rm (II)}: \quad &  \exp \left(-i \omega  \left(-\tilde{a}_1\mdot k+\tilde{a}_1\mdot q_1-b\mdot k-b\mdot q_1+u\right)\right),\nn\\
      {\rm (III)}: \quad & \exp \left(-i \omega  \left(\tilde{a}_1\mdot k-\tilde{a}_1\mdot q_1-b\mdot k-b\mdot q_1+u\right)\right), \nn\\
      {\rm (IV)}: \quad &\exp \left(-i \omega  \left(\tilde{a}_1\mdot k+\tilde{a}_1\mdot q_1-b\mdot k-b\mdot q_1+u\right)\right).
\end{align}
In each sector, we still use the sector-dependent basis in  \eqref{eq:secBasis} and parameterise the $q_1$ variable of  \eqref{qparam} with 
\begin{align}
  \begin{cases}
       \tilde b_{\rm (I)}= \tilde b_{\rm (III)}= -b-\tilde{a}_1\\
       \tilde b_{\rm (II)}= \tilde b_{\rm (IV)}= -b+\tilde{a}_1\\
    \end{cases},&&  \tilde v_{j}=\eps(v_1 v_2 \tilde b_{j}  \bullet)\, ,&&j={\rm I, II,III,IV}.
\end{align}
Using this, the extra  $\delta$-functions in each sector  are
\begin{align}
      {\rm (I)}: \quad &\delta\left(-\tilde{a}_1\mdot k-b\mdot k+\tilde b_{(\rm I)}\mdot q_1+u\right),  & {(\rm III)}: \quad &  \delta\left(\tilde{a}_1\mdot k- b\mdot k+\tilde b_{(\rm III)}\mdot q_1+u\right), \nn\\
     {\rm (II)}: \quad &   \delta\left(-\tilde{a}_1\mdot k-b\mdot k+\tilde b_{(\rm II)}\mdot q_1+u\right), & {\rm (IV)}: \quad &  \delta\left(\tilde{a}_1\mdot k-b\mdot k+\tilde b_{(\rm IV)}\mdot q_1+u\right)\, .
\end{align}
Then the integration  localises onto a hyperplane for each sector and the method is exactly the same as in the last section. The new feature is the appearance of the entire functions $G'_o, G'_e, G^{''}_o, G^{''}_e$. The derivatives will lead to entire functions that are not homogeneous with respect to $\omega$  even while ignoring the exponential factors. Hence the integrand has three different  powers of  $\omega$, schematically 
\begin{align}
	1\times A_{\omega^0} e^{-i\omega (u+a)}+\omega\times A_{\omega^1}e^{-i\omega (u+b)}+\omega^2\times A_{\omega^2}e^{-i\omega (u+c)}\,,
\end{align}
where the $A_{\omega^i}$ and $a,b,c$ do not depend on $\omega$.
Performing the $\omega$ integral leads to a result of the form  
\begin{align}
    1\times \delta(u+a)A_{\omega^0}+i \partial_u  \big (\delta(u+b)A_{\omega^1}\big)-\partial^2_u \big(\delta(u+c)A_{\omega^2}\big)\, .
\end{align}
In practice, our result is obtained from evaluating the $\delta$-functions by integrating over $z_b$ as usual and replacing $\omega$ by $i\partial_u$ at the end, as shown in \eqref{eq:WFclosedForm}. 

 We now perform a numerical check that the result is free of spin-dependent spurious poles. After a random numerical replacement,  the spin-dependent spurious pole  is located at 
\begin{align}
	\tilde{a}_1\cdot v_{\perp}-\frac{42 \tilde{a}_1\cdot b}{5}=\xi\rightarrow 0 \, .
\end{align}
We extract the singular terms at the spurious pole, finding 
\begin{align}
	&-\frac{1323 \sqrt{3} u}{640 \sqrt{-25 u^2-700 u-17444} \xi^3}+\frac{5245317 u}{1100800 \sqrt{-75 u^2-2100 u-58732} \xi^3}\nn\\
	&-\frac{9261 \sqrt{3}}{320 \sqrt{-25 u^2-700 u-17444} \xi^3}+\frac{36717219}{550400 \sqrt{-75 u^2-2100 u-58732} \xi^3}-\frac{277641}{1100800 \xi^3}\nn\\
	&+\cdots\text{1345 more terms}\cdots
\end{align}
After applying the derivative operators and setting $u{=}0$ we get
\begin{align}
	&\frac{0.0123034\, +0.140702 i}{\xi^2}+\frac{0.0205579\, -0.00156661 i}{\xi}\nn\\
	&-\frac{0.0123034\, +0.140702 i}{\xi^2}-\frac{0.0222046\, -0.0033312 i}{\xi}\nn\\
	&+\frac{0.00164663\, -0.00176459 i}{\xi}= 0.
\end{align}
We have also tested that for several  other values of $u$ and find the singular term is always vanishing. This indicates that the final result is free of  spurious poles to any spin order. 

\subsection{Discussion of the resummed spin waveform}
The final result of the waveform has three contributions coming from terms each with up to two $u$-derivatives acting on them,
\begin{align}\label{eq:waveformNonpert}
    h^{\infty}(u)&= \Big(h^{\infty}_0(u)+\omega h^{\infty}_1(u)+\omega^2 h^{\infty}_2(u)\Big)\Big|_{\omega\rightarrow i\partial_u}\, ,
\end{align}
and in terms of $\tilde a, b, v_\perp, k, v_1, v_2$ takes the following schematic form 
\begin{align}
    &\frac{\left(2 y^2-1\right) v_1\mdot F_2\mdot v_2 \left(v_\perp\mdot F_2\mdot v_1 \left(\tilde{a}_1\mdot b-1\right)-\tilde{a}_1\mdot v_\perp b\mdot F_2\mdot v_1\right)}{8 w_1^2 w_2 \left(\left(y^2-1\right) \left(\tilde{a}_1\mdot b-1\right){}^2+\left(\tilde{a}_1\mdot v_\perp\right){}^2\right)}\nn\\
    &+\frac{\left(2 y^2-1\right) v_1\mdot F_2\mdot v_2 \left(\tilde{a}_1\mdot v_\perp b\mdot F_2\mdot v_1-v_\perp\mdot F_2\mdot v_1 \left(\tilde{a}_1\mdot b+1\right)\right)}{8 w_1^2 w_2 \left(\left(y^2-1\right) \left(\tilde{a}_1\mdot b+1\right){}^2+\left(\tilde{a}_1\mdot v_\perp\right){}^2\right)}\nn\\
    &+\cdots \text{ 336 more terms }\cdots
\end{align}
We note that the poles in $w_1, w_2$ correspond to the physical massive poles $1\over v_1\mdot k_2$ and $1\over v_2\mdot k_2$. The singular behaviour on these poles does not depend on the contact terms present in the Compton amplitude, which by definition are free of such poles, and so this behaviour is exact up to any spin order. 
 The explicit result in the case of  $a_1 \mdot v_2 {=}0$   can be found in the {\href{https://github.com/QMULAmplitudes/SpinningWaveformPublicData/tree/main}{GitHub repository}}. In the remainder of this subsection, we focus on the properties of the waveform by plotting its numerical values as a function of the retarded time $u$ and the spin parameter. As in the scalar case, we can choose a frame such that the kinematics are given by
\begin{align}
    v_1&=(1,0,0,0),  ~~~~~~~~~~~~~~~~~~~~~~~~~~~~~~~~~~~~~~~~~~~~~~~~~~~~~~~~v_2=(y,\sqrt{y^2-1},0,0)\nn\\
    \hat{k}&=(1,\sin \theta  \cos \phi ,\sin\theta  \sin \phi ,\cos \theta ), ~~~~~~~~~~~~~~~~~~~~~~~~~~~~~~v_\perp= (0,0,\sqrt{y^2-1},0) \nn\\
   \vareps^{(+)}&=\frac{1}{\sqrt{2}}\big(0,\cos \theta  \cos \phi -i \sin \phi ,\cos \theta  \sin \phi +i \cos\phi ,-\sin \theta  \big)\,,~~  b=(0, 0, 0, 1)
   .
\end{align}
Then we can further parameterise the constrained spin $a_1$ such that $a_1 \mdot v_2 =0$ as
\begin{align}
    a_1&=(0, 0 ,a \cos \psi ,a  \sin \psi ),
\end{align}
where $a>0$ is the magnitude of the spin and $\psi$ the angle of the spin's direction in the plane orthogonal to $v_1$ and $v_2$. 

\begin{figure}
    \centering
     \includegraphics[width=0.4\linewidth]{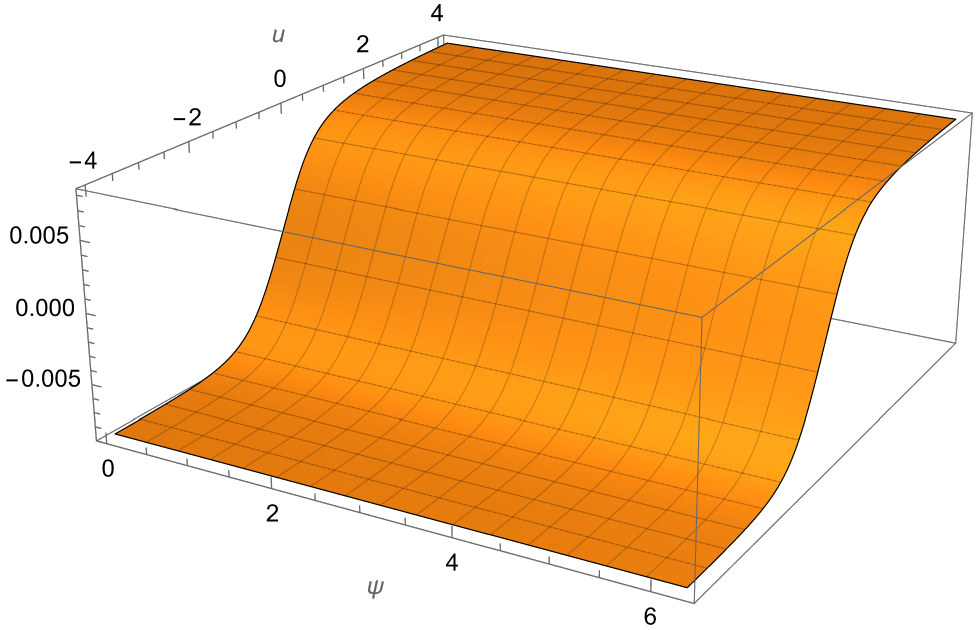}
    \includegraphics[width=0.4\linewidth]{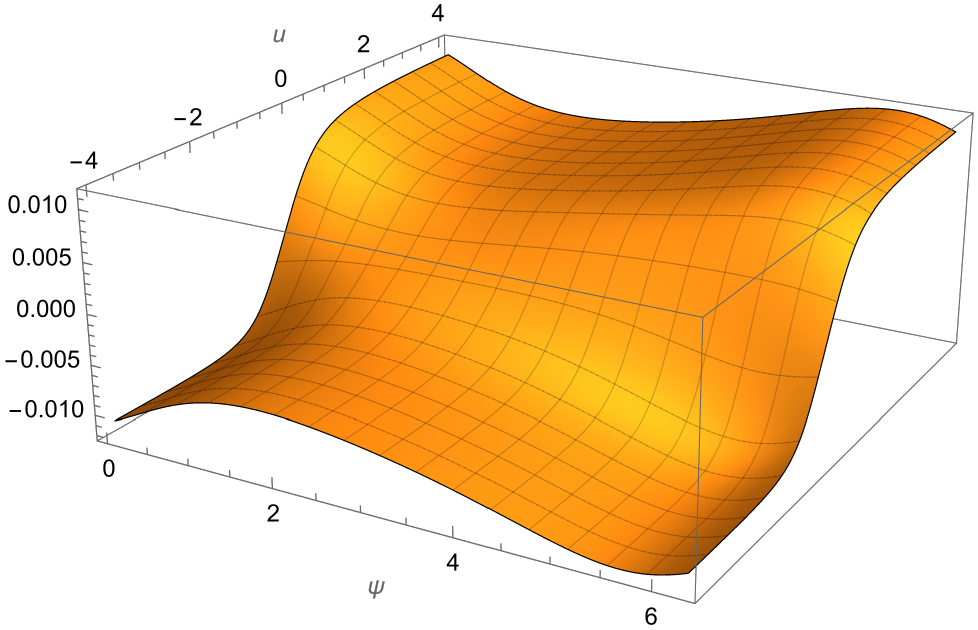}
     \includegraphics[width=0.55\linewidth]{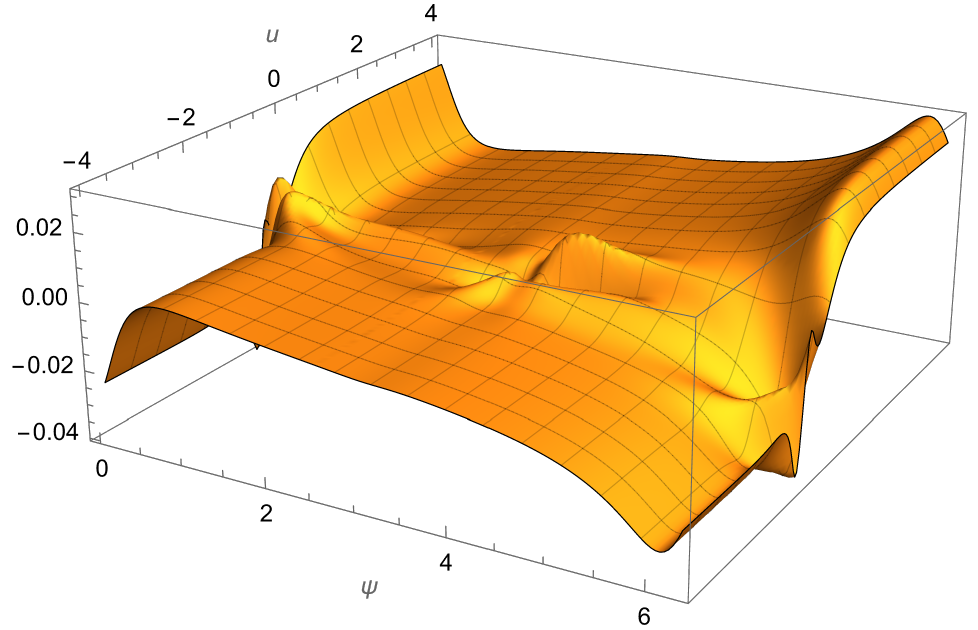}
    \caption{Waveform of $h^{\infty}_+$ at $a/b ={0.0}$, $a/b ={0.2}$, $a/b ={0.65}$.  
    }
    \label{fig:waveform}
\end{figure}

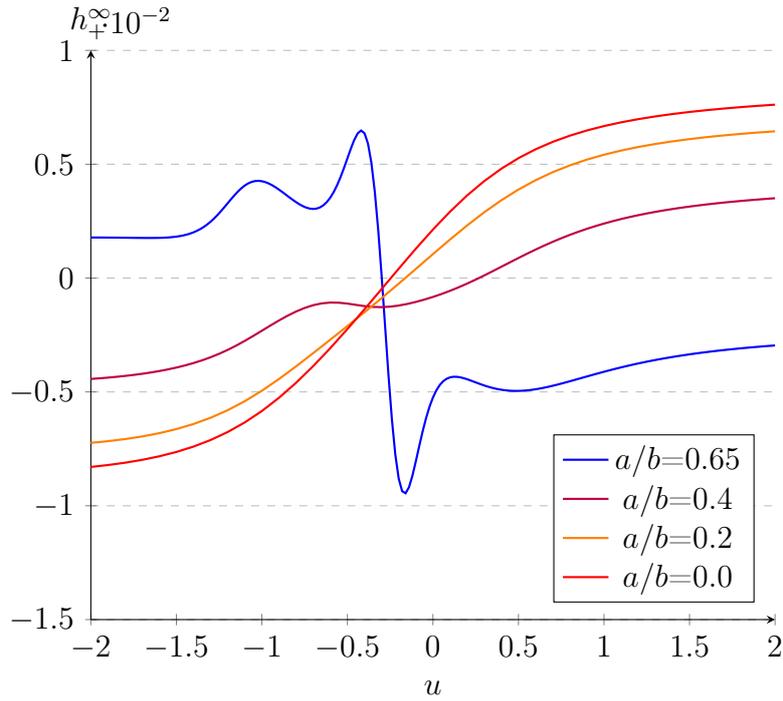
\begin{figure}
    \centering
\begin{tikzpicture}
	\begin{axis}[width=0.7\textwidth,
			xmin=-2., xmax=2,
			ymin=-0.015, ymax=0.01,
			axis lines=left,
			xtick={-2,-1.5,-1,-0.5,0,0.5,1,1.5,2},
			ytick={-0.015,-0.01,-0.005,0,0.005,0.01},
			compat=newest,
			xlabel=$u$, 
			ylabel= ${h}_{+}^{\infty}$, ylabel style={rotate=-90,at={(0,1)},anchor=south},
			legend pos=south east,
			ymajorgrids=true,
			grid style=dashed,
            every axis plot/.append style={thick},
		]
		\addplot[
			color=blue,
			mark=none,
		]
		coordinates {(-2.,0.00178276)(-1.98,0.00178232)(-1.96,0.00178172)(-1.94,0.00178097)(-1.92,0.00178004)(-1.9,0.00177894)(-1.88,0.00177767)(-1.86,0.00177624)(-1.84,0.00177465)(-1.82,0.00177293)(-1.8,0.00177111)(-1.78,0.00176923)(-1.76,0.00176734)(-1.74,0.00176552)(-1.72,0.00176386)(-1.7,0.00176248)(-1.68,0.00176153)(-1.66,0.00176119)(-1.64,0.00176169)(-1.62,0.0017633)(-1.6,0.00176636)(-1.58,0.00177124)(-1.56,0.00177844)(-1.54,0.0017885)(-1.52,0.00180207)(-1.5,0.00181993)(-1.48,0.00184294)(-1.46,0.0018721)(-1.44,0.00190854)(-1.42,0.00195349)(-1.4,0.00200829)(-1.38,0.00207432)(-1.36,0.00215296)(-1.34,0.00224546)(-1.32,0.00235285)(-1.3,0.00247571)(-1.28,0.00261401)(-1.26,0.00276683)(-1.24,0.00293221)(-1.22,0.00310701)(-1.2,0.00328689)(-1.18,0.0034665)(-1.16,0.00363984)(-1.14,0.00380081)(-1.12,0.00394369)(-1.1,0.00406378)(-1.08,0.00415768)(-1.06,0.00422349)(-1.04,0.00426075)(-1.02,0.00427031)(-1.,0.00425398)(-0.98,0.00421429)(-0.96,0.00415422)(-0.94,0.00407697)(-0.92,0.00398587)(-0.9,0.00388424)(-0.88,0.00377539)(-0.86,0.0036626)(-0.84,0.00354916)(-0.82,0.00343839)(-0.8,0.00333372)(-0.78,0.00323875)(-0.76,0.00315726)(-0.74,0.00309332)(-0.72,0.00305129)(-0.7,0.00303585)(-0.68,0.00305197)(-0.66,0.00310486)(-0.64,0.00319978)(-0.62,0.00334175)(-0.6,0.00353515)(-0.58,0.00378295)(-0.56,0.00408578)(-0.54,0.00444048)(-0.52,0.00483831)(-0.5,0.00526261)(-0.48,0.0056863)(-0.46,0.00606931)(-0.44,0.0063569)(-0.42,0.00647989)(-0.4,0.00635838)(-0.38,0.00591066)(-0.36,0.00506814)(-0.34,0.00379514)(-0.32,0.00210947)(-0.3,0.0000963521)(-0.28,-0.00209267)(-0.26,-0.0042608)(-0.24,-0.00620389)(-0.22,-0.00775377)(-0.2,-0.00881151)(-0.18,-0.0093596)(-0.16,-0.00945164)(-0.14,-0.00918722)(-0.12,-0.0086831)(-0.1,-0.00804976)(-0.08,-0.00737724)(-0.06,-0.00673002)(-0.04,-0.00614827)(-0.02,-0.00565234)(0.,-0.00524827)(0.02,-0.00493281)(0.04,-0.00469743)(0.06,-0.00453115)(0.08,-0.00442241)(0.1,-0.00436015)(0.12,-0.00433444)(0.14,-0.00433666)(0.16,-0.00435952)(0.18,-0.00439705)(0.2,-0.00444434)(0.22,-0.00449751)(0.24,-0.00455346)(0.26,-0.00460977)(0.28,-0.00466458)(0.3,-0.00471649)(0.32,-0.00476444)(0.34,-0.00480769)(0.36,-0.00484571)(0.38,-0.0048782)(0.4,-0.00490498)(0.42,-0.00492599)(0.44,-0.0049413)(0.46,-0.00495103)(0.48,-0.00495536)(0.5,-0.00495454)(0.52,-0.00494883)(0.54,-0.00493853)(0.56,-0.00492397)(0.58,-0.00490546)(0.6,-0.00488334)(0.62,-0.00485795)(0.64,-0.00482962)(0.66,-0.00479866)(0.68,-0.00476539)(0.7,-0.00473011)(0.72,-0.00469311)(0.74,-0.00465465)(0.76,-0.004615)(0.78,-0.00457438)(0.8,-0.00453302)(0.82,-0.00449111)(0.84,-0.00444885)(0.86,-0.0044064)(0.88,-0.0043639)(0.9,-0.0043215)(0.92,-0.00427931)(0.94,-0.00423744)(0.96,-0.00419598)(0.98,-0.00415502)(1.,-0.00411462)(1.02,-0.00407485)(1.04,-0.00403574)(1.06,-0.00399735)(1.08,-0.0039597)(1.1,-0.00392282)(1.12,-0.00388674)(1.14,-0.00385146)(1.16,-0.00381699)(1.18,-0.00378335)(1.2,-0.00375052)(1.22,-0.00371851)(1.24,-0.00368732)(1.26,-0.00365693)(1.28,-0.00362734)(1.3,-0.00359852)(1.32,-0.00357048)(1.34,-0.0035432)(1.36,-0.00351666)(1.38,-0.00349084)(1.4,-0.00346572)(1.42,-0.0034413)(1.44,-0.00341755)(1.46,-0.00339446)(1.48,-0.00337201)(1.5,-0.00335017)(1.52,-0.00332894)(1.54,-0.00330829)(1.56,-0.0032882)(1.58,-0.00326867)(1.6,-0.00324967)(1.62,-0.00323119)(1.64,-0.00321321)(1.66,-0.00319571)(1.68,-0.00317868)(1.7,-0.00316211)(1.72,-0.00314598)(1.74,-0.00313027)(1.76,-0.00311497)(1.78,-0.00310008)(1.8,-0.00308557)(1.82,-0.00307143)(1.84,-0.00305766)(1.86,-0.00304423)(1.88,-0.00303115)(1.9,-0.00301839)(1.92,-0.00300595)(1.94,-0.00299381)(1.96,-0.00298197)(1.98,-0.00297042)(2.,-0.00295915)};
  \addplot[
			color=purple,
			mark=none,
		]
		coordinates {(-2.,-0.00443039)(-1.95,-0.00440283)(-1.9,-0.00437179)(-1.85,-0.00433686)(-1.8,-0.00429753)(-1.75,-0.00425324)(-1.7,-0.00420335)(-1.65,-0.00414708)(-1.6,-0.00408354)(-1.55,-0.00401167)(-1.5,-0.00393026)(-1.45,-0.00383786)(-1.4,-0.00373289)(-1.35,-0.00361361)(-1.3,-0.0034783)(-1.25,-0.00332548)(-1.2,-0.00315425)(-1.15,-0.00296475)(-1.1,-0.00275859)(-1.05,-0.00253921)(-1.,-0.00231191)(-0.95,-0.0020835)(-0.9,-0.00186161)(-0.85,-0.00165407)(-0.8,-0.00146841)(-0.75,-0.00131178)(-0.7,-0.00119087)(-0.65,-0.00111126)(-0.6,-0.00107581)(-0.55,-0.00108221)(-0.5,-0.00112121)(-0.45,-0.00117713)(-0.4,-0.00123137)(-0.35,-0.00126774)(-0.3,-0.0012765)(-0.25,-0.00125522)(-0.2,-0.00120676)(-0.15,-0.00113606)(-0.1,-0.00104772)(-0.05,-0.000944811)(0.,-0.000828879)(0.05,-0.000700427)(0.1,-0.00055957)(0.15,-0.000406554)(0.2,-0.00024208)(0.25,-0.0000674367)(0.3,0.000115519)(0.35,0.000304484)(0.4,0.000496892)(0.45,0.000690112)(0.5,0.000881633)(0.55,0.00106921)(0.6,0.00125096)(0.65,0.00142539)(0.7,0.00159145)(0.75,0.00174845)(0.8,0.00189604)(0.85,0.00203414)(0.9,0.00216289)(0.95,0.00228259)(1.,0.00239365)(1.05,0.00249655)(1.1,0.00259182)(1.15,0.00267998)(1.2,0.00276158)(1.25,0.00283713)(1.3,0.00290711)(1.35,0.00297199)(1.4,0.00303221)(1.45,0.00308816)(1.5,0.0031402)(1.55,0.00318867)(1.6,0.00323388)(1.65,0.00327611)(1.7,0.00331559)(1.75,0.00335258)(1.8,0.00338727)(1.85,0.00341984)(1.9,0.00345048)(1.95,0.00347933)(2.,0.00350654)};
    \addplot[
			color=orange,
			mark=none,
		]
		coordinates {(-2.,-0.00723756)(-1.9,-0.00716361)(-1.8,-0.00707126)(-1.7,-0.0069559)(-1.6,-0.00681183)(-1.5,-0.0066323)(-1.4,-0.00640962)(-1.3,-0.00613566)(-1.2,-0.00580303)(-1.1,-0.00540693)(-1.,-0.00494771)(-0.9,-0.00443293)(-0.8,-0.00387692)(-0.7,-0.00329657)(-0.6,-0.00270484)(-0.5,-0.00210647)(-0.4,-0.00149927)(-0.3,-0.000879118)(-0.2,-0.000244395)(-0.1,0.000401899)(0.,0.00105184)(0.1,0.00169319)(0.2,0.00231098)(0.3,0.00289006)(0.4,0.00341816)(0.5,0.00388793)(0.6,0.00429745)(0.7,0.00464928)(0.8,0.00494879)(0.9,0.00520266)(1.,0.00541768)(1.1,0.00560018)(1.2,0.0057557)(1.3,0.0058889)(1.4,0.00600365)(1.5,0.00610311)(1.6,0.00618987)(1.7,0.006266)(1.8,0.0063332)(1.9,0.00639286)(2.,0.00644611)};

  \addplot[
			color=red,
			mark=none,
		]
		coordinates {(-2.,-0.0082948)(-1.9,-0.00821163)(-1.8,-0.00810941)(-1.7,-0.00798355)(-1.6,-0.00782838)(-1.5,-0.00763707)(-1.4,-0.00740162)(-1.3,-0.00711306)(-1.2,-0.00676199)(-1.1,-0.00633962)(-1.,-0.00583921)(-0.9,-0.00525783)(-0.8,-0.00459762)(-0.7,-0.00386607)(-0.6,-0.0030748)(-0.5,-0.0022374)(-0.4,-0.00136772)(-0.3,-0.000479602)(-0.2,0.000411815)(-0.1,0.00128869)(0.,0.00213054)(0.1,0.00291664)(0.2,0.00362986)(0.3,0.00425984)(0.4,0.00480405)(0.5,0.00526655)(0.6,0.00565562)(0.7,0.00598135)(0.8,0.00625395)(0.9,0.00648272)(1.,0.00667565)(1.1,0.00683937)(1.2,0.00697927)(1.3,0.00709966)(1.4,0.00720402)(1.5,0.00729509)(1.6,0.00737511)(1.7,0.00744584)(1.8,0.00750872)(1.9,0.00756493)(2.,0.00761544)};
 
\legend{{\rm $a/b$=0.65},{\rm $a/b$=0.4},{\rm $a/b$=0.2},{\rm $a/b$=0.0}}
	\end{axis}
\end{tikzpicture}
 \caption{Waveform of $h^{\infty}_+\,\,$ for different values of $a/b$ with spin angle $\psi{=}{\pi\over 4}$. }
    \label{fig:diffa}
\end{figure}

In Figures~\ref{fig:waveform} and~\ref{fig:diffa} we show the time-domain waveform $h_{+}$ at $y={3\over 2},\theta={\pi\over 4},\phi={\pi\over 4}$. In all of our graphs, we set $\kappa{=}m_1{=}m_2{=}1$, so each graph is missing a factor of $\kappa^4 m_1 m_2$. Figure~\ref{fig:waveform} shows the waveform dependence on the  retarded time $u$ and angle $\psi$. When the magnitude of the spin is equal to $0.2$, a small spin parameter compared to the magnitude of the impact parameter $|b|$, the time-domain waveform is similar to the scalar case. The spin effect on the waveform can then be taken as a perturbation on top of the spinless cases. However, for a larger magnitude, for example $0.65$, the time-domain waveform is modified greatly due to the effects of spin. To highlight the effect of changing the magnitude of the spin, in the 
Figure~\ref{fig:diffa} we plot the various spinning waveforms at fixed spin angle $\psi=\frac{\pi}{4}$.

From the waveform, we can extract the gravitational memory effect using 
\begin{align}
	\Delta h^{\infty}=h^{\infty}(+\infty)-h^{\infty}(-\infty).
\end{align}
We first study the Taylor expansion around $u\rightarrow \infty$ of the individual pieces $h^{\infty}_i(u)$ which contribute to the waveform in \eqref{eq:waveformNonpert} and find they all have similar behaviour 
\begin{align}
	h^{\infty}_i(u) \sim c_i+  \cO\Big({1 \over u}\Big) . 
\end{align}
The contributions $h_1^{\infty}$ and $h_2^{\infty}$ have the derivative $i \partial_u$ acting on them as such their behaviour in the large $u$ limit is sub-leading and they do not contribute to the memory. The memory can then by computed from the contribution $h_0^{\infty}$ and we find 
\begin{align}
	\Delta h^{\infty}&=\frac{ \kappa^4 m_1 m_2}{8 \pi }\frac{v_1\mdot F_k\mdot v_2 }{4 w_1^2 w_2^2 \sqrt{y^2-1} \left(4 \left(a_1\mdot b\right){}^2+\left(a_1\mdot a_1+1\right){}^2\right)}\nn\\
	&\times\Big(2 a_1\mdot b v_1\mdot F_k\mdot v_2 \left(\left(1-2 y^2\right) a_1\mdot k+w_0 y \left(a_1\mdot a_1-1\right)\right)\nn\\
	&+4 w_2  a_1\mdot b\left(\left(1-2 y^2\right) a_1\mdot F_k\mdot v_1+y \left(a_1\mdot a_1-1\right) v_\perp\mdot F_k\mdot v_1\right)\nn\\
	&-\left(a_1\mdot a_1+1\right) \left(-2 w_2 b\mdot F_k\mdot v_1+w_3 v_1\mdot F_k\mdot v_2\right) \left(2 y v_2\mdot S_1\mdot b+2 y^2-1\right)\Big).
\end{align}
In this compact formula, we notice that all the terms contain at least one pole in $w_1$ and $w_2$. This indicates that contact terms in the Compton amplitude do not contribute to the memory at any order in spin. As such we should expect that the waveform we have computed fully captures the memory to all orders in the spin. In addition, we compute a formula for the tree-level gravitational memory at all orders in spin \eqref{eq: spinning memory} in Section~\ref{sec-memory} below using a classical soft factor. The two formulae are indeed in agreement. We also mention again that we have compared our results to those of \cite{DeAngelis:2023lvf}, finding agreement (see also \cite{Aoude:2023dui}). 

\begin{figure}
    \centering
    \includegraphics[width=0.8\linewidth]{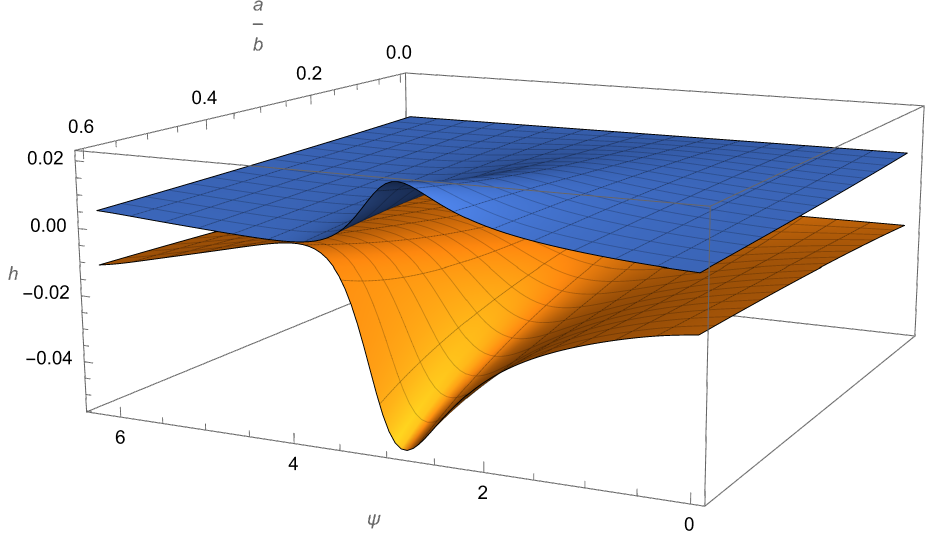}
    \caption{Gravitational memory: the top graph (blue) is the imaginary part, corresponding to $h^{\infty}_{\times}$, and the bottom graph (orange) is the real part, corresponding to~$h^{\infty}_{+}$.}
    \label{fig:memory}
\end{figure}

A graph of the memory, for the same kinematics as before and various values of the magnitude of the spin and direction, is presented in Figure~\ref{fig:memory}. When $|a|$ tends to~1, there are two singular points at $\psi=0,\pi$ corresponding to when the spin vector and impact parameter are orthogonal.

\section{Comparison with the spin-expanded waveforms}
\label{sec: OrderBYOrderInSpin}

If the spin parameter is small with respect to the impact parameter $a\ll |b|$ then we can evaluate the waveform integration order by order in a spin expansion. When we perform such an expansion the tree-level five-point amplitude is free of the spin-dependent spurious poles. One can still work in the $q_1^2$ and $q_2^2$ channels separately, which only contain one spurious pole $\frac{1}{q_1 \mdot k}$. After the usual re-scaling $q_i = \omega \hat q_i$, the waveform integrand is given by
\begin{align}  
\cM_{q_1^2}={1\over\omega^2}\Big(\sum_{i=0}^{\infty}\cM^{(i)}_{q_1^2}\omega^i\Big), && \cM_{q_2^2}={1\over\omega^2}\Big(\sum_{i=0}^{\infty}\cM^{(i)}_{q_2^2}\omega^i\Big).
\end{align} 
We still integrate over the frequency first but now after expanding in the spin parameter there is only one sector per cut. Thus the integrand contains the same delta functions as in the scalar case
\begin{align}\label{eq:scalar}
\begin{cases}
    \quad \delta \left(-b\mdot \hat{q}_1+u\right) & q_1^2\text{-channel,}\\
    \quad \delta \left(-b\mdot \hat{q}_2+u\right) &q_2^2\text{-channel}.
\end{cases}
\end{align}
The extra powers of $\omega$ become derivatives in the retarded time, $i\partial_u$, as before.
Now using the original parameterisation \eqref{eq: ScalarBasis}, after we localise $z_b$ each term in the integrand belongs to one of the following general expressions
\begin{align}
{c_0+c_1z_v+c_2z_v^2+c_3z_v^3+\cdots\over (z_v+\tilde c)(z_v^2+\bar c)}, && {\bar c_0+\bar c_1z_v+\bar c_2z_v^2+\bar c_3z_v^3+\cdots\over (z_v^2+\bar c)}\,,
\end{align}
where the $c$'s are functions of the external kinematics. The ${1\over z_v^2+c'_2}$ is the physical $q_1^2$ (or $q_2^2$) pole  and ${1\over z_v+c'_1}$ is the spurious pole at $q_1\Cdot k$.  Since the waveform only receives contributions from the physical pole, we can use polynomial division to reduce the numerators. Explicitly, we perform polynomial division over the physical pole, and obtain 
\begin{align}
	{c'_0+c'_1z_v\over (z_v+\tilde c)(z_v^2+\bar c)}+(\text{terms without physical poles}), \nn\\
	 {\bar c'_0+\bar c'_1z_v\over (z_v^2+\bar c)}+(\text{terms without physical poles}).
\end{align}
Terms without physical poles correspond to contributions that are proportional to delta functions in $b$ (and derivatives thereof) and hence do not contribute to the long-range waveform. Thus we only have the following two types of master integrals after performing partial fractions over the spurious pole
\begin{align}
	&{c^{''}_0\over (z_v+\tilde c)(z_v^2+\bar c)}\,\nn,\\
 &{\bar c^{''}_0+\bar c^{''}_1z_v\over (z_v^2+\bar c)}={\bar c^{''}_0\over (z_v^2+\bar c)}+\text{(terms that integrate to zero)} .
\end{align} \black
The two master integrals can then be evaluated by calculating the residue on the physical pole. The final result of the $q_1$ integral is of the form 
\begin{align}
	 h^{\infty}_{\rm expanded}(u)&=\Big( \sum_{i=0}^{\infty} \omega^i h^{\infty}_{i}(u)\Big)|_{\omega\rightarrow i\partial_u}\nn\\
  &\sim\Bigg[\frac{\omega ^4 a_1\mdot a_1 a_1\mdot \hat{k}\,  v_\perp\mdot F_{\hat k}\mdot v_1 \, a_1\mdot F_{\hat k}\mdot v_1}{24 \left(y^2-1\right)}-\frac{i \omega ^3 a_1\mdot a_1 \, v_\perp\mdot F_{\hat{k}}\mdot v_1 \text{tr}\left(F_{\hat{k}},S_1\right)}{48 \left(y^2-1\right)}\nn\\
  &-\frac{\hat{w}_0 \left(2 y^2-1\right) \left(v_1\mdot F_{\hat k}\mdot v_2\right){}^2}{8 \hat w_1^2 \hat w_2^2 \left(y^2-1\right)}-\frac{\left(2 y^2-1\right) \left(v_1\mdot F_{\hat k}\mdot v_2\right){}^2 }{8 \hat w_1^2 \hat{w}_2^2 \left(y^2-1\right) \sqrt{-u^2 \left(y^2-1\right)-\hat{w}_2^2}}\nn\\
  &\times\left(-\hat{w}_0 \sqrt{u^2 \left(-y^2\right)+u^2-\hat{w}_2^2}+u \hat{w}_3 y^2-u \hat{w}_3+\hat{w}_1 \hat{w}_2 y-\hat{w}_2^2\right)\nn\\
  &+\cdots \text{more terms}\cdots\Bigg]\Bigg|_{\omega\rightarrow i\partial_u}.
\end{align}
 The full waveform result expanded in the spin parameter up to $a^4$ order is included in the {\href{https://github.com/QMULAmplitudes/SpinningWaveformPublicData/tree/main}{GitHub repository}}. Our result contains contributions at orders beyond $a^4$ but these will in general be incomplete until possible additional contact terms are included in the Compton amplitude. 
 
\begin{figure}[t]
    \centering
     \includegraphics[width=0.47\linewidth]{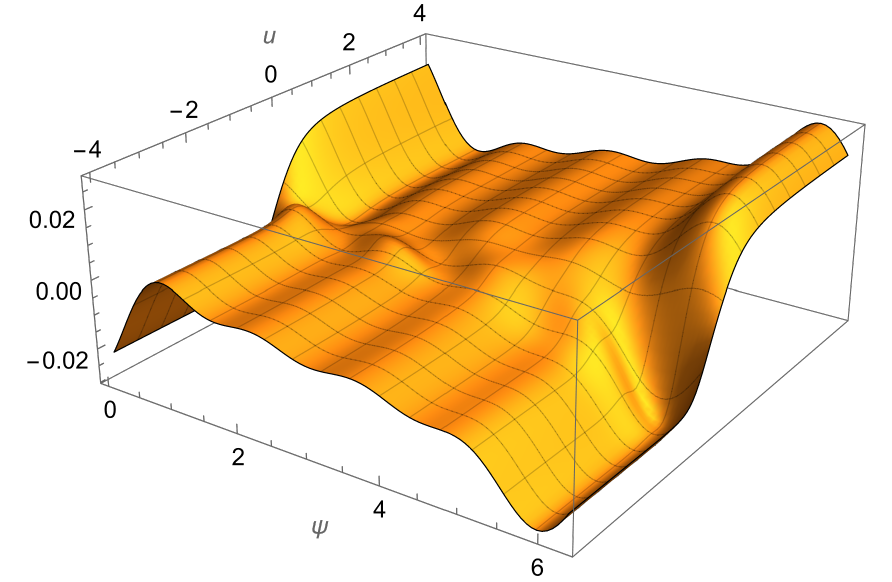}
     \includegraphics[width=0.47\linewidth]{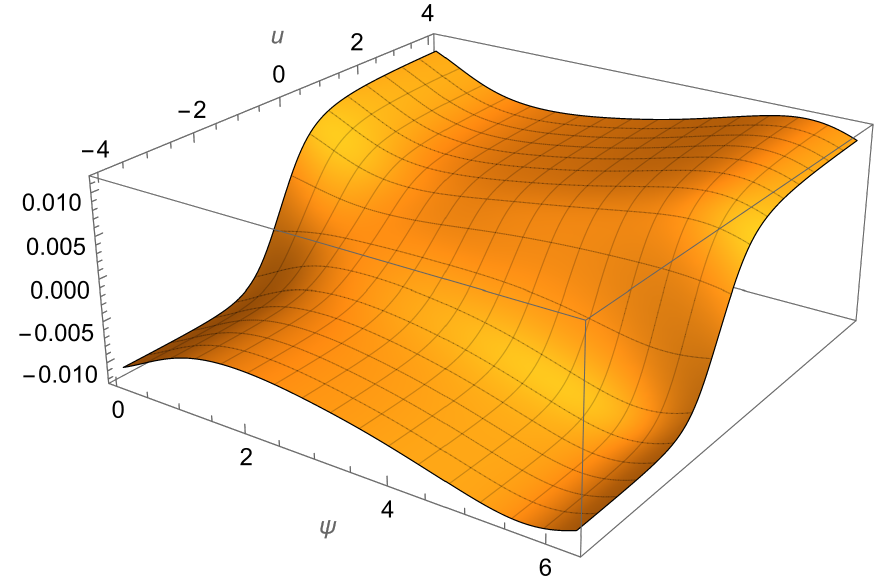}
    \caption{Waveform of $h^+$ at $a/b ={0.65}$, $a/b ={0.2}$.}
    \label{fig:waveformSeries}
    
\end{figure}

 We now comment on the difference between the resummed spinning waveform versus the 
 spin-expanded waveform truncated at $\cO(a^4)$.   
 To do so, we illustrate the 
 spin-expanded waveform at $a/b=0.2$ and $a/b=0.65$ in Figure~\ref{fig:waveformSeries}. Comparing with the resummed result shown for the same values in Figure~\ref{fig:waveform}, we see that for $a/b=0.2$ the spin-expanded result at $\cO(a^4)$ is accurate. However, at $a/b=0.65$ the spin expansion breaks down and the perturbative result is no longer valid.

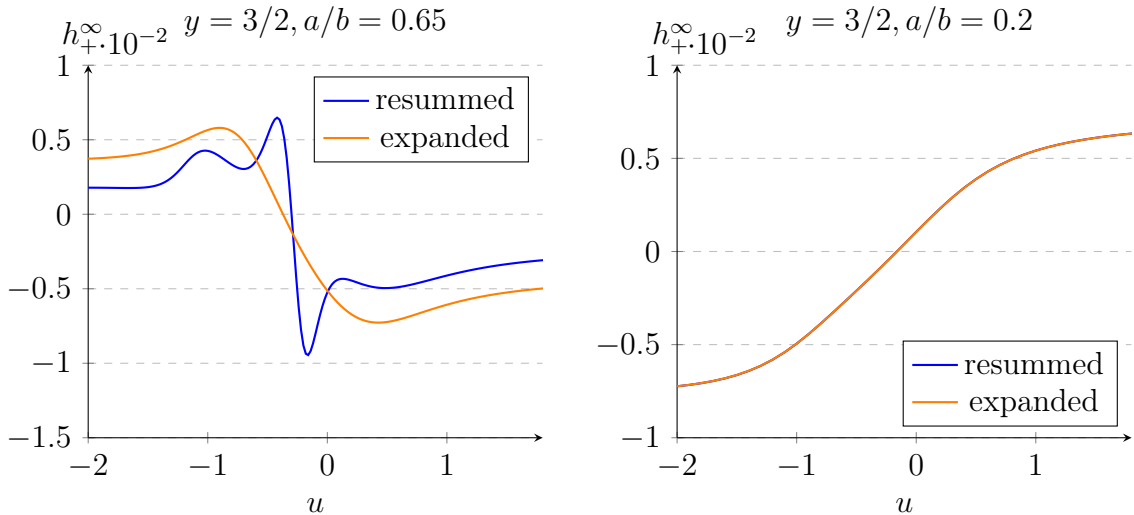
\begin{figure}[b]
\centering
\begin{tikzpicture}
	\begin{axis}[width=0.5\textwidth,
			title={$y=3/2, a/b=0.65$},
			xmin=-2., xmax=1.8,
			ymin=-0.015, ymax=0.01,
			axis lines=left,
			xtick={-2,-1,0,1,2},
			ytick={-0.015,-0.01,-0.005,0,0.005,0.01},
			compat=newest,
			xlabel=$u$, 
			ylabel= ${h}_{+}^{\infty}\,\,$, ylabel style={rotate=-90,at={(0,1)},anchor=south},
			legend pos=north east,
			ymajorgrids=true,
			grid style=dashed,
            every axis plot/.append style={thick},
		]
		\addplot[
			color=blue,
			mark=none,
		]
		coordinates {(-2.,0.00178276)(-1.98,0.00178232)(-1.96,0.00178172)(-1.94,0.00178097)(-1.92,0.00178004)(-1.9,0.00177894)(-1.88,0.00177767)(-1.86,0.00177624)(-1.84,0.00177465)(-1.82,0.00177293)(-1.8,0.00177111)(-1.78,0.00176923)(-1.76,0.00176734)(-1.74,0.00176552)(-1.72,0.00176386)(-1.7,0.00176248)(-1.68,0.00176153)(-1.66,0.00176119)(-1.64,0.00176169)(-1.62,0.0017633)(-1.6,0.00176636)(-1.58,0.00177124)(-1.56,0.00177844)(-1.54,0.0017885)(-1.52,0.00180207)(-1.5,0.00181993)(-1.48,0.00184294)(-1.46,0.0018721)(-1.44,0.00190854)(-1.42,0.00195349)(-1.4,0.00200829)(-1.38,0.00207432)(-1.36,0.00215296)(-1.34,0.00224546)(-1.32,0.00235285)(-1.3,0.00247571)(-1.28,0.00261401)(-1.26,0.00276683)(-1.24,0.00293221)(-1.22,0.00310701)(-1.2,0.00328689)(-1.18,0.0034665)(-1.16,0.00363984)(-1.14,0.00380081)(-1.12,0.00394369)(-1.1,0.00406378)(-1.08,0.00415768)(-1.06,0.00422349)(-1.04,0.00426075)(-1.02,0.00427031)(-1.,0.00425398)(-0.98,0.00421429)(-0.96,0.00415422)(-0.94,0.00407697)(-0.92,0.00398587)(-0.9,0.00388424)(-0.88,0.00377539)(-0.86,0.0036626)(-0.84,0.00354916)(-0.82,0.00343839)(-0.8,0.00333372)(-0.78,0.00323875)(-0.76,0.00315726)(-0.74,0.00309332)(-0.72,0.00305129)(-0.7,0.00303585)(-0.68,0.00305197)(-0.66,0.00310486)(-0.64,0.00319978)(-0.62,0.00334175)(-0.6,0.00353515)(-0.58,0.00378295)(-0.56,0.00408578)(-0.54,0.00444048)(-0.52,0.00483831)(-0.5,0.00526261)(-0.48,0.0056863)(-0.46,0.00606931)(-0.44,0.0063569)(-0.42,0.00647989)(-0.4,0.00635838)(-0.38,0.00591066)(-0.36,0.00506814)(-0.34,0.00379514)(-0.32,0.00210947)(-0.3,0.0000963521)(-0.28,-0.00209267)(-0.26,-0.0042608)(-0.24,-0.00620389)(-0.22,-0.00775377)(-0.2,-0.00881151)(-0.18,-0.0093596)(-0.16,-0.00945164)(-0.14,-0.00918722)(-0.12,-0.0086831)(-0.1,-0.00804976)(-0.08,-0.00737724)(-0.06,-0.00673002)(-0.04,-0.00614827)(-0.02,-0.00565234)(0.,-0.00524827)(0.02,-0.00493281)(0.04,-0.00469743)(0.06,-0.00453115)(0.08,-0.00442241)(0.1,-0.00436015)(0.12,-0.00433444)(0.14,-0.00433666)(0.16,-0.00435952)(0.18,-0.00439705)(0.2,-0.00444434)(0.22,-0.00449751)(0.24,-0.00455346)(0.26,-0.00460977)(0.28,-0.00466458)(0.3,-0.00471649)(0.32,-0.00476444)(0.34,-0.00480769)(0.36,-0.00484571)(0.38,-0.0048782)(0.4,-0.00490498)(0.42,-0.00492599)(0.44,-0.0049413)(0.46,-0.00495103)(0.48,-0.00495536)(0.5,-0.00495454)(0.52,-0.00494883)(0.54,-0.00493853)(0.56,-0.00492397)(0.58,-0.00490546)(0.6,-0.00488334)(0.62,-0.00485795)(0.64,-0.00482962)(0.66,-0.00479866)(0.68,-0.00476539)(0.7,-0.00473011)(0.72,-0.00469311)(0.74,-0.00465465)(0.76,-0.004615)(0.78,-0.00457438)(0.8,-0.00453302)(0.82,-0.00449111)(0.84,-0.00444885)(0.86,-0.0044064)(0.88,-0.0043639)(0.9,-0.0043215)(0.92,-0.00427931)(0.94,-0.00423744)(0.96,-0.00419598)(0.98,-0.00415502)(1.,-0.00411462)(1.02,-0.00407485)(1.04,-0.00403574)(1.06,-0.00399735)(1.08,-0.0039597)(1.1,-0.00392282)(1.12,-0.00388674)(1.14,-0.00385146)(1.16,-0.00381699)(1.18,-0.00378335)(1.2,-0.00375052)(1.22,-0.00371851)(1.24,-0.00368732)(1.26,-0.00365693)(1.28,-0.00362734)(1.3,-0.00359852)(1.32,-0.00357048)(1.34,-0.0035432)(1.36,-0.00351666)(1.38,-0.00349084)(1.4,-0.00346572)(1.42,-0.0034413)(1.44,-0.00341755)(1.46,-0.00339446)(1.48,-0.00337201)(1.5,-0.00335017)(1.52,-0.00332894)(1.54,-0.00330829)(1.56,-0.0032882)(1.58,-0.00326867)(1.6,-0.00324967)(1.62,-0.00323119)(1.64,-0.00321321)(1.66,-0.00319571)(1.68,-0.00317868)(1.7,-0.00316211)(1.72,-0.00314598)(1.74,-0.00313027)(1.76,-0.00311497)(1.78,-0.00310008)(1.8,-0.00308557)(1.82,-0.00307143)(1.84,-0.00305766)(1.86,-0.00304423)(1.88,-0.00303115)(1.9,-0.00301839)(1.92,-0.00300595)(1.94,-0.00299381)(1.96,-0.00298197)(1.98,-0.00297042)(2.,-0.00295915)};
		    \addplot[
			color=orange,
			mark=none,
		]
		coordinates {(-2.,0.00372652)(-1.98,0.00373317)(-1.96,0.00374031)(-1.94,0.00374797)(-1.92,0.00375618)(-1.9,0.00376497)(-1.88,0.0037744)(-1.86,0.0037845)(-1.84,0.00379532)(-1.82,0.00380691)(-1.8,0.00381931)(-1.78,0.00383258)(-1.76,0.00384678)(-1.74,0.00386198)(-1.72,0.00387824)(-1.7,0.00389564)(-1.68,0.00391423)(-1.66,0.00393412)(-1.64,0.00395537)(-1.62,0.00397809)(-1.6,0.00400235)(-1.58,0.00402826)(-1.56,0.00405592)(-1.54,0.00408542)(-1.52,0.00411688)(-1.5,0.0041504)(-1.48,0.00418609)(-1.46,0.00422405)(-1.44,0.00426439)(-1.42,0.0043072)(-1.4,0.00435257)(-1.38,0.00440058)(-1.36,0.0044513)(-1.34,0.00450476)(-1.32,0.00456098)(-1.3,0.00461996)(-1.28,0.00468165)(-1.26,0.00474595)(-1.24,0.00481273)(-1.22,0.00488179)(-1.2,0.00495285)(-1.18,0.00502559)(-1.16,0.00509956)(-1.14,0.00517426)(-1.12,0.00524908)(-1.1,0.00532328)(-1.08,0.00539605)(-1.06,0.00546643)(-1.04,0.00553339)(-1.02,0.00559577)(-1.,0.0056523)(-0.98,0.00570166)(-0.96,0.00574243)(-0.94,0.00577315)(-0.92,0.00579232)(-0.9,0.00579847)(-0.88,0.00579015)(-0.86,0.00576597)(-0.84,0.00572465)(-0.82,0.00566507)(-0.8,0.00558625)(-0.78,0.00548744)(-0.76,0.0053681)(-0.74,0.00522796)(-0.72,0.00506701)(-0.7,0.0048855)(-0.68,0.00468397)(-0.66,0.00446318)(-0.64,0.00422417)(-0.62,0.00396816)(-0.6,0.00369658)(-0.58,0.00341098)(-0.56,0.00311302)(-0.54,0.00280445)(-0.52,0.00248702)(-0.5,0.00216248)(-0.48,0.00183253)(-0.46,0.00149882)(-0.44,0.00116287)(-0.42,0.000826116)(-0.4,0.000489837)(-0.38,0.000155195)(-0.36,-0.00017679)(-0.34,-0.000505231)(-0.32,-0.000829373)(-0.3,-0.00114858)(-0.28,-0.00146231)(-0.26,-0.00177014)(-0.24,-0.0020717)(-0.22,-0.00236671)(-0.2,-0.00265493)(-0.18,-0.00293614)(-0.16,-0.00321019)(-0.14,-0.00347692)(-0.12,-0.00373616)(-0.1,-0.00398777)(-0.08,-0.00423159)(-0.06,-0.00446745)(-0.04,-0.00469514)(-0.02,-0.00491447)(0.,-0.00512522)(0.02,-0.00532714)(0.04,-0.00551999)(0.06,-0.00570351)(0.08,-0.00587744)(0.1,-0.00604155)(0.12,-0.0061956)(0.14,-0.00633939)(0.16,-0.00647275)(0.18,-0.00659554)(0.2,-0.00670768)(0.22,-0.00680914)(0.24,-0.00689996)(0.26,-0.00698021)(0.28,-0.00705007)(0.3,-0.00710974)(0.32,-0.00715949)(0.34,-0.00719966)(0.36,-0.00723063)(0.38,-0.00725282)(0.4,-0.0072667)(0.42,-0.00727276)(0.44,-0.00727151)(0.46,-0.0072635)(0.48,-0.00724925)(0.5,-0.00722931)(0.52,-0.00720422)(0.54,-0.0071745)(0.56,-0.00714066)(0.58,-0.00710319)(0.6,-0.00706256)(0.62,-0.00701921)(0.64,-0.00697356)(0.66,-0.00692599)(0.68,-0.00687687)(0.7,-0.00682653)(0.72,-0.00677527)(0.74,-0.00672337)(0.76,-0.00667106)(0.78,-0.00661858)(0.8,-0.00656613)(0.82,-0.00651387)(0.84,-0.00646197)(0.86,-0.00641055)(0.88,-0.00635974)(0.9,-0.00630962)(0.92,-0.0062603)(0.94,-0.00621183)(0.96,-0.00616427)(0.98,-0.00611767)(1.,-0.00607207)(1.02,-0.00602749)(1.04,-0.00598396)(1.06,-0.00594148)(1.08,-0.00590007)(1.1,-0.00585971)(1.12,-0.00582042)(1.14,-0.00578218)(1.16,-0.00574498)(1.18,-0.00570881)(1.2,-0.00567365)(1.22,-0.00563949)(1.24,-0.0056063)(1.26,-0.00557406)(1.28,-0.00554275)(1.3,-0.00551235)(1.32,-0.00548284)(1.34,-0.00545418)(1.36,-0.00542637)(1.38,-0.00539937)(1.4,-0.00537315)(1.42,-0.0053477)(1.44,-0.005323)(1.46,-0.00529901)(1.48,-0.00527572)(1.5,-0.00525311)(1.52,-0.00523114)(1.54,-0.00520981)(1.56,-0.00518909)(1.58,-0.00516895)(1.6,-0.00514939)(1.62,-0.00513038)(1.64,-0.00511189)(1.66,-0.00509393)(1.68,-0.00507646)(1.7,-0.00505947)(1.72,-0.00504294)(1.74,-0.00502687)(1.76,-0.00501122)(1.78,-0.00499599)(1.8,-0.00498117)(1.82,-0.00496674)(1.84,-0.00495268)(1.86,-0.00493899)(1.88,-0.00492565)(1.9,-0.00491265)(1.92,-0.00489998)(1.94,-0.00488763)(1.96,-0.00487559)(1.98,-0.00486385)(2.,-0.00485239)};
 
\legend{{\rm resummed},{\rm expanded}}
	\end{axis}
\end{tikzpicture} ~~
\begin{tikzpicture}
	\begin{axis}[width=0.5\textwidth,
			title={\ $y=3/2, a/b=0.2$},
			xmin=-2., xmax=1.8,
			ymin=-0.01, ymax=0.01,
			axis lines=left,
			xtick={-2,-1,0,1,2},
			ytick={-0.01,-0.005,0,0.005,0.01},
			compat=newest,
			xlabel=$u$, 
			ylabel=${h}_{+}^{\infty\,\,}$, ylabel style={rotate=-90,at={(0,1)},anchor=south},
			legend pos=south east,
			ymajorgrids=true,
			grid style=dashed,
            every axis plot/.append style={thick},
		]
		\addplot[
			color=blue,
			mark=none,
		]
		coordinates {(-2.,-0.00723756)(-1.9,-0.00716361)(-1.8,-0.00707126)(-1.7,-0.0069559)(-1.6,-0.00681183)(-1.5,-0.0066323)(-1.4,-0.00640962)(-1.3,-0.00613566)(-1.2,-0.00580303)(-1.1,-0.00540693)(-1.,-0.00494771)(-0.9,-0.00443293)(-0.8,-0.00387692)(-0.7,-0.00329657)(-0.6,-0.00270484)(-0.5,-0.00210647)(-0.4,-0.00149927)(-0.3,-0.000879118)(-0.2,-0.000244395)(-0.1,0.000401899)(0.,0.00105184)(0.1,0.00169319)(0.2,0.00231098)(0.3,0.00289006)(0.4,0.00341816)(0.5,0.00388793)(0.6,0.00429745)(0.7,0.00464928)(0.8,0.00494879)(0.9,0.00520266)(1.,0.00541768)(1.1,0.00560018)(1.2,0.0057557)(1.3,0.0058889)(1.4,0.00600365)(1.5,0.00610311)(1.6,0.00618987)(1.7,0.006266)(1.8,0.0063332)(1.9,0.00639286)(2.,0.00644611)};
    \addplot[
			color=orange,
			mark=none,
		]
		coordinates {(-2.,-0.00723577)(-1.9,-0.00716178)(-1.8,-0.00706935)(-1.7,-0.00695381)(-1.6,-0.00680946)(-1.5,-0.00662951)(-1.4,-0.0064063)(-1.3,-0.00613185)(-1.2,-0.00579906)(-1.1,-0.0054036)(-1.,-0.00494615)(-0.9,-0.00443397)(-0.8,-0.00388032)(-0.7,-0.00330074)(-0.6,-0.00270769)(-0.5,-0.00210689)(-0.4,-0.00149774)(-0.3,-0.000876978)(-0.2,-0.000242842)(-0.1,0.000402301)(0.,0.00105107)(0.1,0.00169151)(0.2,0.00230874)(0.3,0.00288759)(0.4,0.00341566)(0.5,0.00388551)(0.6,0.00429515)(0.7,0.00464709)(0.8,0.00494669)(0.9,0.00520061)(1.,0.00541568)(1.1,0.00559822)(1.2,0.00575376)(1.3,0.00588697)(1.4,0.00600173)(1.5,0.00610121)(1.6,0.00618797)(1.7,0.00626411)(1.8,0.00633131)(1.9,0.00639098)(2.,0.00644422)};
 
\legend{{\rm resummed},{\rm expanded}}
	\end{axis}
\end{tikzpicture} 
\caption{Comparison of the expanded and resummed waveforms $h_+$ at $a/b =0.65$ and $a/b =0.2$ with spin angle $\psi={\pi\over 4}$. For $a/b =0.2$ the graphs are indistinguishable. \black}
\label{fig:waveformpsiSeriesComp}
\end{figure}

To see more clearly the difference between the resummed
\black spin result and the perturbative spin result truncated at  $\cO(a^4)$, we also fix $\psi={\pi\over 4}$. For lower values of spin, for example $a/b={0.2}$, the expanded and resummed  waveforms are nearly identical, as shown in the right-hand side of Figure~\ref{fig:waveformpsiSeriesComp}. Conversely,  for large values of the spin, for example $a/b={0.65}$, the expanded and resummed results are markedly different, although their limiting values as $u\rightarrow \pm \infty$ are similar.

The above comparisons between our expanded and resummed waveforms require a number of considerations. For physical black holes we require $a/Gm\ll 1$, and additionally, in the PM expansion we require $Gm/b\ll 1$. For the case of large spin $a$, for example $a/b=0.65$ plotted above, it is clear that only one of the ratios, $Gm/b$ or $a/Gm$, can be taken as small. If we consider physical black holes,  $a/Gm\ll 1$,  then $Gm/b$ is no longer small, and we require higher orders in the PM expansion to reliably reproduce the physical waveform. Thus the plots in Figure~\ref{fig:waveformpsiSeriesComp}  would then change significantly once we include such terms.  Alternatively, we could consider again the case where $a {\sim} b$ but now require that $Gm/b\ll1$ such that we only need consider low orders in the PM expansion. In this case, we must again resum in the spin parameter $a/b$, but now we are in fact considering super-extremal Kerr, $a/Gm\gg1$.  Figure~\ref{fig:waveformpsiSeriesComp} much more accurately reproduces the waveform in this regime, and we see that resuming in spin substantially changes the waveform.

Finally, we also remind the reader that the results presented in this paper are valid up to $\cO(a^4)$, as discussed in Section~\ref{sec:Compton}.

\section{Gravitational memory}

\label{sec-memory}

\subsection{General strategy}

An elegant way to compute the memory was discussed in \cite{Herderschee:2023fxh} for the spinless case, and we adapt it to the case of  spinning heavy particles. Given a function \begin{align}
    f(u)\coloneqq \int_{-\infty}^{+\infty}\!\frac{d\omega}{2\pi}e^{- i \omega u} \tilde{f}(\omega)
\end{align}
of the retarded time $u$, the memory is defined as 
\begin{align}
\begin{split}
\label{deltaomega}
    \Delta f &\coloneqq f(u\to +\infty) - f (u\to - \infty) = \int_{-\infty}^{+\infty} \!du \, \frac{d}{du} f(u)
    \\ & = - i \int_{-\infty}^{+\infty}\!{d\omega}\, \delta( {\omega} ) \big[ \omega \tilde{f}(\omega) \big] \, , 
\end{split}
\end{align}
showing that it is determined by the  pole at $\omega{=}0$, i.e.~its soft limit, as observed by~\cite{Strominger:2014pwa}.

We now apply  \eqref{deltaomega} to \eqref{combinedres} to compute the gravitational memory, getting
\begin{align}
    \Delta (h^{\infty}_{+} \pm i  h^{\infty}_{\times}) = -i \frac{\kappa}{2} \left[ \lim_{\omega\to 0^+} \big[ \omega  W\big(b, k^\pm\big)\big]_{k= \omega (1, \mathbf{\hat{x}})}  +\lim_{\omega\to 0^-}\big[ \omega  W^\ast \big(b, k^\mp\big)\big]_{k= -\omega (1, \mathbf{\hat{x}})}\right]\, .
\end{align}
From this relation we see that the  memory effect arises from the leading soft behaviour of the five-point amplitude, which   factorises into a soft factor times a four-point amplitude, schematically
\begin{align}
    \cM_5\rightarrow  {\rm Soft}\times \cM_4.
\end{align}
Correspondingly, 
as $\omega\to 0$ the waveform tends to its leading soft limit,   
\begin{align}
    W_{\rm soft} \big(b, k^h\big) 
    = 
    -i\int\!d\mu^{(D)} e^{iq\Cdot b}S_{\rm W}^{\rm HEFT}(k, q; h)  \cM_4^{\rm HEFT} (q)\, , 
\end{align}
where \cite{Brandhuber:2023hhy}
\begin{align}
\begin{split}
	\label{SWHEFT}
	S_{\rm W}^{\rm HEFT} &= - \frac{\kappa}{2}\,\varepsilon_{\mu \nu}^{(h)}(k) \left[
		\frac{p_1^\mu q^\nu +p_1^\nu q^\mu}{p_1\Cdot k} - p_1^\mu p_1^\nu \frac{q\Cdot k}{(p_1\Cdot k)^2}\, - \, 1\leftrightarrow 2\right]
  \\ &=
   - \frac{\kappa}{2}\,\frac{1}{\omega}\varepsilon_{\mu \nu}^{(h)}(k) \left[
		\frac{p_1^\mu q^\nu +p_1^\nu q^\mu}{p_1\Cdot \hat{k}} - p_1^\mu p_1^\nu \frac{q\Cdot \hat{k}}{(p_1\Cdot \hat{k})^2}\, - \, 1\leftrightarrow 2\right]\, , 
  \, 
\end{split}
\end{align}
is the classical Weinberg soft factor for the emission of a graviton with momentum $k{=}\omega \hat{k}$ and helicity $h$, with  $q{=}q_1{=}-q_2$ in the soft limit and $\hat{k}{=}(1, \mathbf{\hat{x}})$ (see \cite{Brandhuber:2023hhy} for a derivation of the classical soft factor and a discussion of classical limits in the HEFT context).

We then change integration variables $q{\to} -q$, and use 
\begin{align}
\label{softone}
S_{\rm W}^{\rm HEFT} (-k, -q; h) &=  S_{\rm W}^{\rm HEFT} (k, q; h)\, ,\\ \cr
\label{softtwo}
\big[S_{\rm W}^{\rm HEFT}(k, q, -h)\big]^\ast  &=  S_{\rm W}^{\rm HEFT}(k, q, h) \, ,
\end{align}
also noting that,  at tree level in the spinning (and spinless) case,%
\footnote{In the spinless case we further have  $\cM_4^{\rm HEFT}(-q) = \cM_4^{\rm HEFT} (q)$. This is no longer true in the presence of spin.}
\begin{align}
    -i \cM_4^{\rm HEFT}(q) = \big[-i \cM_4^{\rm HEFT} (-q)\big]^\ast\, , 
\end{align}
which can be checked from the explicit expression derived later in \eqref{fourfinn}. With these observations, we get   \begin{align}
\begin{split}
   \left. W_{\rm soft}^\ast \big(b, k^{-h}\big)\right|_{k = -\omega (1, \hat{\mathbf{x}})} &= 
    \int\!d\mu^{(D)} e^{-iq\Cdot b}\big[ S_{\rm W}^{\rm HEFT}(-k, q; -h)\big]^\ast  (-i\cM_4^{\rm HEFT})^\ast (q)\\ 
    &= 
    \int\!d\mu^{(D)} e^{iq\Cdot b}S_{\rm W}^{\rm HEFT}(-k, -q; h)  (-i\cM_4^{\rm HEFT})^\ast (-q)
    \\ 
    &= 
    \int\!d\mu^{(D)} e^{iq\Cdot b}S_{\rm W}^{\rm HEFT}(k, q; h)  (-i\cM_4^{\rm HEFT}) (q)
    \, . 
\end{split}
\end{align}
 Hence we can write 
\black 
\begin{align}
\begin{split}
   & \lim_{\omega\to 0^+} \big[ \omega W\big(b, k^h\big)\big]_{k= \omega (1, \mathbf{\hat{x}})}  +\lim_{\omega\to 0^-}\big[ \omega W^\ast \big(b, k^{-h}\big)\big]_{k= -\omega (1, \mathbf{\hat{x}})}\\ 
   & = \int\!d\mu^{(D)} e^{iq_\Cdot b}S_{\rm W}^{\rm HEFT} (\hat{k}, q; h)
    (-i \cM_4^{\rm HEFT}) \, . 
\end{split}
\end{align}
In conclusion 
\begin{align}
\begin{split}
\label{mem-final-old}
    \Delta (h^{\infty}_{+} \pm i  h^{\infty}_{\times}) &= 
  - i\, \kappa \int\!d\mu^{(D)} e^{i q_\Cdot b}S_{\rm W}^{\rm HEFT} (\hat{k}, q; \pm)\, 
    \big(-i \cM_4^{\rm HEFT}\big)(q) \\ &
   = -i \, \kappa S_{\rm W}^{\rm HEFT} \Big(\hat{k}, -i \frac{\partial}{\partial b}; \pm\Big)\int\!d\mu^{(D)} e^{iq\Cdot b}\, 
   \, \big( -i \cM_4^{\rm HEFT}\big) (q) 
   \\ &
   = -i \, \kappa S_{\rm W}^{\rm HEFT} \Big(\hat{k}, -i \frac{\partial}{\partial b}; \pm\Big)\, \delta_{\rm HEFT}
   \, ,  
\end{split}
\end{align}
or
\begin{align}
\begin{split}
\label{mem-final}
    \Delta (h^{\infty}_{+} \pm i  h^{\infty}_{\times}) &= 
  -i \, \kappa S_{\rm W}^{\rm HEFT} \Big(\hat{k}, -i \frac{\partial}{\partial b}; \pm\Big)\, \delta_{\rm HEFT}
   \, ,  
\end{split}
\end{align}
where
\begin{align}
\label{deltaHEFT}
  \delta_{\rm HEFT}\coloneqq  
  \int\!d\mu^{(D)} e^{iq_\Cdot b} \, \big( -i \cM_4^{\rm HEFT}\big)(q)\, , 
\end{align}
and $S_{\rm W}^{\rm HEFT}$ is given in \eqref{SWHEFT}, and we also recall that $k{=}\omega\hat{k}$.
Note that $\delta_{\rm HEFT}$ is real because of the property \eqref{softtwo}. 

In the spinless case, one can further simplify this result by noticing that 
\begin{align}
    \frac{\partial}{\partial b_\mu} {=} - P \hat b^\mu \frac{\partial}{\partial  J}\, , 
\end{align}
where $J = P \sqrt{-b^2}$,  and the relation between the scattering angle and the real part of the HEFT phase 
\begin{align}
\label{scatang}
    -\frac{\partial}{\partial  J}\,{\rm Re}\, 
    \delta_{\rm HEFT} = \chi,
\, 
\end{align}
which itself is already a real quantity at tree level. Using these one finds  
\begin{align}
    S_{\rm W}^{\rm HEFT}\Big(\hat{k}, -i \frac{\partial}{\partial b_\mu}; h\Big)\, \delta_{\rm HEFT} = - i P S_{\rm W}^{\rm HEFT}(\hat{k}, \hat b^\mu; h) \chi\, , 
\end{align}
leading  to the compact relation, valid in the spinless case, 
\begin{align}
\begin{split}
    \Delta (h^{\infty}_{+} \pm i  h^{\infty}_{\times}) &= -\frac{\kappa^2}{2} P \vareps_{\rho\lambda}^{\pm\pm} s^{\rho \lambda} (\hat{k}, \hat b)\, \chi \, , 
    \end{split}
\end{align}
where we have set 
\begin{align}
\begin{split}
\label{SWW}
    S_{\rm W}^{\rm HEFT}&\coloneqq \frac{\kappa}{2}\vareps_{\rho\lambda}^{\pm\pm} s^{\rho \lambda} (\hat{k}, \hat b) \\ 
    s^{\mu \nu} (\hat{k}, \hat b)&=
    -\Big[
		\frac{p_1^\mu \hat b^\nu +p_1^\nu \hat b^\mu}{p_1\Cdot \hat{k}} - p_1^\mu p_1^\nu \frac{\hat b\Cdot \hat{k}}{(p_1\Cdot \hat{k})^2}\, - \, 1\leftrightarrow 2\Big]\, , 
\end{split}
\end{align}
and we recall that $\hat{k}=(1, \mathbf{\hat{x}})$ and $\hat b = b/\sqrt{-b^2}$. 
In the spinning case we do not have a simple relation such as \eqref{scatang}  and we will instead make use of  \eqref{mem-final}.
To compute the gravitational memory in the spinning case we will then use \eqref{mem-final} and \eqref{deltaHEFT}. 

We now move on to compute the tree-level-four-point amplitude that features in~\eqref{deltaHEFT}.

\subsection{Four-point two-to-two spinning amplitude}
 In this section we derive the  tree-level amplitude for the two-to-two scattering of two heavy particles with spin vectors $a_1$ and $a_2$ to all orders in the spin. We will then  compute
 its  Fourier transform to impact parameter space needed in~\eqref{deltaHEFT}. 

We can derive the four-point amplitude using the HEFT BCFW method first described in \cite{Brandhuber:2023hhy}, to which we refer the reader for further details. There is a single diagram in the $q^2$-channel for which we glue two of the three point amplitudes \eqref{threepointgrav} with the BCFW-shifted momenta described in \cite{Brandhuber:2023hhy}. We find that the  four-point tree-level amplitude $\cM_4$ is then
\begin{align}
\label{fourone}
\begin{split}
    \cM_4&=-{i\frac{\kappa^2}{q^2}}m_1^2 m_2^2\Big[ \Big( y^2-\frac{1}{2}\Big) \cosh \left(a\Cdot q\right)\\
    &+ \, i\, y\Big( \sinh \left(a_2\Cdot q\right) \cosh \left(a_1\Cdot q\right) 
    \frac{\epsilon\left(a_2qv_1v_2\right)}{a_2\Cdot q} + \sinh \left(a_1\Cdot q\right) \cosh \left(a_2\Cdot q\right) 
 \frac{\epsilon\left(a_1qv_1v_2\right)}{a_1\Cdot q}\Big)\Big] \\ & +\cM_{4, c}\, , 
\end{split}
\end{align}
where $a{\coloneqq}a_1+a_2$, and    the contact term $\cM_{4, c}$ is 
\begin{equation}
\label{ct22}
   \cM_{4, c}\coloneqq -i \kappa^2 m_1^2 m_2^2 \Big[y\, a_1 \mdot v_2\, a_2 \mdot v_1-a_1 \mdot a_2 \Big( y^2 -{1 \over 2} \Big) \Big]{\frac{\sinh(a_1 \mdot q)}{a_1 \mdot q}}{\frac{\sinh(a_2 \mdot q)}{a_2 \mdot q}}\, .
\end{equation}
 We note however that contact terms   play no  role for the computation of the memory, since they only contribute delta-function supported terms after Fourier transforming to impact parameter space. We will then drop them from now on (denoting the contact terms as $\mathcal{O}(1)$). 

We now simplify the expression  \eqref{fourone} for the four-point amplitude making use of the    new spin vectors  \cite{Vines:2017hyw,Guevara:2019fsj}  
\begin{align}
\label{eq: atildedefinition}
\mathfrak{a}^\mu_i\coloneqq\frac{\eps(a_i\mu v_1v_2)}{\sqrt{y^2-1}}\, , \quad i=1,2, 
\end{align}
which are orthogonal to both $v_1$ and $v_2$. 
     These quantities also satisfy the following Gram determinant relations 
\begin{align}\label{eq: a and atilde}
    (a_i\mdot q)^2=
({i\mathfrak{a}_i \mdot q})^2+\cO(q^2)\, , 
\end{align}
which are
proven in Appendix~\ref{Appendix-A}, and their ``square rooted'' form 
\begin{align}
    a_i\mdot q=
\pm {i\mathfrak{a}_i \mdot q}\, , 
\end{align}
valid up to terms of  $\cO(q^2)$, that is $q$ on-shell and so necessarily complex.
Furthermore, as both $\cosh(a_i\mdot q)$ and ${\sinh(a_i\Cdot q)\over  a_i\Cdot q }$ are 
parity-even functions  of $a_i\mdot q$,  the sign ambiguity drops out and the amplitude can be simplified to 
\begin{align}
\begin{split}
\label{fourfinn}
   \cM_4&= -{i\kappa^2\over q^2}{m_1^2 m_2^2}\Big\{ \Big( y^2-\frac{1}{2}\Big) \cosh({i\mathfrak a \mdot q})+ y {\sqrt{y^2-1}}\sinh({i\mathfrak{a} \mdot q})
\Big\}+\mathcal{O}(1)\, \\
&=-{i\kappa^2\over q^2}{m_1^2 m_2^2 \over 2}\Big\{ \Big( y^2-\frac{1}{2}+ y{\sqrt{y^2-1}}\Big) e^{{i\mathfrak{a} \mdot q}}+\Big( y^2-\frac{1}{2}- y{\sqrt{y^2-1}}\Big) e^{-{i\mathfrak{a} \mdot q}}
\Big\}
\\ & +\mathcal{O}(1)\, , 
\end{split}
\end{align}
where 
\begin{align}
    \mathfrak{a}{\coloneqq} \mathfrak{a}_1 {+} \mathfrak{a}_2\ . 
\end{align}
Note the nontrivial fact that at  tree level the pole part of the amplitude that we have considered so far depends only on the  sum  $\mathfrak{a}$ of the spins  of the two heavy particles. We also remark that the contact term \eqref{ct22} does not have this property.

\subsection{Fourier transform to impact parameter space}

Having cast the amplitude (up to contact terms) in the form \eqref{fourfinn}, 
we can perform the Fourier transform to impact parameter space to all orders in the spin, which will trivially shift  $b^\mu\to  b^\mu\pm \mathfrak{a}$, as was seen in \cite{Guevara:2019fsj}. 
We have 
\begin{align}
    \widetilde{\cM}_4 = \int\!\frac{d^Dq}{(2\pi)^{D-2}}\delta (2p_1\Cdot q)
    \delta (2p_2\Cdot q)\, e^{i q\Cdot b}\, \cM_4= \frac{1}{4m_1 m_2\sqrt{y^2-1}}\int\!\frac{d^{D-2}q_\perp}{(2\pi)^{D-2}}e^{- i \vec{q}_{\perp}\Cdot \vec{b}} \, {\cM}_4\, , 
\end{align}
where $q_\perp\Cdot p_{1,2}{=}0$ and 
\begin{align}
    \cM_4 = f_{+} (y) \frac{e^{i \mathfrak{a} \Cdot q}}{q^2} +  f_{-} (y) \frac{e^{-i \mathfrak{a} \Cdot q}}{q^2}\,  , 
\end{align}
with 
\begin{align}
\label{fpm}
f_{\pm}(y) \coloneqq  \ -i\,  \frac{\kappa^2 m_1^2 m_2^2}{2}\left(
    y^2 -\frac{1}{2}\pm y \sqrt{y^2-1}\right)\, .
\end{align}
Thus, we have to compute the Fourier transform 
\begin{align}
   \widetilde{\cM}_4= \int\!\frac{d^Dq}{(2\pi)^{D-2}}\delta (2p_1\Cdot q)
    \delta (2p_2\Cdot q)\, \left[ e^{i q\Cdot (b + \mathfrak{a})}f_{+}(y) +e^{i q\Cdot (b - \mathfrak{a})}f_{-}(y) \right] \, . 
\end{align}
We use
\begin{align}
\label{FTinvq}
\int\!\frac{d^dq}{(2\pi)^{d}} e^{- i \vec{q}\Cdot \vec{b}}
|\vec{q}\, |^p = \frac{2^p \pi^{-\frac{d}{2}} \Gamma\left( \frac{d+p}{2}\right)}{\Gamma\left( - 
\frac{p}{2}\right)} \frac{1}{|\vec{b}\, |^{d+p}}\, ,  
\end{align}
 which in our case gives the result, as $D\to 4$, 
 \begin{align}\label{eq: scalarintegral}
     \int\!\frac{d^{D-2}q}{(2\pi)^{D-2}} \frac{e^{-i \vec{q}\Cdot \vec{b}}}{\vec{q}^{\, 2}}=  \frac{\Gamma\left(\frac{D-4}{2}\right)}{4 \pi^{\frac{D-2}{2}}|\vec{b}\, |^{D-4}}\xrightarrow[D\to 4]{}  - \frac{1}{2\pi} \log (|\vec{b}\, |)+\cdots
 \, , 
 \end{align}
where the dots stand for $b$-independent terms. This leads to
\begin{align}\label{eq: fourpointIPSfinal}
\widetilde{\cM}_4= \frac{1}{8 \pi m_1 m_2\sqrt{y^2-1}}\Big[ f_{+}(y) \log ( |b+\mathfrak{a}|) + f_{-} (y) \log (|b-\mathfrak{a}|) 
\Big] +\cdots \, , 
    \end{align}
where we observe that the vector $\mathfrak{a}{=}\mathfrak{a}_1 + \mathfrak{a}_2$  lives in the same two-dimensional subspace orthogonal to $p_1$ and $p_2$ as $b$. \eqref{eq: fourpointIPSfinal} agrees with (51) of \cite{Guevara:2019fsj}.


\subsection{Result for the gravitational memory}
Finally, to compute the gravitational memory we use 
\eqref{mem-final} and 
   \eqref{deltaHEFT}. Introducing the two vectors 
   \begin{align}
       b_{\pm} \coloneqq b\pm \mathfrak{a}\, , \qquad \text{with} \qquad 
   \hat{b}_{\pm} \coloneqq \frac{b_\pm}{|b_{\pm}|}\, , 
   \end{align}
   we have at once 
\begin{align}\label{eq: spinning memory}
    \begin{split}
\Delta (h^{\infty}_{+} \pm i  h^{\infty}_{\times}) &= -\frac{i\kappa}{8 \pi m_1 m_2\sqrt{y^2-1}} \Big[ 
S_{\rm W}^{\rm HEFT}(\hat{k}, \hat{b}_{+}; \pm) \frac{f_{+}(y)}{|b_+|} + S_{\rm W}^{\rm HEFT}(\hat{k}, \hat{b}_{-}; \pm) \frac{f_{-}(y)}{|b_{-}|}\Big]    
    \end{split}, 
\end{align}
which is the final result for the memory, with $S_{\rm W}^{\rm HEFT}$ defined in \eqref{SWW} and 
$f_{\pm}$  in \eqref{fpm}. This result is exact to all orders in the spin vector $a$. 
One can expand it to various order in $a$,  and doing so one finds perfect agreement with  the result of \cite{Jakobsen:2021lvp}  for the memory in the aligned spin case up to $\cO(a^2)$. 

We also note that  in the spinless case,  the previous formula becomes 
\begin{align}
    \begin{split}
\left.\Delta (h^{\infty}_{+} \pm i  h^{\infty}_{\times})\right|_{a=0} &= -\frac{i\kappa}{8 \pi m_1 m_2\sqrt{y^2-1}}   (-i \kappa^2 m_1^2 m_2^2 )\frac{1}{|b|}\Big(y^2 - \frac{1}{2}\Big)
S_{\rm W}^{\rm HEFT}(\hat{k}, \hat{b}; \pm) \\
& = 
-\frac{\kappa^3}{8\pi} \frac{m_1 m_2}{\sqrt{y^2-1} }\frac{1}{|b|}\Big(y^2-\frac{1}{2}\Big) S_{\rm W}^{\rm HEFT}(\hat{k}, \hat{b}; \pm)
\, , 
    \end{split}
\end{align}
in agreement with known results (see e.g.~\cite{Jakobsen:2021lvp}).

\newpage
\section*{Acknowledgements}

We would like to thank  Massimo Bianchi, 
 Emil Bjerrum-Bohr, Stefano De Angelis, Thibault Damour, Claudio Gambino,  Paolo Pichini, Fabio Riccioni and Marcos Skowronek
for several interesting conversations. 
GT thanks the  Physics Department at the University of Rome ``Tor Vergata'' for their warm hospitality and support.
This work was supported by the Science and Technology Facilities Council (STFC) Consolidated Grants ST/P000754/1 \textit{``String theory, gauge theory \& duality''} and  ST/T000686/1 \textit{``Amplitudes, strings  \& duality''}.
The work of GRB and JG  is supported by an STFC quota studentship.   GC has received funding from the European Union's Horizon 2020 research and innovation program under the Marie Sk\l{}odowska-Curie grant agreement No.~847523 ``INTERACTIONS''. 
No new data were generated or analysed during this study.

\appendix

\section{Simplifying the four-point amplitude}
\label{Appendix-A}

In the main text we have defined a new spin vector \eqref{eq: atildedefinition},  and  $\mathfrak{a} \coloneqq \mathfrak{a}_1 + \mathfrak{a}_2$ which are all orthogonal to both $v_1$ and $v_2$. Now from the square of the Levi-Civita tensor we obtain a Gram determinant, for $i=1,2$, 
\begin{align}
    \big[\epsilon(a_i q v_1 v_2)\big]^2=
{(a_i \mdot q)^2 \over 1-y^2}+q^2 \Big[(a_i \mdot v_1)^2  +  (a_i \mdot v_2)^2-2 y\, a_i \mdot v_1 a_i \mdot v_2 +a_i^2(y^2 -1)\Big]\, ,
\end{align}
valid with the HEFT constraints $v_{1,2}\Cdot q {=}0$. Using that $v_{i}\Cdot a_i {=}0$ we then find
\begin{align}
\begin{split}
    \big[\epsilon\left(a_1 q v_1 v_2\right)\big]^2&=
-{(a_1 \mdot q)^2 \over y^2-1}+q^2 \Big((a_1 \mdot v_2)^2+a_1^2(y^2 -1)\Big)\, ,\\
\big[\epsilon\left(a_2 q v_1 v_2\right)\big]^2&=
-{(a_2 \mdot q)^2 \over y^2-1}+q^2 \Big((a_2 \mdot v_1)^2+a_2^2(y^2 -1)\Big)\, .
\end{split}
\end{align}
In the calculation in impact parameter space,  $\cO(q^2)$ terms do not contribute,  giving \eqref{eq: a and atilde} in the main text. 

To simplify the four-point amplitude we actually used the square root of the above relations, that is
\begin{align}  
\label{bigsim}
\frac{\epsilon\left(a_2qv_1 v_2\right)}{a_2\Cdot q}= \frac{\epsilon\left(a_1qv_1 v_2\right)}{a_1\Cdot q}=\mp i \sqrt{y^2-1}+\mathcal{O}(q^2)\, , 
\end{align}
which are again valid up to terms order $\cO(q^2)$. Since the amplitude is parity even, the sign ambiguity in these relations drops out. 

One might ask what determines the sign on the right-hand side of \eqref{bigsim}. A simple way to answer this question  is to go to the rest frame of particle one, and show that the $\mp$ sign in \eqref{bigsim} follows the particular choice of the on-shell momentum $q$. 
We can set
\begin{align}
    v_1= (1, 0,0,0)\, , \qquad v_2= (y, 0, 0, \sqrt{y^2-1})\, . 
\end{align}
Now, $q$ is on-shell, $q^2{=}0$, and satisfies the usual constraints $q\Cdot v_1 {=}q\Cdot v_2=0$, hence it  must have the form 
\begin{align}
\label{q}
q= (0, r, \pm i r, 0)\, .
\end{align}
Finally, using $a_1\Cdot v_1=0$ and $a_2\Cdot v_2=0$, the spin vectors are of the form 
\begin{align}
\begin{split}
 a_1& = (0, a_{1x}, a_{1y}, a_{1z})\, ,\\
    a_2& = \Big({s \over y}, a_{2x}, a_{2y}, {s \over \sqrt{y^2-1}}\Big)\, ,  
    \end{split}
\end{align}
and hence  we can evaluate, for $i=1,2$,
\begin{align}
\begin{split}
    \epsilon(v_1 v_2 a_i q) &= \epsilon( \vec{v}_2 \vec{a}_i \vec{q} )= \pm \sqrt{y^2-1}\, r\,  (i a_{ix} \mp a_{iy})\, ,
    \\
    a_i\Cdot q &= i r (i a_{ix} \mp a_{iy})\, . 
    \end{split}
\end{align}
In conclusion
\begin{align}
\label{nicebis}
\frac{\epsilon(v_1 v_2 a_i q)}{a_i\Cdot q} = \mp i \sqrt{y^2-1}\, ,
    \end{align}
with the same plus or minus sign appearing for $a_1$ or $a_2$ and following  from  the solution \eqref{q} chosen for the on-shell momentum $q$.  
Finally note that the right-hand side of \eqref{bigsim} is manifestly imaginary, and the above discussion shows that the left-hand side of that equation is too.

\section{More on the integrand}\label{app: integrnadcoeffs} 

In this appendix we provide the $c_i$ coefficients used in the formulae for the residues in each channel which appear in Sections \ref{sec:q1^2channel} and \ref{sec:q2^2channel}. As in Section~\ref{sec:6}, in this appendix the $k, q_i$ and $w_i$ should be understood as the hatted quantities with all $\omega$ dependence scaled out and once again we drop these hats purely for conciseness. The $q^2_1$-channel coefficients in Section~\ref{sec:q1^2channel} are
\begin{align}
\begin{split}
    &c_1\to -\frac{1}{2} \left(v_1\mdot F_k\mdot v_2\right){}^2,
    \ c_2\to \frac{1-2 y^2}{4 w_1^2},\ c_3\to \frac{i y}{2 w_1^2},\ c_4\to \frac{i y v_1\mdot F_k\mdot v_2}{2 w_1^2},\\
    &c_5\to -\frac{i y v_1\mdot F_k\mdot v_2}{2 w_1^2},\ 
    c_6\to \frac{y v_1\mdot F_k\mdot v_2}{w_1}, \ 
    c_7\to -\frac{i v_1\mdot F_k\mdot v_2}{2 w_1}, \ 
    c_8\to -\frac{i \left(v_1\mdot F_k\mdot v_2\right){}^2}{2 w_1},\\
    &c_9\to -\frac{\left(2 y^2-1\right) \left(v_1\mdot F_k\mdot v_2\right){}^2}{4 w_1^2 w_2^2},\ 
    c_{10}\to -\frac{\left(2 y^2-1\right) v_1\mdot F_k\mdot v_2}{2 w_1^2 w_2},\ 
    c_{11}\to \frac{i y \left(v_1\mdot F_k\mdot v_2\right){}^2}{2 w_1^2 w_2^2},\\
    &c_{12}\to \frac{i y v_1\mdot F_k\mdot v_2}{w_1^2 w_2},\ 
    c_{13}\to \frac{i y \left(v_1\mdot F_k\mdot v_2\right){}^2}{2 w_1^2 w_2},\ 
    c_{14}\to \frac{y \left(v_1\mdot F_k\mdot v_2\right){}^2}{w_1 w_2},\\
    &c_{15}\to -\frac{i \left(v_1\mdot F_k\mdot v_2\right){}^2}{2 w_1 w_2},\ 
    c_{16}\to -\frac{i w_2 y}{2 w_1^2},\ 
    c_{17}\to \frac{i w_2 v_1\mdot F_k\mdot v_2}{2 w_1} \, .
\end{split}
\end{align}
The coefficients in the $q^2_2$-channel, which appear in Section~\ref{sec:q2^2channel}, are listed below:
\begin{align}
    c_2&\to -\frac{1}{2} a_1\mdot a_1 \left(v_1\mdot F_k\mdot v_2\right)^2 \nn\\
   c_4&\to \left(a_1\mdot v_2\right){}^2 \left(a_1\mdot F_k\mdot v_1\right){}^2-\frac{1}{4} a_1\mdot a_1 a_1\mdot F_k\mdot v_1 \left(a_1\mdot F_k\mdot v_1-2 a_1\mdot v_2 v_1\mdot F_k\mdot v_2\right)\nn\\  
    c_5&\to \frac{1}{16} (-3) i a_1\mdot k \left(a_1\mdot a_1-2 \left(a_1\mdot v_2\right){}^2\right) \text{tr}\left(F_k\mdot S_1\right),\nn\\
   c_6&\to\frac{3}{16} i \left(a_1\mdot a_1-2 \left(a_1\mdot v_2\right){}^2\right) \text{tr}\left(F_k\mdot S_1\right) a_1\mdot F_k\mdot v_1,\nn\\
   c_7&\to\frac{y \left(a_1\mdot v_2 a_1\mdot F_k\mdot v_1+a_1\mdot a_1 v_1\mdot F_k\mdot v_2\right)}{2 w_1^3},\nn\\
   c_{10}&\to\frac{a_1\mdot F_k\mdot v_1 \left(2 y a_1\mdot k a_1\mdot v_2+w_1 \left(\left(a_1\mdot v_2\right){}^2+\left(y^2-1\right) a_1\mdot a_1\right)\right)}{4 w_1^3},\nn\\
   c_{11}&\to-\frac{\left(a_1\mdot F_k\mdot v_1\right){}^2+2 a_1\mdot v_2 v_1\mdot F_k\mdot v_2 a_1\mdot F_k\mdot v_1+2 a_1\mdot a_1 \left(v_1\mdot F_k\mdot v_2\right){}^2}{4 w_1^2},\nn\\
    c_{12}&\to\frac{i \left(2 y^2-1\right) \text{tr}\left(F_k\mdot S_1\right)}{8 w_1^2},~~~~
   c_{13}\to\frac{\left(1-2 y^2\right) \text{tr}\left(F_k\mdot S_1\right) a_1\mdot F_k\mdot v_1}{16 w_1^2},\nn\\
   c_{14}&\to-\frac{w_1 a_1\mdot k \left(\left(a_1\mdot v_2\right){}^2+\left(y^2-1\right) a_1\mdot a_1\right)+y \left(a_1\mdot k\right){}^2 a_1\mdot v_2+w_1^2 y a_1\mdot a_1 a_1\mdot v_2}{4 w_1^3},\nn\\
   c_{15}&\to\frac{a_1\mdot F_k\mdot v_1 \left(y a_1\mdot k+w_1 a_1\mdot v_2\right)}{4 w_1^2},\nn\\
   c_{16}&\to\frac{y a_1\mdot F_k\mdot v_1 \left(2 a_1\mdot v_2 a_1\mdot F_k\mdot v_1+a_1\mdot a_1 v_1\mdot F_k\mdot v_2\right)}{2 w_1},\nn\\
   c_{19}&\to-\frac{i \left(w_1 a_1\mdot v_2 v_1\mdot F_k\mdot v_2 \text{tr}\left(F_k\mdot S_1\right)+a_1\mdot F_k\mdot v_1 k\mdot S_1\mdot F_k\mdot v_1\right)}{4 w_1},\nn\\
    c_{20}&\to\frac{\text{tr}\left(F_k\mdot S_1\right) \left(\left(2 y^2-1\right) a_1\mdot k+2 w_1 y a_1\mdot v_2\right)}{16 w_1^2},\nn\\
   c_{21}&\to\frac{\text{tr}\left(F_k\mdot S_1\right) \left(2 y a_1\mdot k a_1\mdot v_2-w_1 \left(a_1\mdot a_1-2 \left(a_1\mdot v_2\right){}^2\right)\right)}{16 w_1},\nn\\
   c_{22}&\to-\frac{y a_1\mdot v_2 \left(a_1\mdot F_k\mdot v_1\right){}^2}{4 w_1^3},c_{25}\to\frac{i y a_1\mdot F_k\mdot v_1}{2 w_1^2},c_{26}\to-\frac{y \left(a_1\mdot F_k\mdot v_1\right)^2}{4 w_1^2},\nn\\
   c_{29}&\to-\frac{i y v_1\mdot F_k\mdot S_1\mdot v_2}{2 w_1^2},c_{32}\to\frac{a_1\mdot F_k\mdot v_1}{8 w_1},c_{33}\to-\frac{i a_1\mdot F_k\mdot v_1}{4 w_1},c_{34}\to-\frac{a_1\mdot k a_1\mdot F_k\mdot v_1}{8 w_1},\nn\\
   c_{35}&\to-\frac{\left(a_1\mdot F_k\mdot v_1\right){}^2}{8 w_1},c_{40}\to\frac{v_1\mdot F_k\mdot v_2 a_1\mdot F_k\mdot v_1}{4 w_1},c_{41}\to-\frac{i v_1\mdot F_k\mdot v_2 a_1\mdot F_k\mdot v_1}{2 w_1},\nn\\
   c_{43}&\to\frac{i v_1\mdot F_k\mdot v_2 v_1\mdot F_k\mdot S_1\mdot v_2}{2 w_1},c_{44}\to\frac{3 i a_1\mdot k a_1\mdot F_k\mdot v_1}{8 w_1},c_{45}\to\frac{3 i a_1\mdot k a_1\mdot v_2 a_1\mdot F_k\mdot v_1}{4 w_1},\nn\\
   c_{46}&\to\frac{3 i \left(a_1\mdot F_k\mdot v_1\right){}^2}{8 w_1},c_{47}\to-\frac{3 i a_1\mdot v_2 \left(a_1\mdot F_k\mdot v_1\right)^2}{4 w_1},
   c_{48}\to\frac{i y a_1\mdot v_2 \text{tr}\left(F_k\mdot S_1\right)}{4 w_1},\nn\\
   c_{49}&\to-\frac{y a_1\mdot v_2 \text{tr}\left(F_k\mdot S_1\right) a_1\mdot F_k\mdot v_1}{8 w_1},c_{50}\to\frac{y v_1\mdot F_k\mdot v_2 \text{tr}\left(F_k\mdot S_1\right)}{8 w_1},c_{51}\to-\frac{i y v_1\mdot F_k\mdot v_2 \text{tr}\left(F_k\mdot S_1\right)}{4 w_1},\nn\\
   c_{52}&\to\frac{3 i y a_1\mdot k a_1\mdot v_2 \text{tr}\left(F_k\mdot S_1\right)}{8 w_1}, c_{53}\to-\frac{3 i y a_1\mdot v_2 \text{tr}\left(F_k\mdot S_1\right) a_1\mdot F_k\mdot v_1}{8 w_1},\nn\\
  c_{57}&\to\frac{1}{4} a_1\mdot F_k\mdot v_1 \left(a_1\mdot a_1 \left(a_1\mdot k+2 w_2 a_1\mdot v_2\right)-4 a_1\mdot k \left(a_1\mdot v_2\right){}^2\right),\nn\\
   c_{58}&\to-\left(a_1\mdot v_2\right) a_1\mdot F_k\mdot v_1 \left(w_2 a_1\mdot F_k\mdot v_1+a_1\mdot k v_1\mdot F_k\mdot v_2\right),\nn\\
   c_{59}&\to-\frac{1}{4} a_1\mdot a_1 v_1\mdot F_k\mdot v_2 \left(w_2 a_1\mdot F_k\mdot v_1+a_1\mdot k v_1\mdot F_k\mdot v_2\right),\nn\\
   c_{60}&\to\frac{1}{16} \Big(a_1\mdot k \left(a_1\mdot F_k\mdot v_1 \left(\text{tr}\left(F_k\mdot S_1\right)-\frac{2 k\mdot S_1\mdot F_k\mdot v_1}{w_1}\right)-2 a_1\mdot v_2 v_1\mdot F_k\mdot v_2 \text{tr}\left(F_k\mdot S_1\right)\right)\nn\\
   &-2 w_2 a_1\mdot v_2 \text{tr}\left(F_k\mdot S_1\right) a_1\mdot F_k\mdot v_1\Big),\nn\\
   c_{61}&\to\frac{3 i \left( \text{tr}\left(F_k\mdot S_1\right) \left(2 w_2 a_1\mdot v_2 a_1\mdot F_k\mdot v_1-a_1\mdot k \left(a_1\mdot F_k\mdot v_1-2 a_1\mdot v_2 v_1\mdot F_k\mdot v_2\right)\right)\right)}{16}\nn\\
   &+\frac{3 i  \text{tr}\left(F_k\mdot S_1\right) \left(2 a_1\mdot k a_1\mdot F_k\mdot v_1 k\mdot S_1\mdot F_k\mdot v_1\right)}{16 w_1},\nn\\
   c_{62}&\to-\frac{y \left(a_1\mdot k a_1\mdot v_2+a_1\mdot a_1 \left(w_1 y-w_2\right)\right)}{2 w_1^3},c_{64}\to\frac{\left(w_1-2 w_2 y\right) a_1\mdot F_k\mdot v_1-2 y a_1\mdot k v_1\mdot F_k\mdot v_2}{4 w_1^3},\nn\\
   c_{63}&\to\frac{a_1\mdot F_k\mdot v_1 \left(v_1\mdot F_k\mdot v_2 \left(y a_1\mdot k+w_1 a_1\mdot v_2\right)+w_2 y a_1\mdot F_k\mdot v_1\right)}{4 w_1^3},\nn\\
   c_{66}&\to-\frac{v_1\mdot F_k\mdot v_2 \left(2 w_1 a_1\mdot k a_1\mdot v_2+y \left(a_1\mdot k\right){}^2+w_1^2 y a_1\mdot a_1\right)+w_2 a_1\mdot F_k\mdot v_1 \left(y a_1\mdot k+w_1 a_1\mdot v_2\right)}{4 w_1^3},\nn\\
   c_{67}&\to\frac{a_1\mdot F_k\mdot v_1 \left(-\left(v_1\mdot F_k\mdot v_2\right) \left(a_1\mdot k a_1\mdot v_2+w_1 y a_1\mdot a_1\right)-w_2 a_1\mdot v_2 a_1\mdot F_k\mdot v_1\right)}{4 w_1^2},\nn\\
   c_{68}&\to\frac{2 a_1\mdot a_1 \left(2 w_1 y-w_2\right) v_1\mdot F_k\mdot v_2+a_1\mdot k \left(a_1\mdot F_k\mdot v_1+2 a_1\mdot v_2 v_1\mdot F_k\mdot v_2\right)}{4 w_1^2},\nn\\
   c_{69}&\to-\frac{v_1\mdot F_k\mdot v_2 \left(w_2 a_1\mdot F_k\mdot v_1+a_1\mdot k v_1\mdot F_k\mdot v_2\right)}{4 w_1^2},\nn\\
   c_{70}&\to\frac{v_1\mdot F_k\mdot v_2 \left(w_2 a_1\mdot F_k\mdot v_1+a_1\mdot k v_1\mdot F_k\mdot v_2\right)}{2 w_1^2},\nn\\
   c_{71}&\to\frac{v_1\mdot F_k\mdot v_2 \left(\left(a_1\mdot k\right){}^2 a_1\mdot v_2+2 w_1 y a_1\mdot a_1 a_1\mdot k+w_1^2 a_1\mdot a_1 a_1\mdot v_2\right)+w_2 a_1\mdot F_k\mdot v_1 \left(a_1\mdot k a_1\mdot v_2+w_1 y a_1\mdot a_1\right)}{4 w_1^2},\nn\\
   c_{72}&\to-\frac{a_1\mdot F_k\mdot v_1 \left(4 y a_1\mdot k a_1\mdot v_2+a_1\mdot a_1 \left(w_1-2 w_2 y\right)\right)}{4 w_1},\nn\\
   c_{73}&\to-\frac{a_1\mdot F_k\mdot v_1 \left(w_2 a_1\mdot F_k\mdot v_1+a_1\mdot k v_1\mdot F_k\mdot v_2\right)}{4 w_1},\nn\\
   c_{74}&\to\frac{a_1\mdot F_k\mdot v_1 \left(\left(w_1-2 w_2 y\right) a_1\mdot F_k\mdot v_1-2 y a_1\mdot k v_1\mdot F_k\mdot v_2\right)}{2 w_1},\nn\\
   c_{76}&\to\frac{3 i a_1\mdot F_k\mdot v_1 \left(w_2 a_1\mdot F_k\mdot v_1+a_1\mdot k v_1\mdot F_k\mdot v_2\right)}{4 w_1},\nn\\
   c_{80}&\to\frac{2 v_1\mdot F_k\mdot v_2 \text{tr}\left(F_k\mdot S_1\right) \left(w_1 a_1\mdot v_2-y a_1\mdot k\right)+a_1\mdot F_k\mdot v_1 \left(\left(w_1-2 w_2 y\right) \text{tr}\left(F_k\mdot S_1\right)+2 k\mdot S_1\mdot F_k\mdot v_1\right)}{16 w_1},\nn\\
   c_{81}&\to-\frac{3 i \text{tr}\left(F_k\mdot S_1\right) \left(\left(w_1-2 w_2 y\right) a_1\mdot F_k\mdot v_1-2 y a_1\mdot k v_1\mdot F_k\mdot v_2\right)}{16 w_1},\nn\\
   c_{82}&\to\frac{\left(\left(a_1\mdot k\right){}^2+w_1^2 a_1\mdot a_1\right) \left(v_1\mdot F_k\mdot v_2\right){}^2+2 w_2 a_1\mdot k v_1\mdot F_k\mdot v_2 a_1\mdot F_k\mdot v_1+w_2^2 \left(a_1\mdot F_k\mdot v_1\right){}^2}{4 w_1^2}\, .
\end{align}

\bibliographystyle{JHEP}
\bibliography{ScatEq}

\end{document}